\documentclass[fleqn,10pt]{wlscirep}
\usepackage{url}
\usepackage{graphicx}
\usepackage{graphicx}
\usepackage{dcolumn}
\usepackage{bm}
\usepackage{lipsum}
\usepackage[maxfloats=256]{morefloats}
\usepackage{lineno}
\begin{document}
\title{Market states: A new understanding}
\author[1,*]{Hirdesh K. Pharasi}
\author[2]{Eduard Seligman}
\author[1,3]{Thomas H. Seligman}
\affil[1]{Instituto de Ciencias F\'{i}sicas, Universidad Nacional Aut\'{o}noma de M\'{e}xico, Cuernavaca-62210, M\'{e}xico}
\affil[2]{Swiss Air Trainer S.A. Payerne, Vaud, Switzerland}
\affil[3]{Centro Internacional de Ciencias, Cuernavaca-62210, M\'{e}xico}
\affil[*]{hirdeshpharasi@gmail.com}
\date{\today}
\begin{abstract}
We present the clustering analysis of the financial markets of S\&P 500 (USA) and Nikkei 225 (JPN) markets over a period of 2006-2019 as an example of a complex system. We investigate the statistical properties of correlation matrices constructed from the sliding epochs. The correlation matrices can be classified into different clusters, named as ``market states" based on the similarity of correlation structures. We cluster the S\&P 500 market into four and Nikkei 225 into six market states by optimizing the value of intracluster distances. The market shows transitions between these market states and the statistical properties of the transitions to critical market states can indicate likely precursors to the catastrophic events. We also analyze the same clustering technique on surrogate data constructed from average correlations of market states and the fluctuations arise due to the white noise of short time series. We use the correlated Wishart orthogonal ensemble for the construction of surrogate data whose average correlation equals the average of the real data. 
\end{abstract}


\maketitle
\section*{Introduction}
Some time ago a proposal to consider correlation matrices to define market states was put forward and found some resonance~\cite{Munnix_2012}. A noise suppression technique was applied directly to the correlation matrices~\cite{Guhr_2003} and thereafter a clustering tree was established and eight market states were established with a certain degree of arbitrariness. 
Since then ideas from this paper have resonated inside and outside our group. Alternative techniques to handle non-stationary time series through spectral properties have been developed in~\cite{schafer_2013,prosen_2014,chetalova_2015}. Other approaches to behave at critical or catastrophic moments have been put forward by various authors~\cite{gidea_2020,rings_2019,muhlbacher_2018,vyas_2018,jurczyk_2017}. More recently two of us participated in a proposal~\cite{Pharasi_2018} to improve the choice of cluster by simultaneously optimizing the clustering process and the noise suppression parameter. We expanded the analysis from the original study of S\&P 500 market by including the Nikkei 225 market.  The results were essentially a dynamics consistent with the master equation and transitions to the highest correlation cluster mainly from the previous one, which thus allowed to determine possible precursors. The study had the disadvantage of counting to a large extent old data of last century. We here propose to refine the methods further and to improve the criteria to reduce the number of precursors in a given time period and also to get a better understanding of the transition dynamics between states. We also address a basic weakness of the previous studies that implied roughly that the average correlation or equivalently the largest eigenvalues of the correlation matrix, largely determine to which cluster a given correlation matrix belongs. In the Japanese market we found an interesting essential non-linear evolution, as well as a smaller basis of precursor states, thus reducing the risk base. Also we develop a technique of surrogate data based on correlated orthogonal Wishart ensembles (CWOE) draping white noise around the average correlation of each state. This gives a strong indication as to the interrelation of correlation and noise that we see. 

In the present work we shall make actual connection to this matrix by noise to the correlation matrices pertaining to the market states by forming a correlated Wishart ensemble~\cite{vinayak_2014,pandey_2010,Burda_pre2005} from the average correlation matrix representing each market state. We shall find that the clusters obtained for each market state are reasonably related to the original clusters if we chose the same time frame for the added noise. 
We extend the analysis to the Japanese market to confirm our findings as far as the validity of the clustering is concerned and we finally present the result, that a highly turbulent market-phase started near the Lehman Brother crash and terminated around 2016. The latter result is more clear for the US market than for the Japanese market. This likely results from two facts: The data basis (number of stocks) is smaller for the Japanese market and the Japanese market has six states. Both properties make more noise. On the other hand the fact that two intermediate states of the Japanese market display practically the same average correlation eliminates the disturbing fact mentions in \cite{Munnix_2012} that the average correlation actually dominates the definition of market states to large extent. We thus give a solid base to the concept of market states, as we show, how the clusters, i.e., the market states are made up from average correlations and noise.
\section*{Methodology and results}
We begin with the identification of market states as a clusters of similar correlation matrices. We have used the adjusted daily closure prices of the stocks making up the two indices S\&P 500 of US market (USA) and Nikkei 225 of Japanese market (JPN). We have considered $N = 350$ stocks of the S\&P 500 index and $N = 156$ stocks of the Nikkei 225 index traded in the 14-year period from January 2006 to December 2019 which correspond to $T=3523$ and $T = 3459$ trading days, respectively. Here we include only those stocks which were present for the entire duration of 14 years. We have also shown in the supplementary material of $N = 368$ stocks of the S\&P 500 index and $N = 173$ stocks of the Nikkei 225 index for 13-year period from January 2006 to December 2018 which correspond to $T=3270$ and $T = 3219$ trading days, respectively. The list of the stock considered for the analysis are also listed in the supplementary material.
\begin{figure}[b!]
\centering
\includegraphics[width=0.99\linewidth]{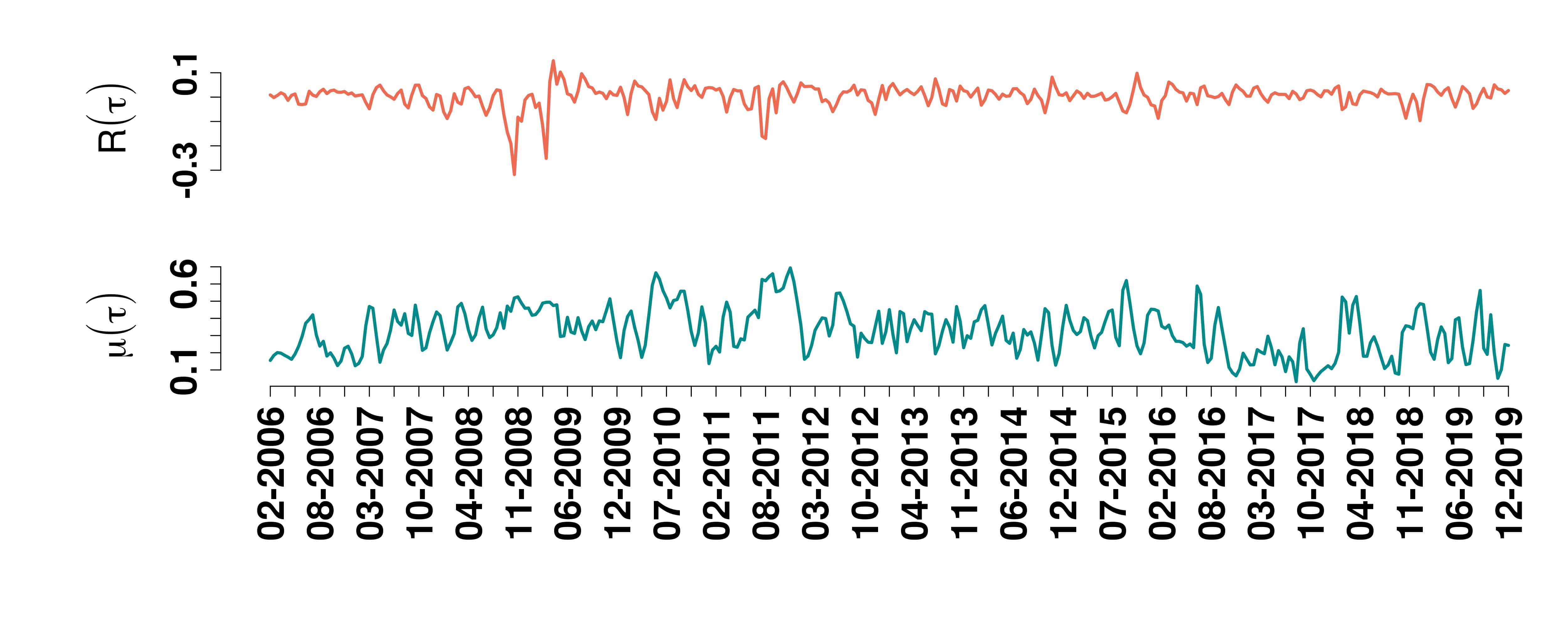}\llap{\parbox[b]{6.6in}{\textbf{{\Large (a)}}\\\rule{0ex}{2.6in}}}
\includegraphics[width=0.99\linewidth]{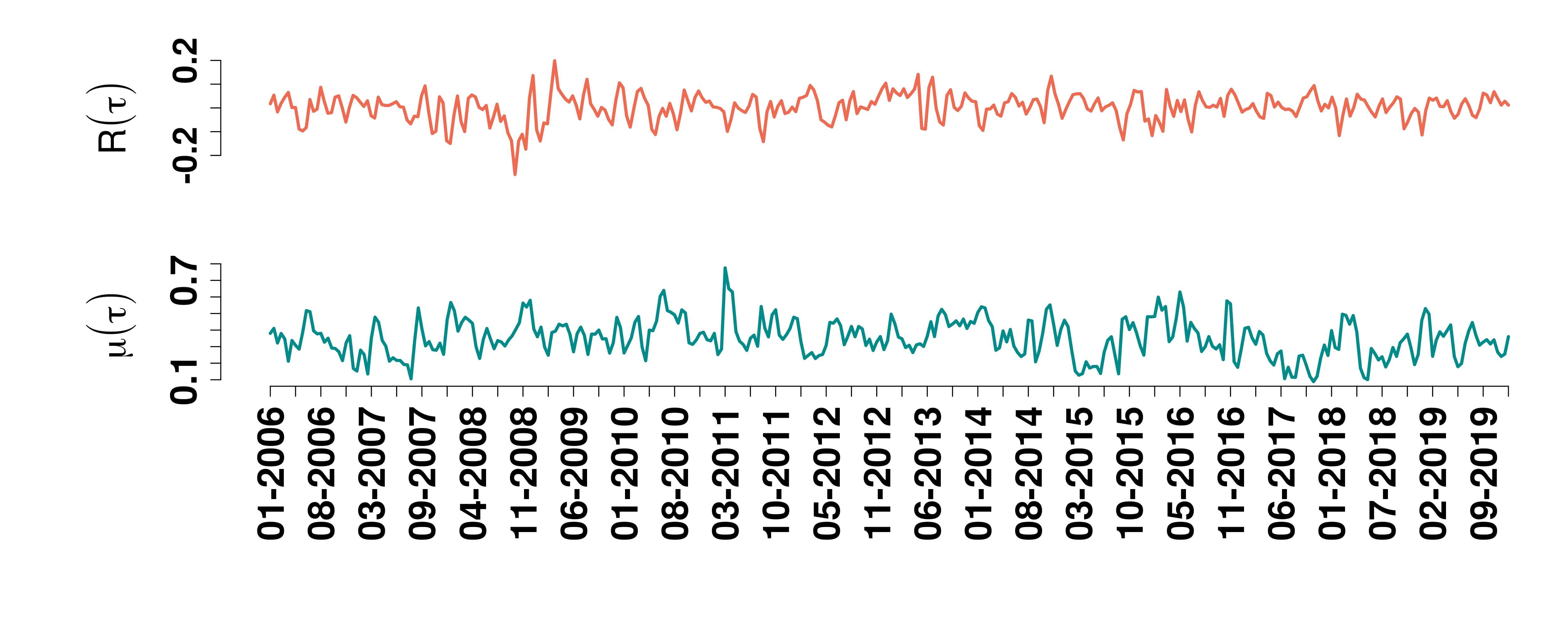}\llap{\parbox[b]{6.6in}{\textbf{{\Large (b)}}\\\rule{0ex}{2.6in}}}
\caption{Temporal evolution of (\textbf{a}) S\&P 500 and (\textbf{b}) Nikkei 225 markets over a period of 2006-2019. The returns  of the two market indices $R(\tau)$ as well as the corresponding average correlations of the stock returns $\mu(\tau)$ are shown in the plots.}\label{fig:corr_minlambda}
\end{figure} 
\begin{figure}[ht!]
	\centering
	\includegraphics[width=0.45\linewidth]{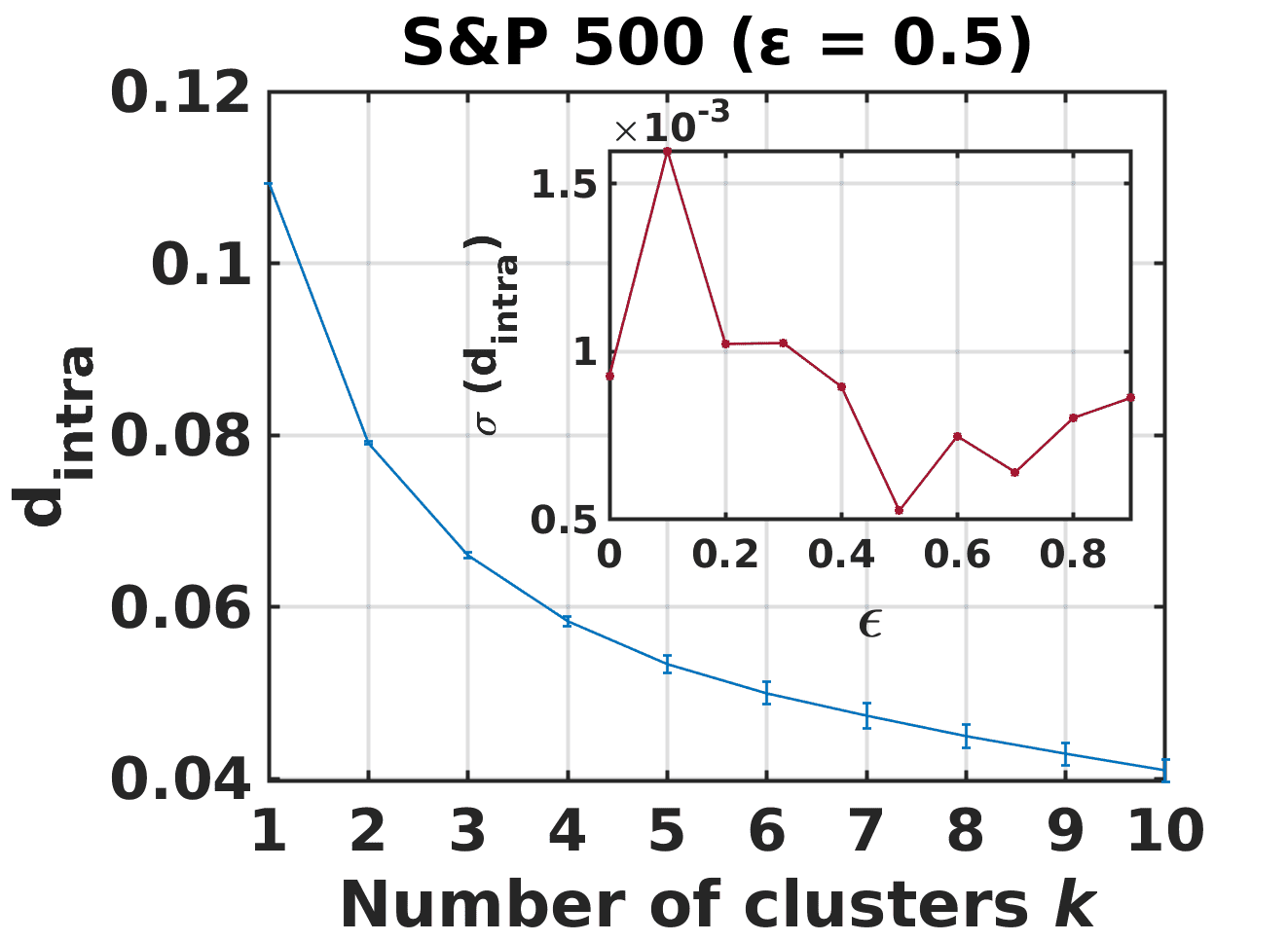}\llap{\parbox[b]{3.2in}{\textbf{{\Large (a)}}\\\rule{0ex}{2.2in}}}
	\includegraphics[width=0.45\linewidth]{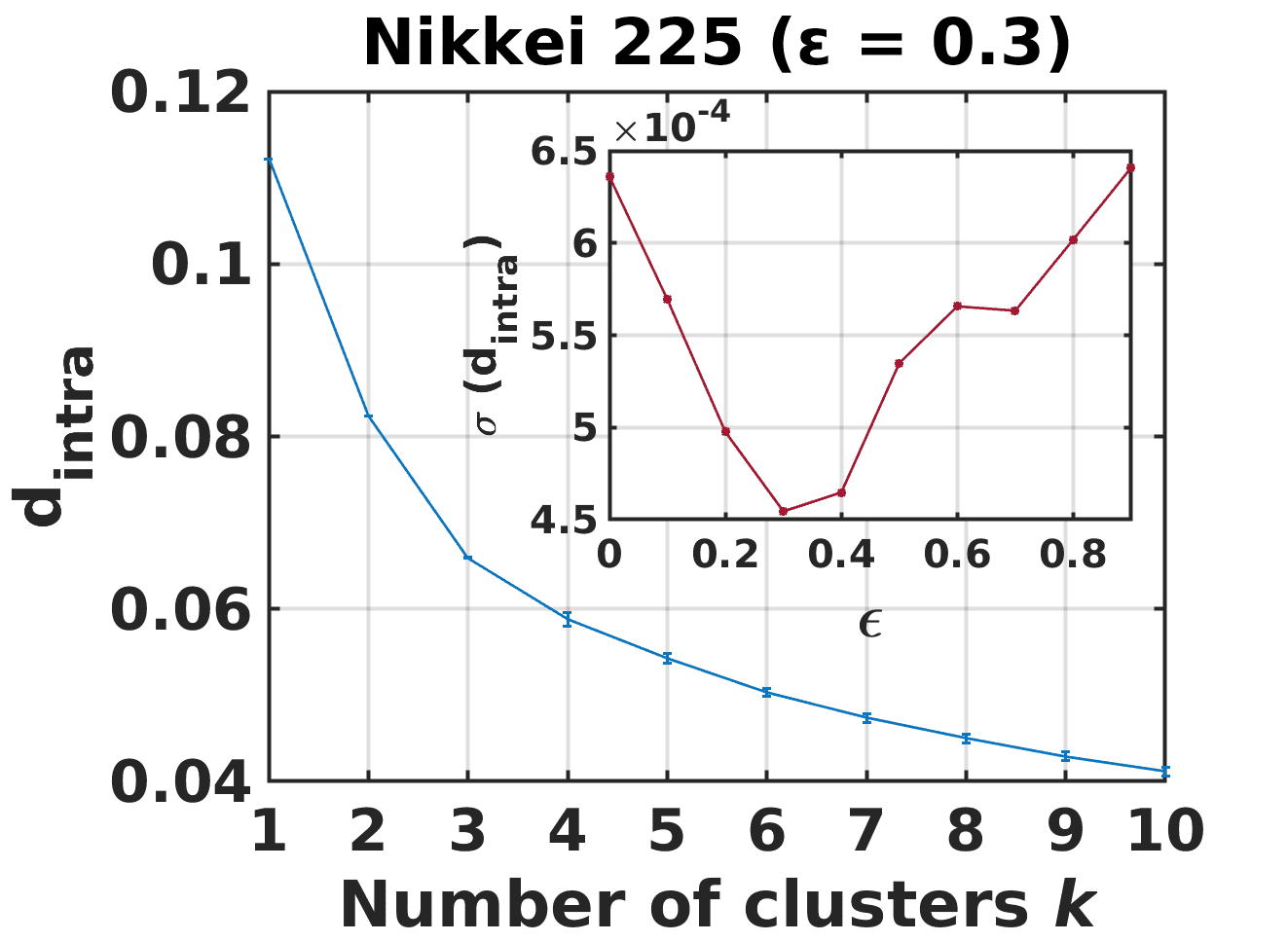}\llap{\parbox[b]{3.2in}{\textbf{{\Large (b)}}\\\rule{0ex}{2.2in}}}\\
	\includegraphics[width=0.45\linewidth]{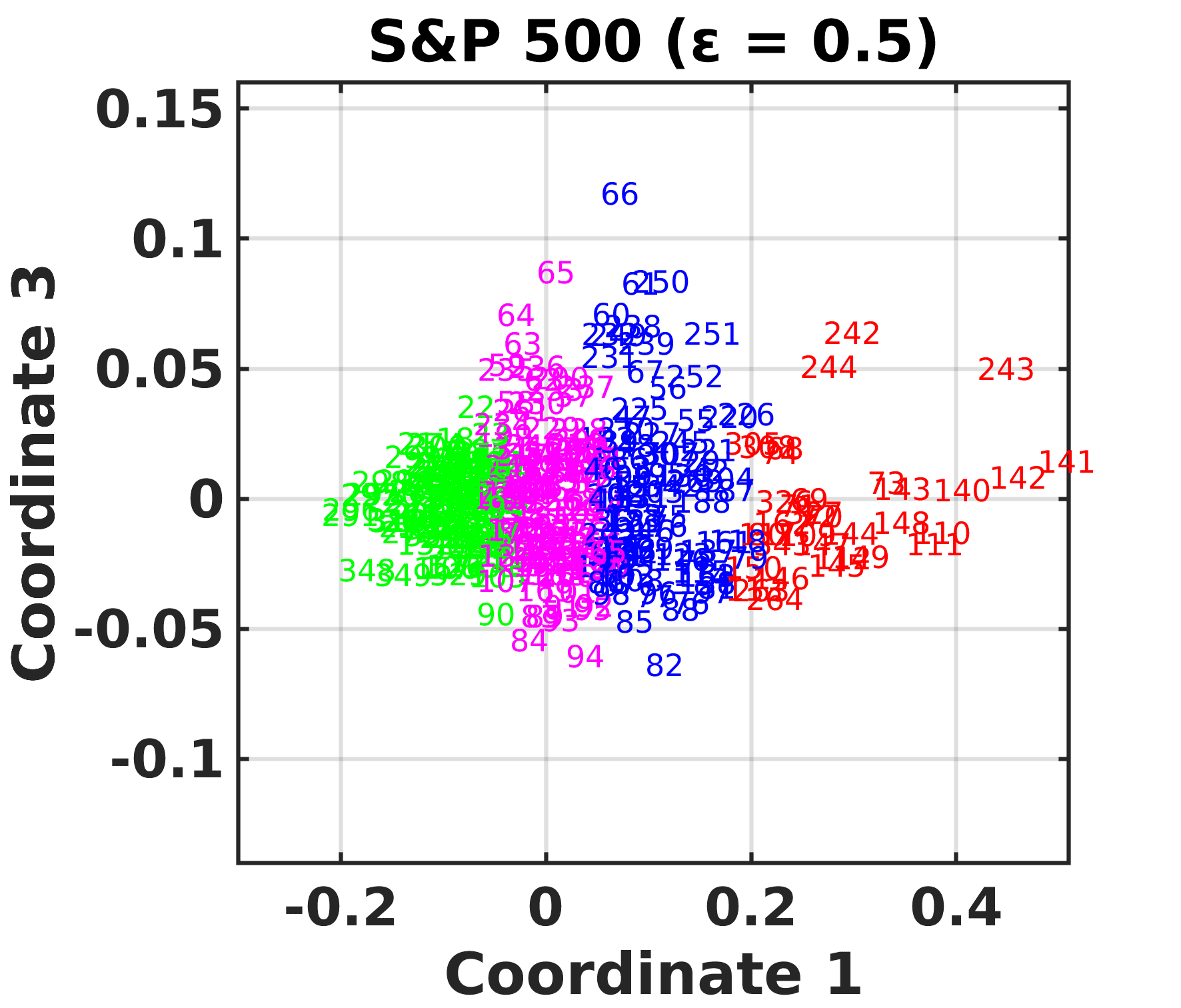}\llap{\parbox[b]{3.2in}{\textbf{{\Large (c)}}\\\rule{0ex}{2.5in}}}	
	\includegraphics[width=0.45\linewidth]{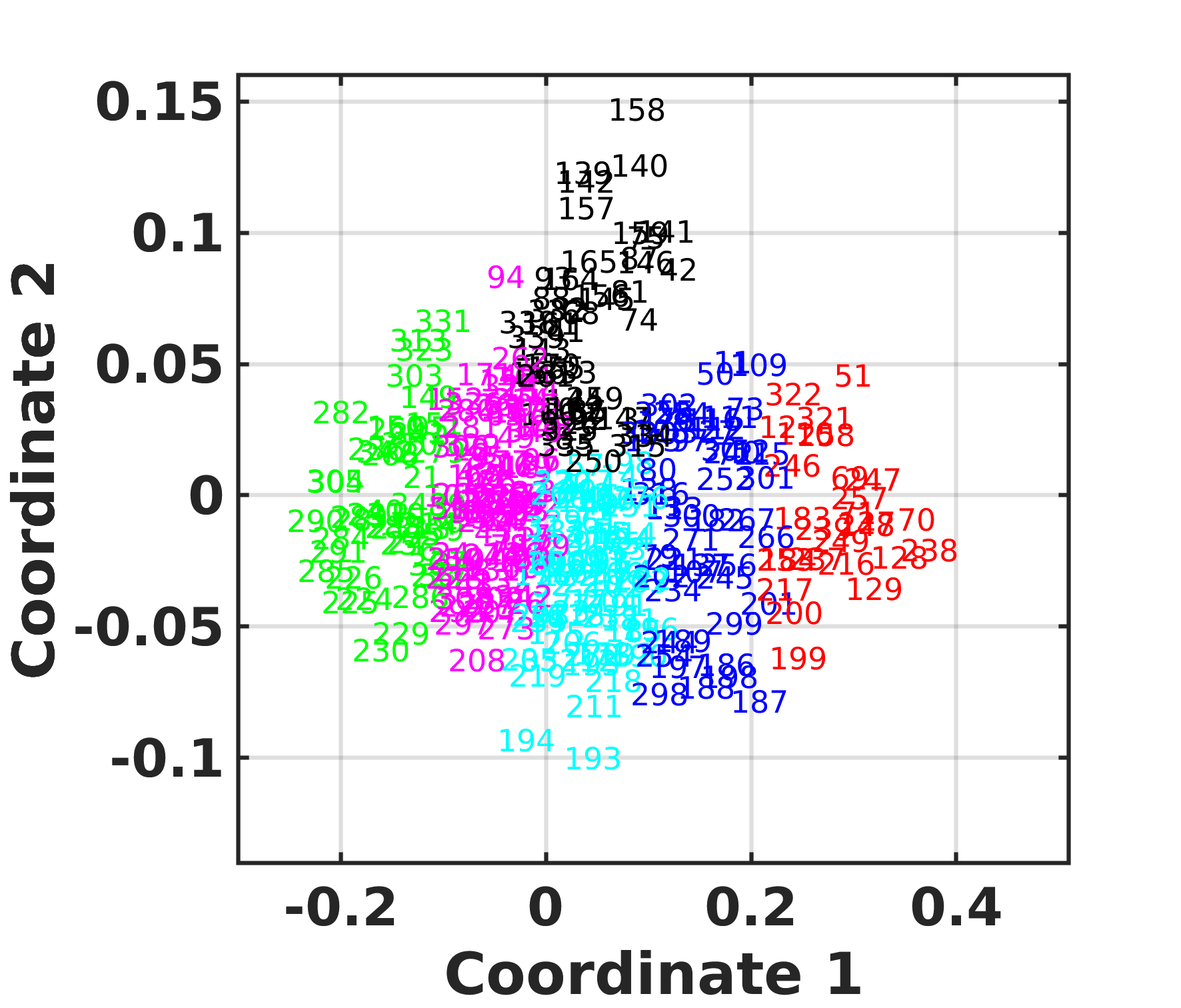}\llap{\parbox[b]{3.2in}{\textbf{{\Large (d)}}\\\rule{0ex}{2.5in}}}	
	\caption{Classification of market states based on optimal intra cluster distances for S\&P 500 and Nikkei 225. The measure of the intra cluster distance $d_{intra}$ calculated for different number of clusters is shown in (a) and (b) for S\&P 500 and Nikkei 225, respectively. The three dimensional (3D) $k$-means clustering is performed on $351$ noise-suppressed correlation frames of (c) S\&P 500 and $344$ noise-suppressed correlation frames of (d) Nikkei 225. Here, we use best 2D projection of 3D plots. The errorbar in the plot shows the deviations of the intra cluster distances  calculated for 1000 different initial conditions. The plots  show the minima of standard deviations at $k=4$ for S\&P 500 and $k=6$ for Nikkei 225, respectively. Inset: Plots of $d_{intra}$ measured for differnet noise-suppression parameters $\epsilon$ and shows minima at $\epsilon=0.5$ and $\epsilon=0.3$ for S\&P 500 and Nikkei 225, respectively.}\label{intracluster}
\end{figure}
\begin{figure}[!t]
	\centering
		\includegraphics[width=0.99\textwidth]{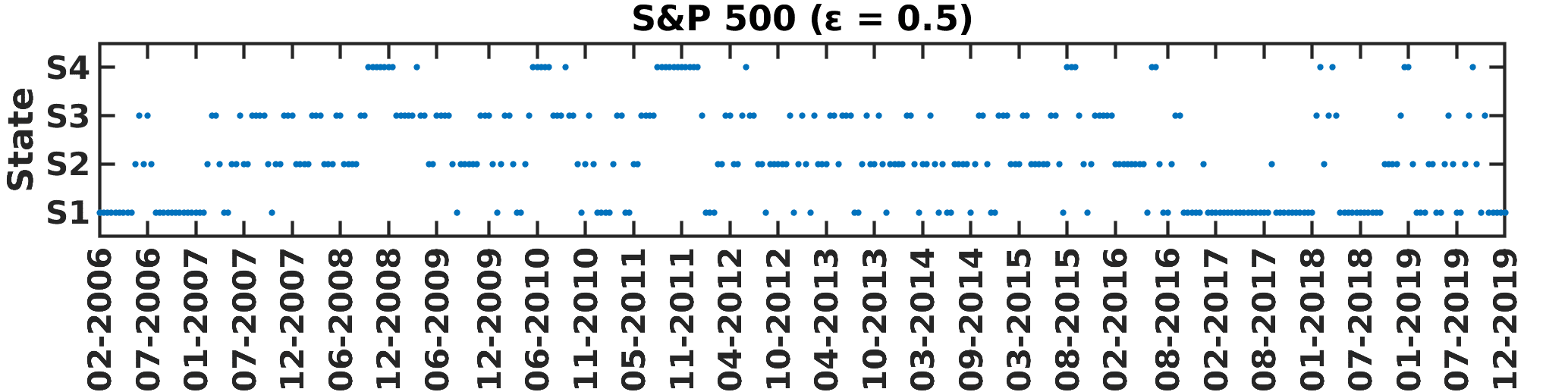}\llap{\parbox[b]{6.9in}{\textbf{{\Large (a)}}\\\rule{0ex}{1.7in}}}
		\includegraphics[width=0.99\textwidth]{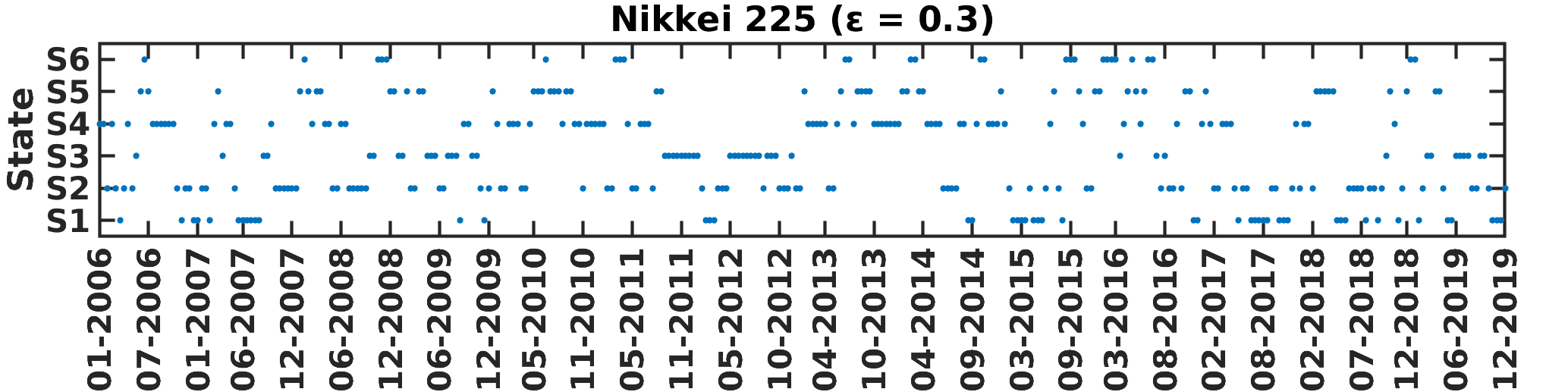}\llap{\parbox[b]{6.9in}{\textbf{{\Large (b)}}\\\rule{0ex}{1.7in}}}
	\caption{Plot shows dynamics of the S\&P 500 and Nikkei 225 markets. (a) Evolution of S\&P 500 market through the transitions between four different characterized states $S1, S2, S3$ and $S4$ over the period of $2006-2019$. (b) Evolution of Nikkei 225 market though the transitions between six different characterized states $S1, S2, S3$, $S4$, $S5$ and $S6$ for the period of $2006-2019$. US market, relative to Japanese market, is relative calm after 2016 and stays more in lower states. }\label{MS_evolution}
\end{figure}
We construct a Pearson cross-correlation matrix (equal-time) of the returns $r_i(\tau)$ of stock $i$:   $C_{ij}(\tau) = ({\langle r_i r_j \rangle - \langle r_i \rangle \langle r_j \rangle})/{\sigma_i \sigma_j},$ where the epoch average $\langle \dots \rangle $ and the standard deviations $\sigma $ are computed over that epoch of size $20$ days with $i,j=1,...,N$ and $\tau$ is the end date of the epoch using daily returns. Note that we do not compensate for weekends or holidays! However, the short time series the correlation matrices become highly singular~\cite{Potters_prl_1999,Plerou_prl_1999}. We use the power map method~\cite{Guhr_2003,vinayak_2014,Pharasi_2018,Pharasi_2019}, for noise-suppression. In this method, a nonlinear distortion is given to each cross-correlation coefficient within an epoch by: $C_{ij} = sign(C_{ij})|C_{ij}|^{1+\epsilon}$ , where $\epsilon \in (0,1)$ is the noise-suppression parameter. We then study the evolution of the cross-correlation structures $C(\tau)$ of returns for epochs of $20$ days with 10-day overlap.

Figs.~\ref{fig:corr_minlambda} (a) and (b) show the plots of the index returns $R(\tau)$ and mean market correlation $\mu(\tau)$ for S\&P 500 and Nikkei 225, respectively. The rather different behavior of the two markets, which we will elaborate on later, already show in these simple measures.

\begin{center}
	\begin{figure*}[ht!]
	\centering
	\includegraphics[width=6cm]{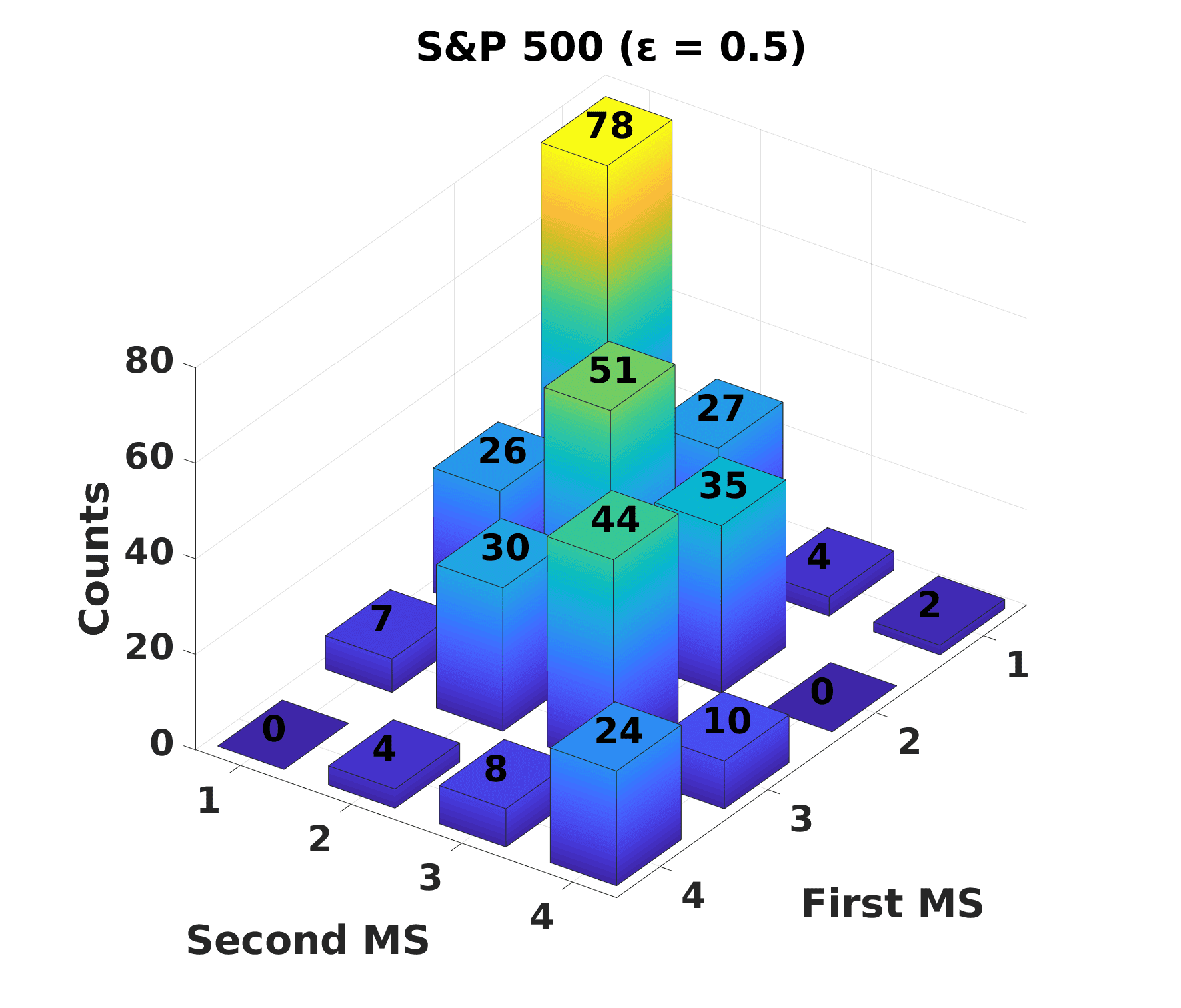}\llap{\parbox[b]{2.4in}{\textbf{{\Large (a)}}\\\rule{0ex}{1.6in}}}
	\includegraphics[width=6cm]{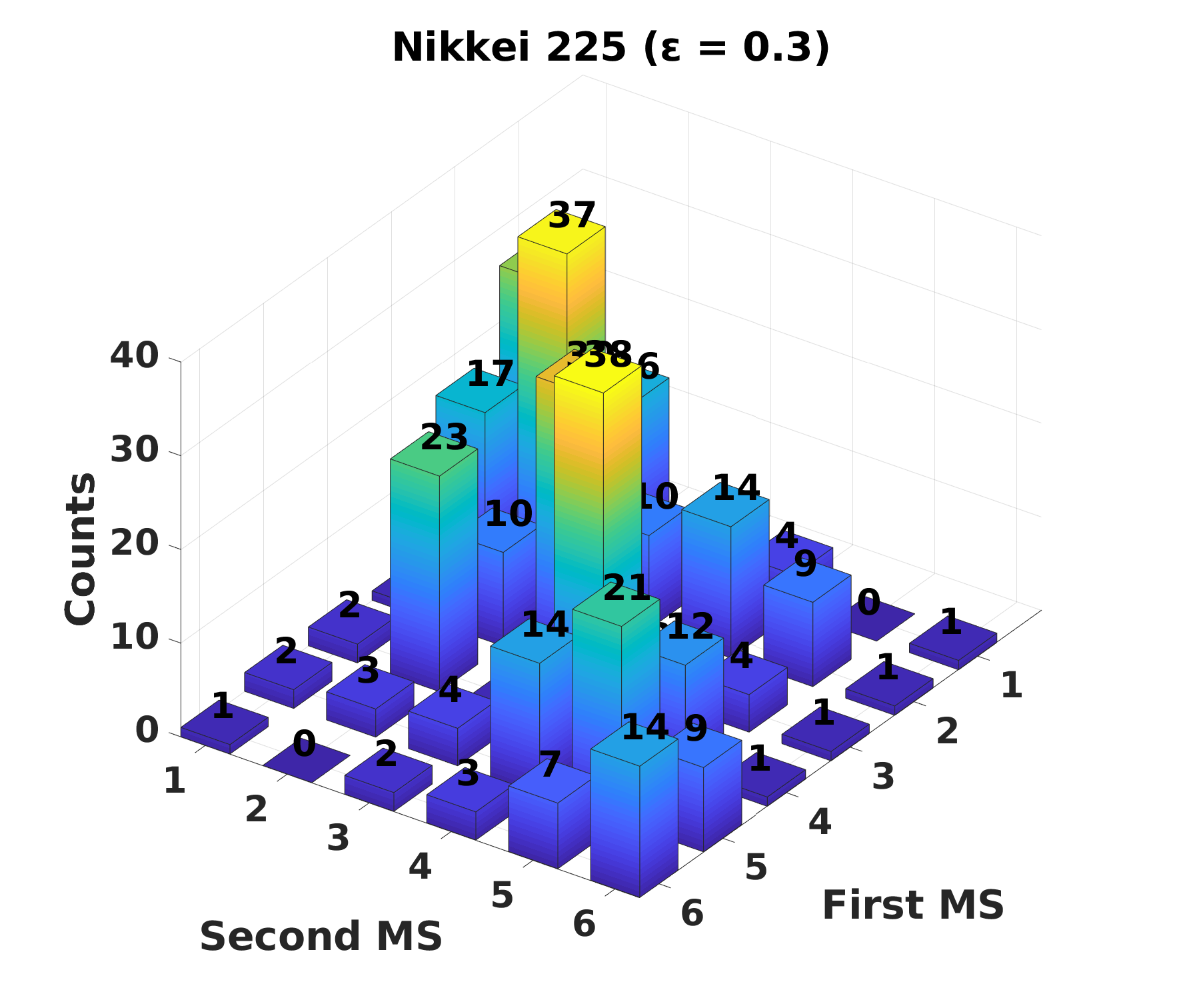}\llap{\parbox[b]{2.4in}{\textbf{{\Large (b)}}\\\rule{0ex}{1.6in}}}	\\
	\includegraphics[width=7cm]{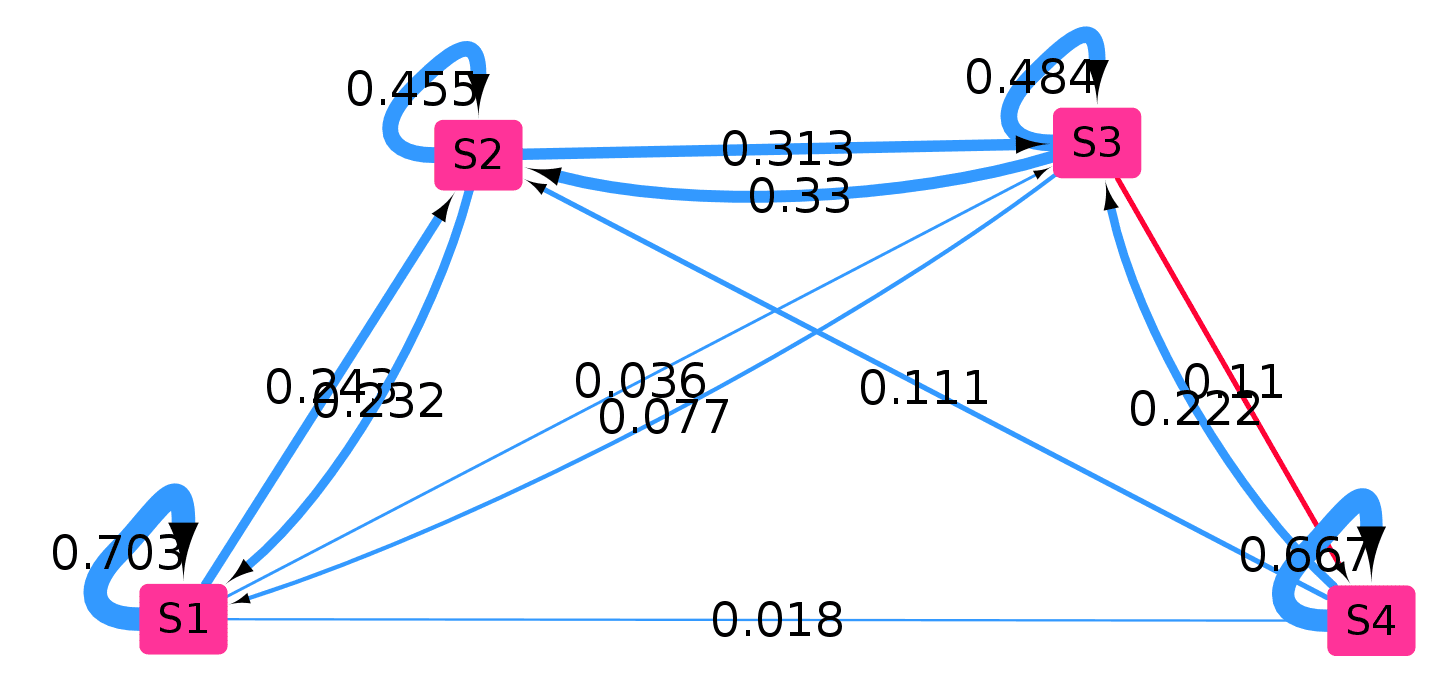}\llap{\parbox[b]{2.6in}{\textbf{{\Large (c)}}\\\rule{0ex}{1.2in}}}
	\includegraphics[width=7cm]{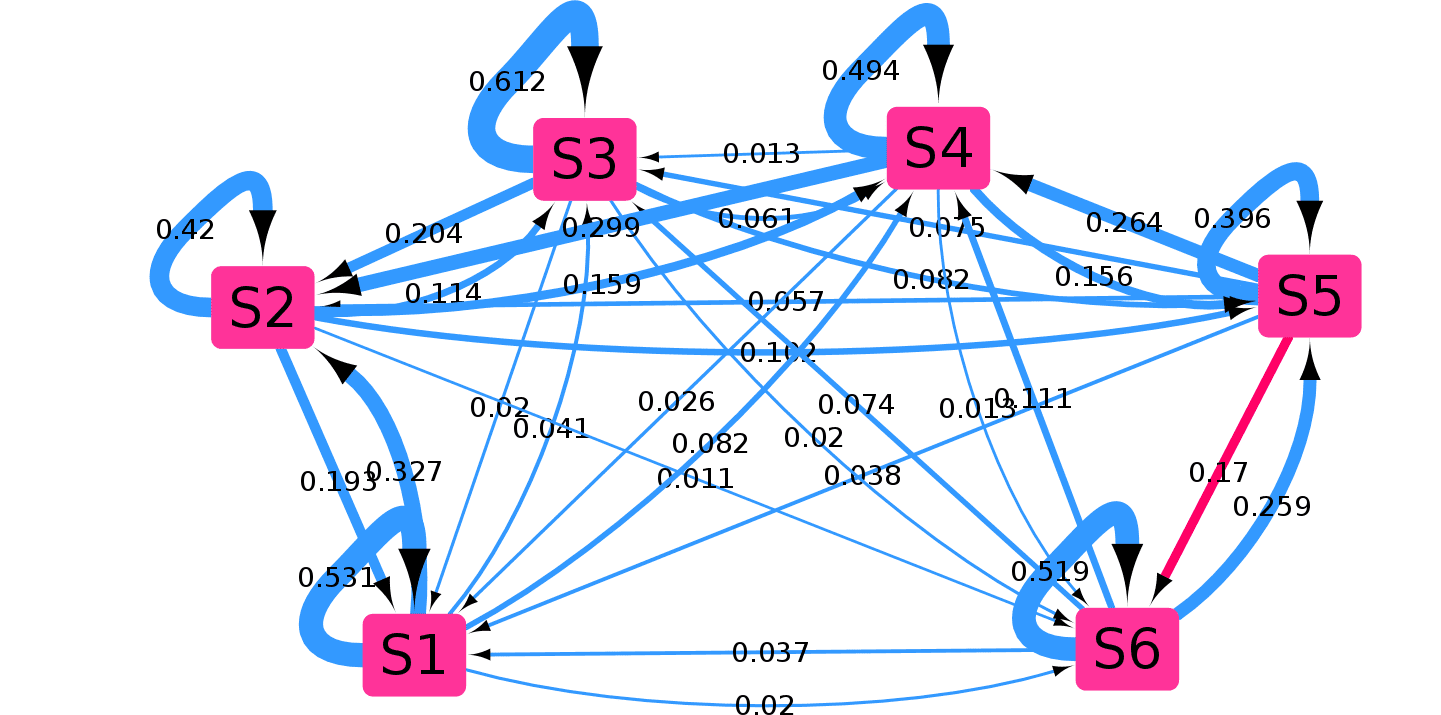}\llap{\parbox[b]{2.6in}{\textbf{{\Large (d)}}\\\rule{0ex}{1.2in}}}	
	\caption{Bar plots show the transition counts (frequencies) of paired market states (MS) for (a) S\&P 500 and (b) Nikkei 225 markets, respectively. The market show back and forth transitions between these states. Sometime the market remain in a particular state for a long time and sometime it jumps shortly to another state and jumps back or evolve further. Transitions to the nearby state are high probable. The networks plot of transition probabilities (see, Tables S1 and Tables S2) between different states of S\&P 500 and Nikkei 225 are shown in (c) and (d) respectively. The probability of market state transition of  $S3$ to $S4$ is $11\%$ for S\&P 500 market, and similarly, for Nikkei 225, the probability of market state transition of  $S5$ to $S6$ is $17\%$.}\label{transition_probabilities}
	\end{figure*}
\end{center}
\begin{figure*}[h!]
\centering
\includegraphics[width=3.6cm]{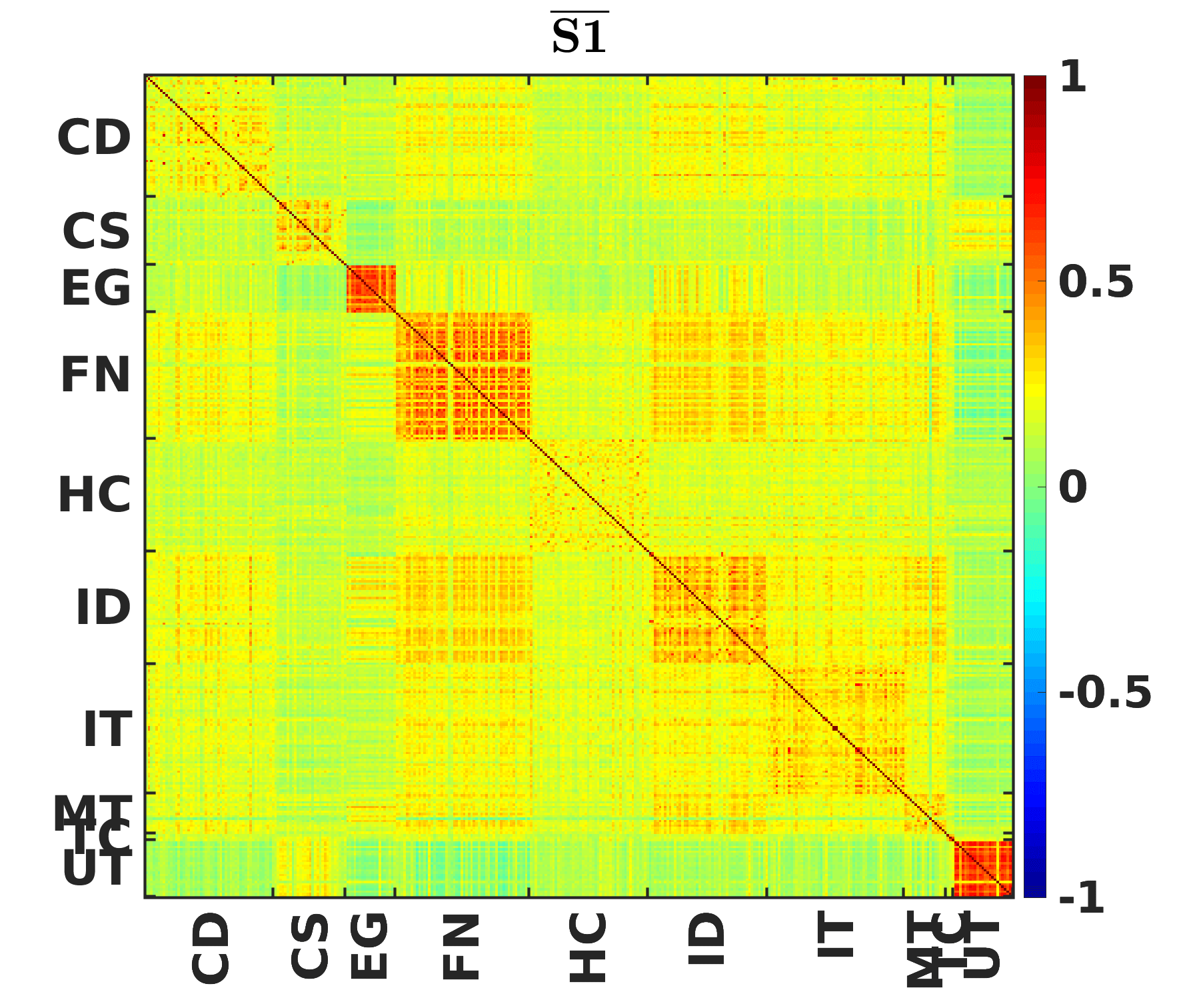}\llap{\parbox[b]{1.5in}{\textbf{{\large (a)}}\\\rule{0ex}{1.1in}}}
\includegraphics[width=3.6cm]{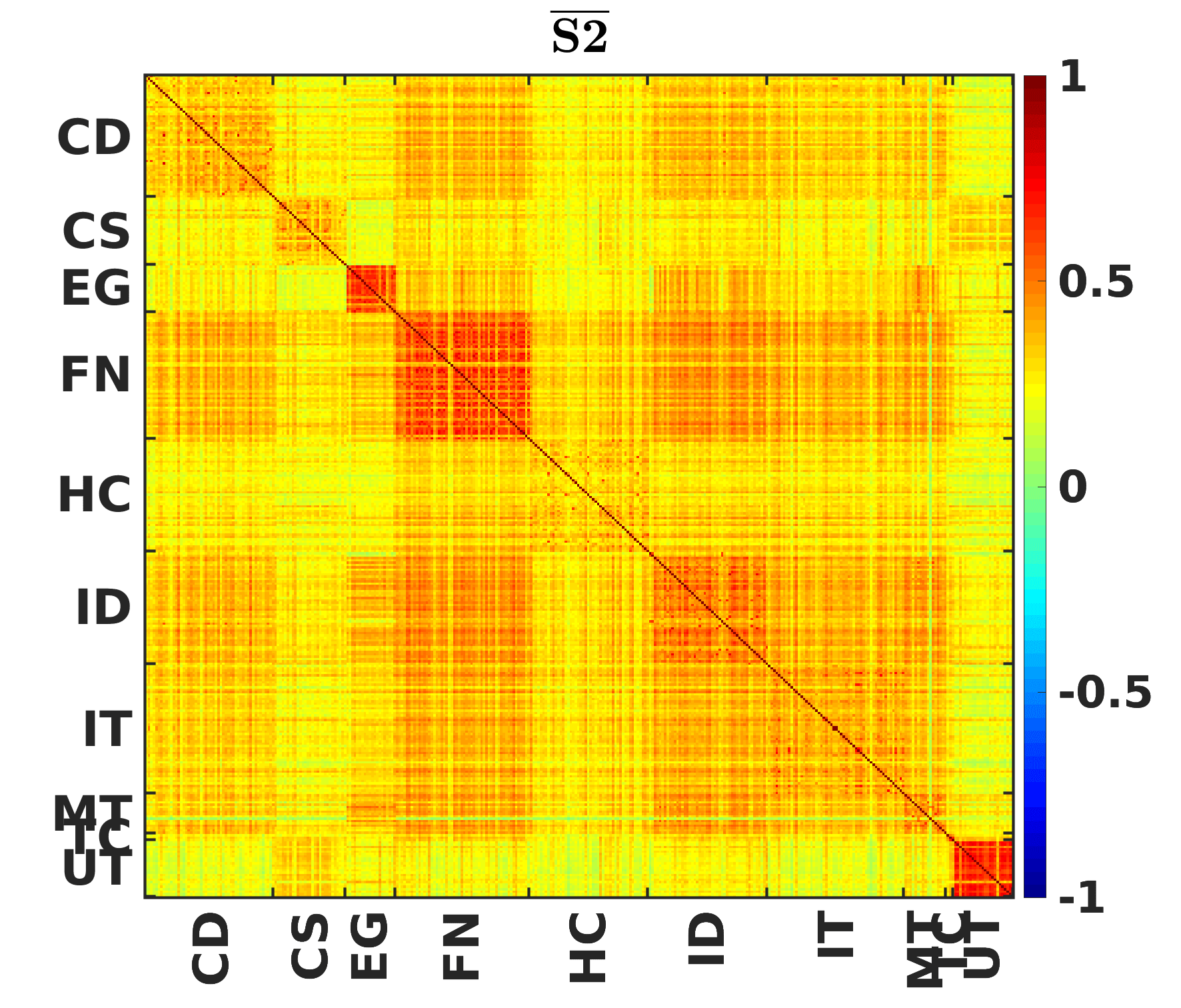}\llap{\parbox[b]{1.5in}{\textbf{{\large (b)}}\\\rule{0ex}{1.1in}}}
\includegraphics[width=3.6cm]{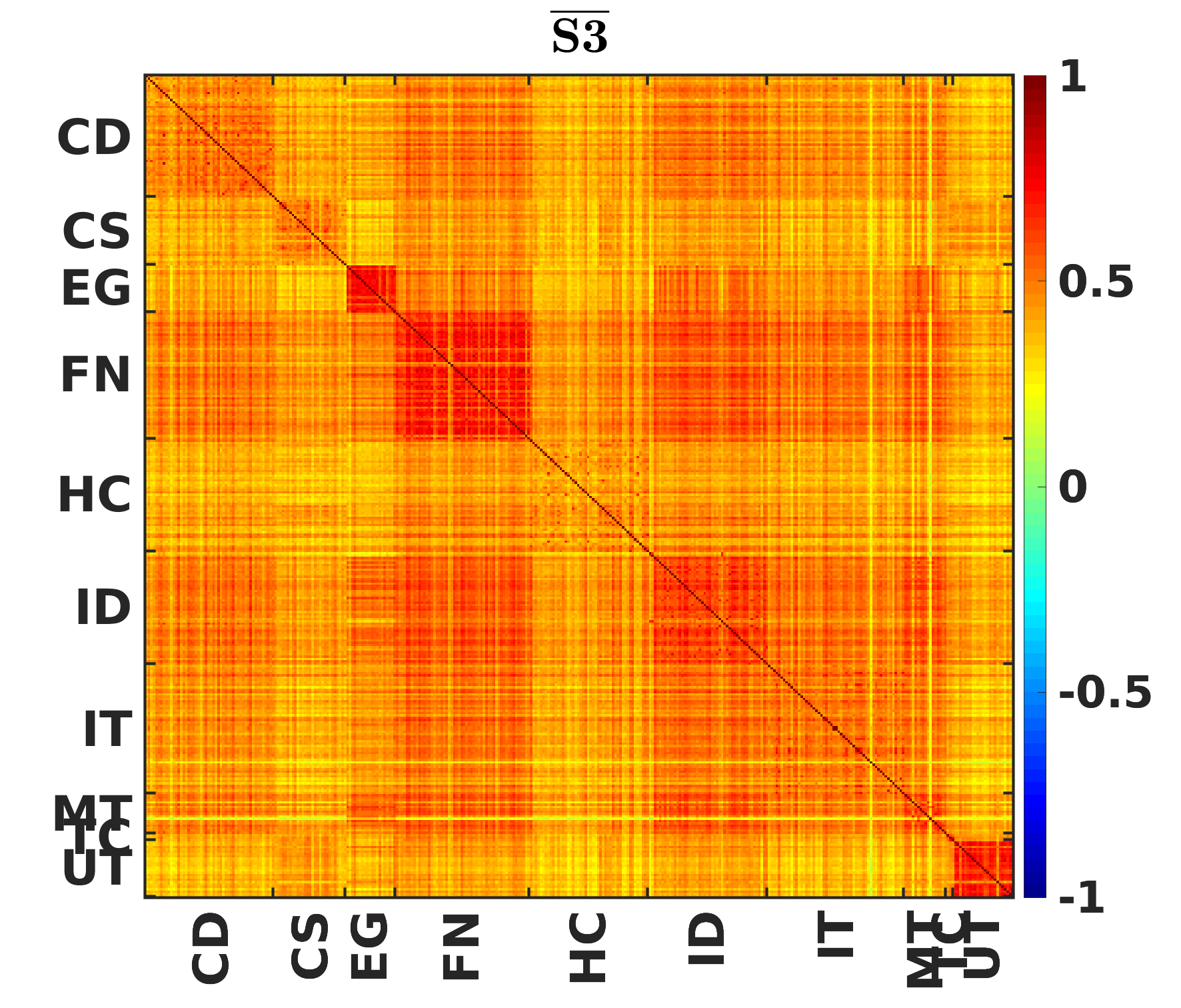}\llap{\parbox[b]{1.5in}{\textbf{{\large (c)}}\\\rule{0ex}{1.1in}}}
\includegraphics[width=3.6cm]{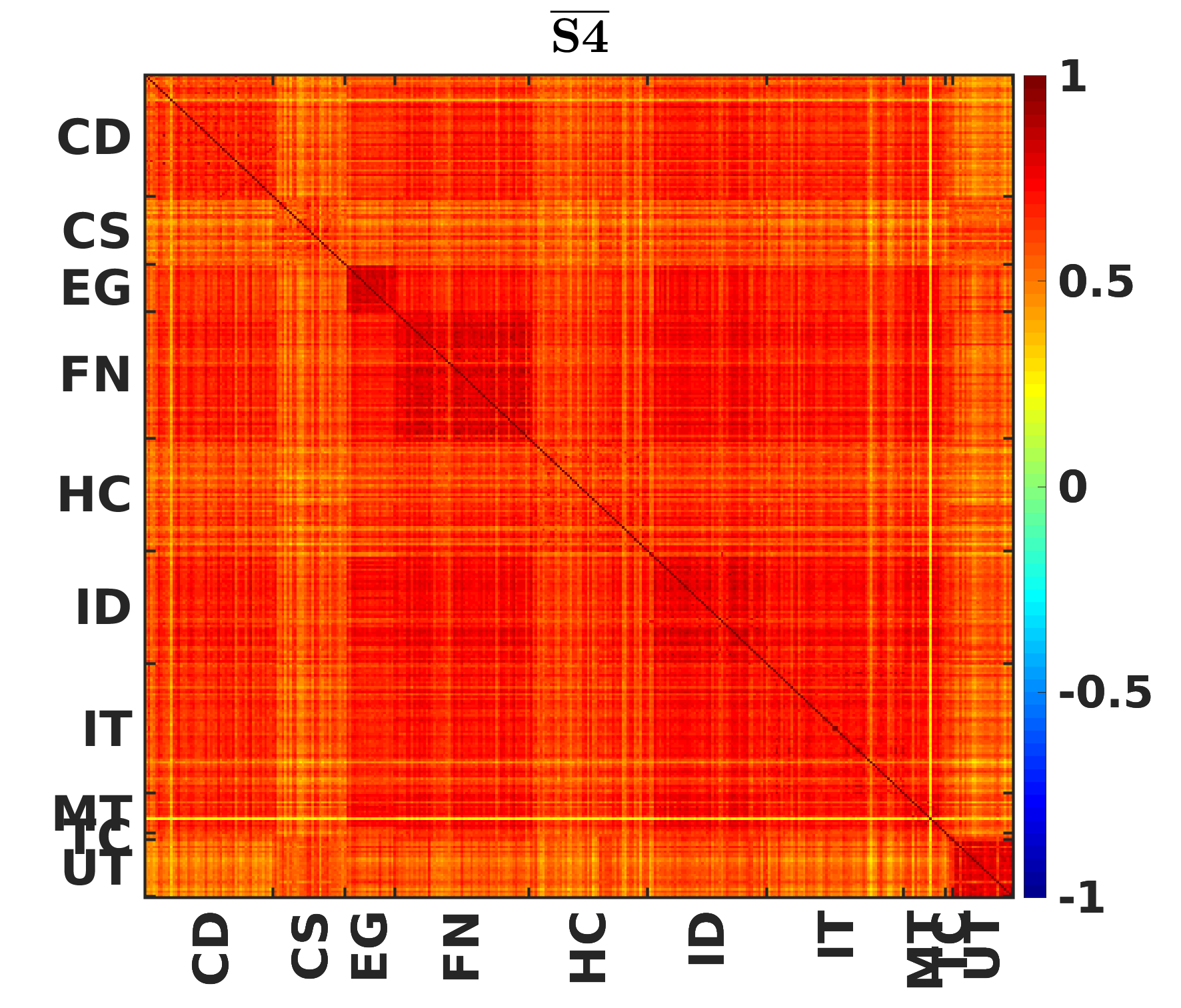}\llap{\parbox[b]{1.5in}{\textbf{{\large (d)}}\\\rule{0ex}{1.1in}}}
\includegraphics[width=4cm]{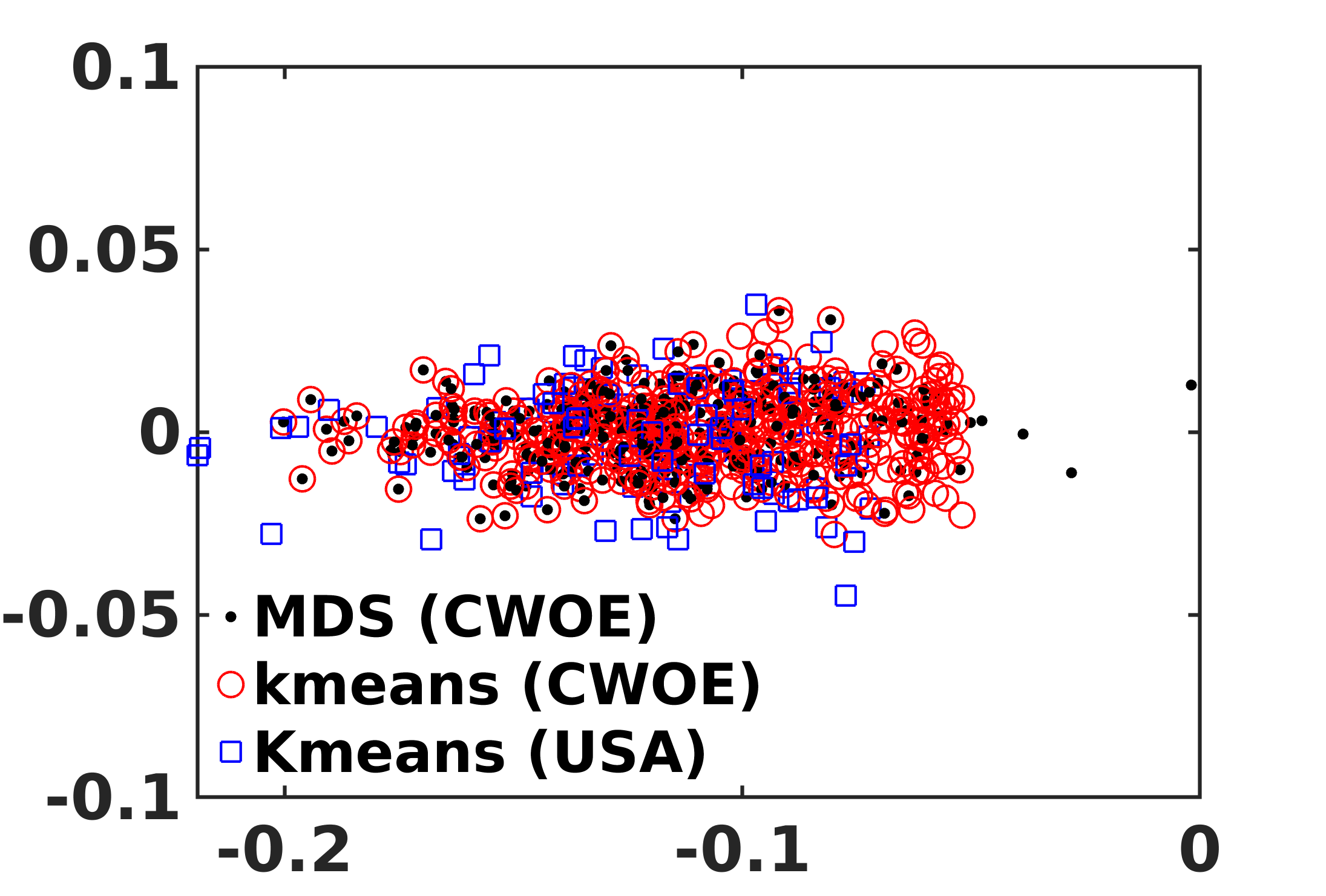}\llap{\parbox[b]{1.58in}{\textbf{{\large (e)}}\\\rule{0ex}{1.1in}}}
\includegraphics[width=4cm]{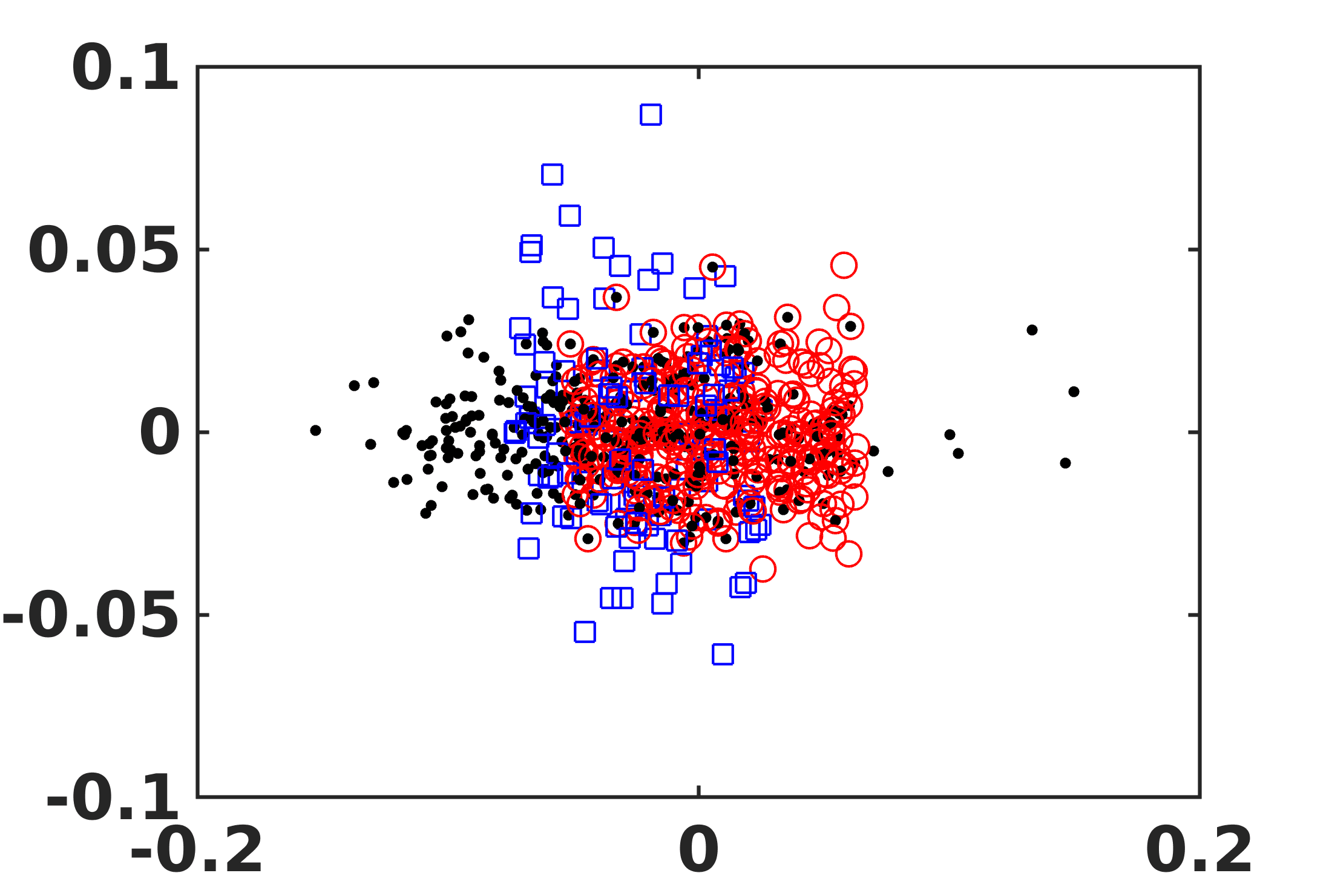}\llap{\parbox[b]{1.58in}{\textbf{{\large (f)}}\\\rule{0ex}{1.1in}}}
\includegraphics[width=4cm]{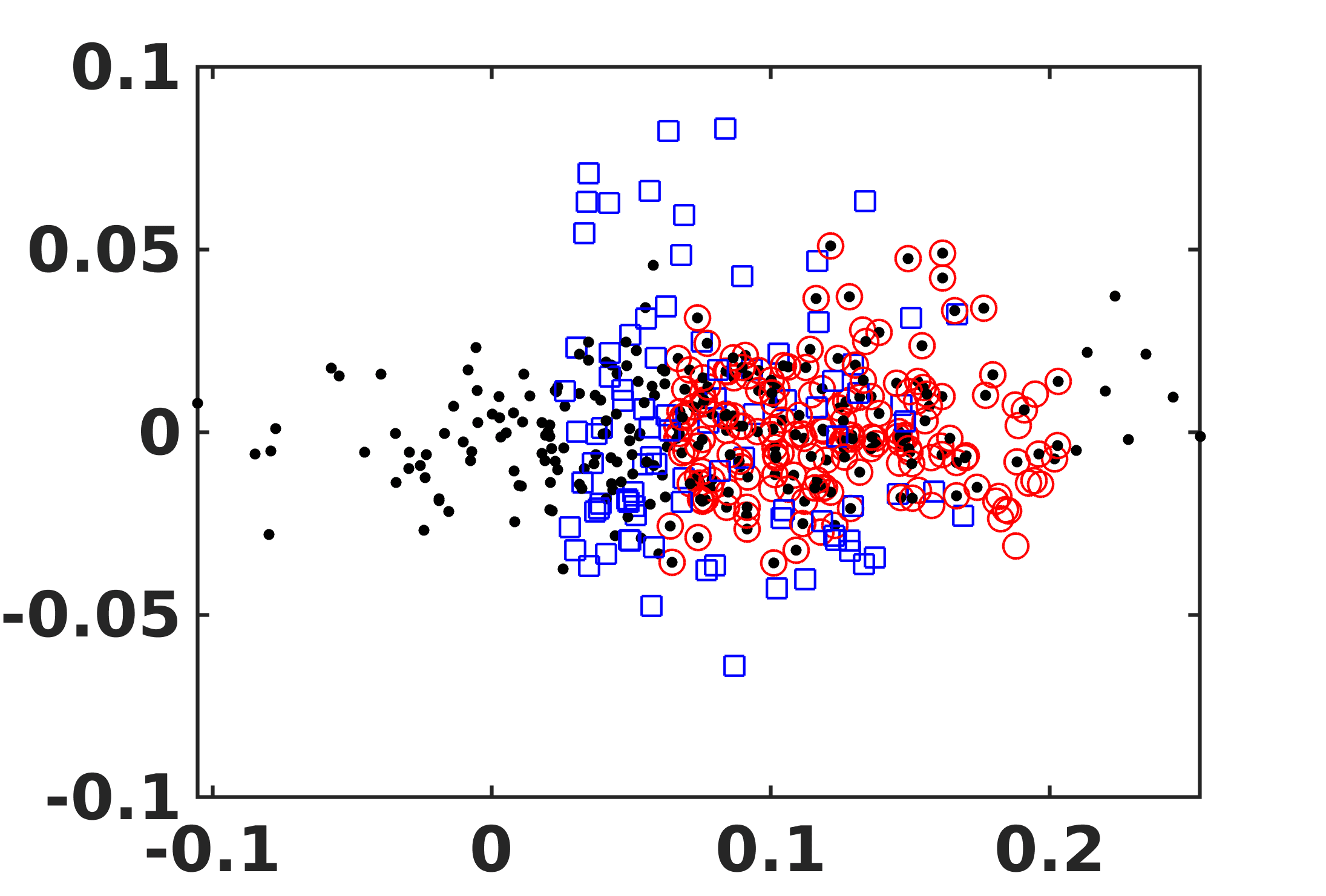}\llap{\parbox[b]{1.58in}{\textbf{{\large (g)}}\\\rule{0ex}{1.1in}}}
\includegraphics[width=4cm]{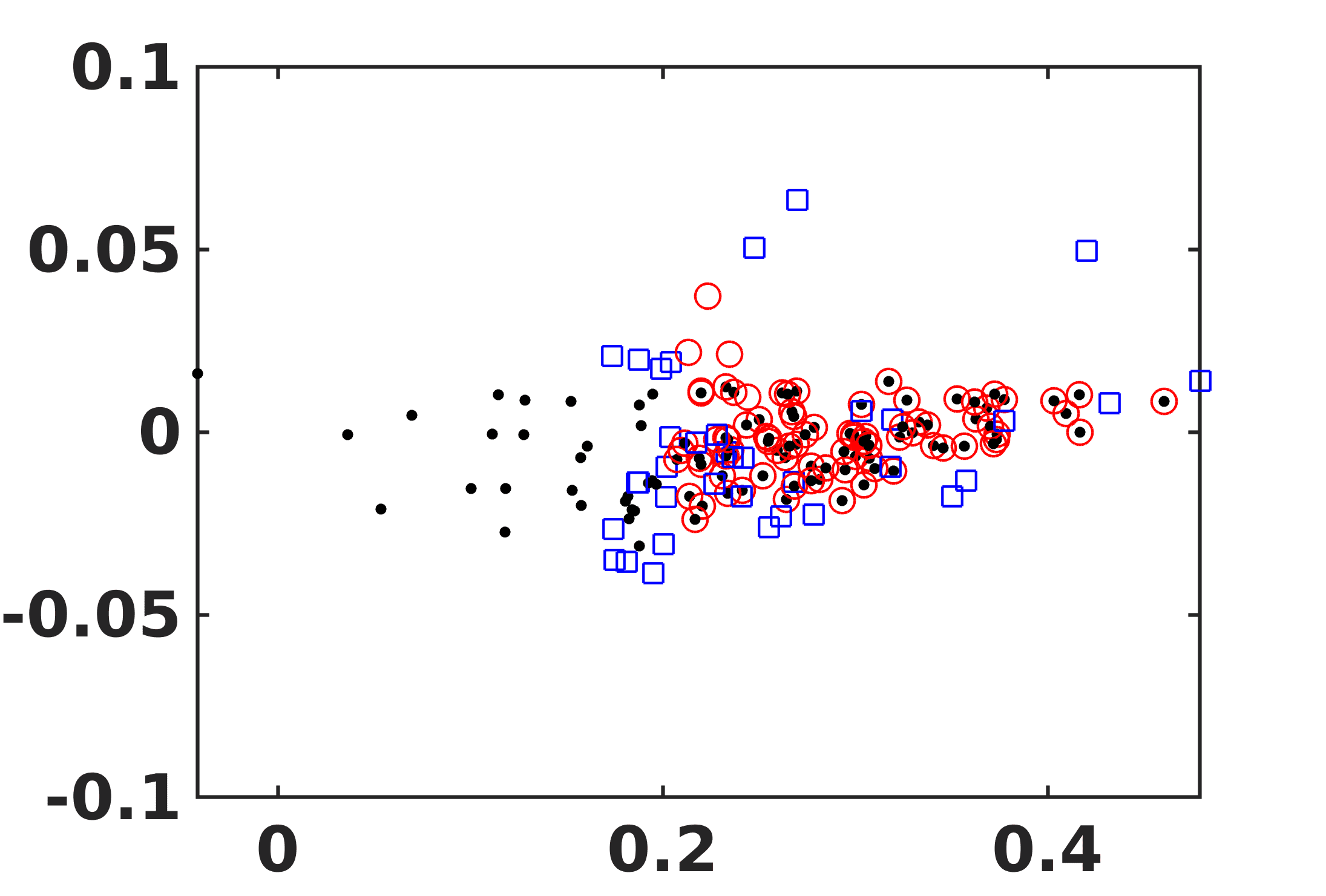}\llap{\parbox[b]{1.58in}{\textbf{{\large (h)}}\\\rule{0ex}{1.1in}}}
\caption{Plots of average correlation matrices of each market state of S\&P 500 market and clustering analysis on surrogate data (CWOE). (a-d) The mean correlation matrices $\overline{Si}$ evaluated over all the correlation frames correspond to each market state $S1,S2,S3,\&~S4$ but for the original correlation matrices ($\epsilon=0$). It shows the average behavior of each market states of S\&P 500 over a period of 14 years (2006-2019). (e) Black dots (MDS (CWOE)) in plot show the MDS map of CWOE using mean correlations martix as $\overline{S1}$ and construction of three times bigger ensemble than $S1$ market state of S\&P 500 market (see, Fig.~\ref{MS_evolution} (a)) with the same noise-suppression $\epsilon=0.5$. Red circles (k-means (CWOE)) in plot show the points of the first cluster of $k$-means clustering performed on CWOE (dots). Blue squares ($k$-means (USA)) in the plot show the $k$-means clustering on the emperical data of S\&P 500. $k$-means clustering on the CWOE and S\&P 500 data shows a qualitative similar behavior. (f), (g), and (h) show the same for mean correlation matrices $\overline{S2}, \overline{S3}$, and $\overline{S4}$,, respectively.}\label{mean_MS_usa}
\end{figure*}
\begin{figure*}[ht!]
\centering
\includegraphics[width=5cm]{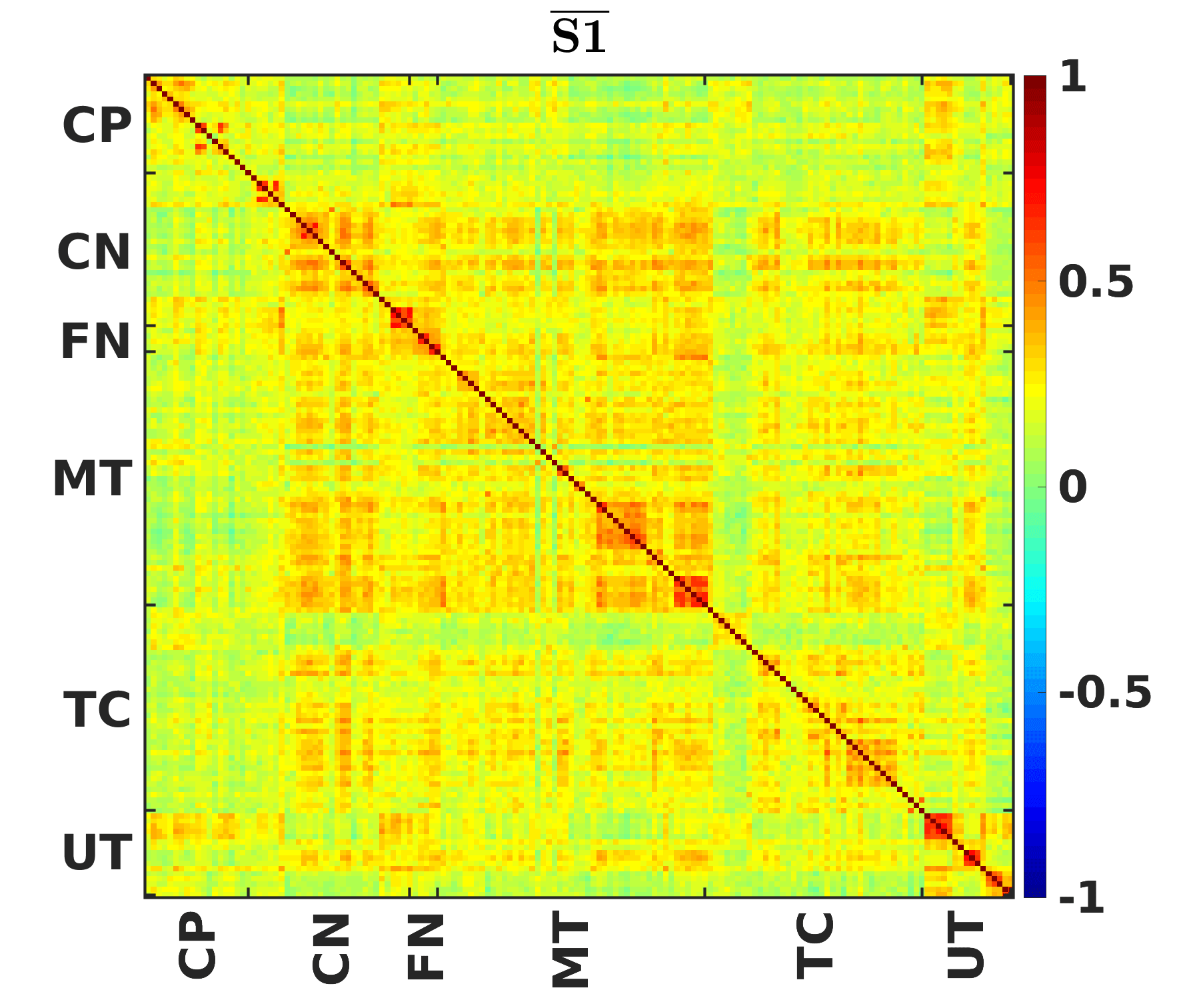}\llap{\parbox[b]{2.in}{\textbf{{\large (a)}}\\\rule{0ex}{1.6in}}}
\includegraphics[width=5cm]{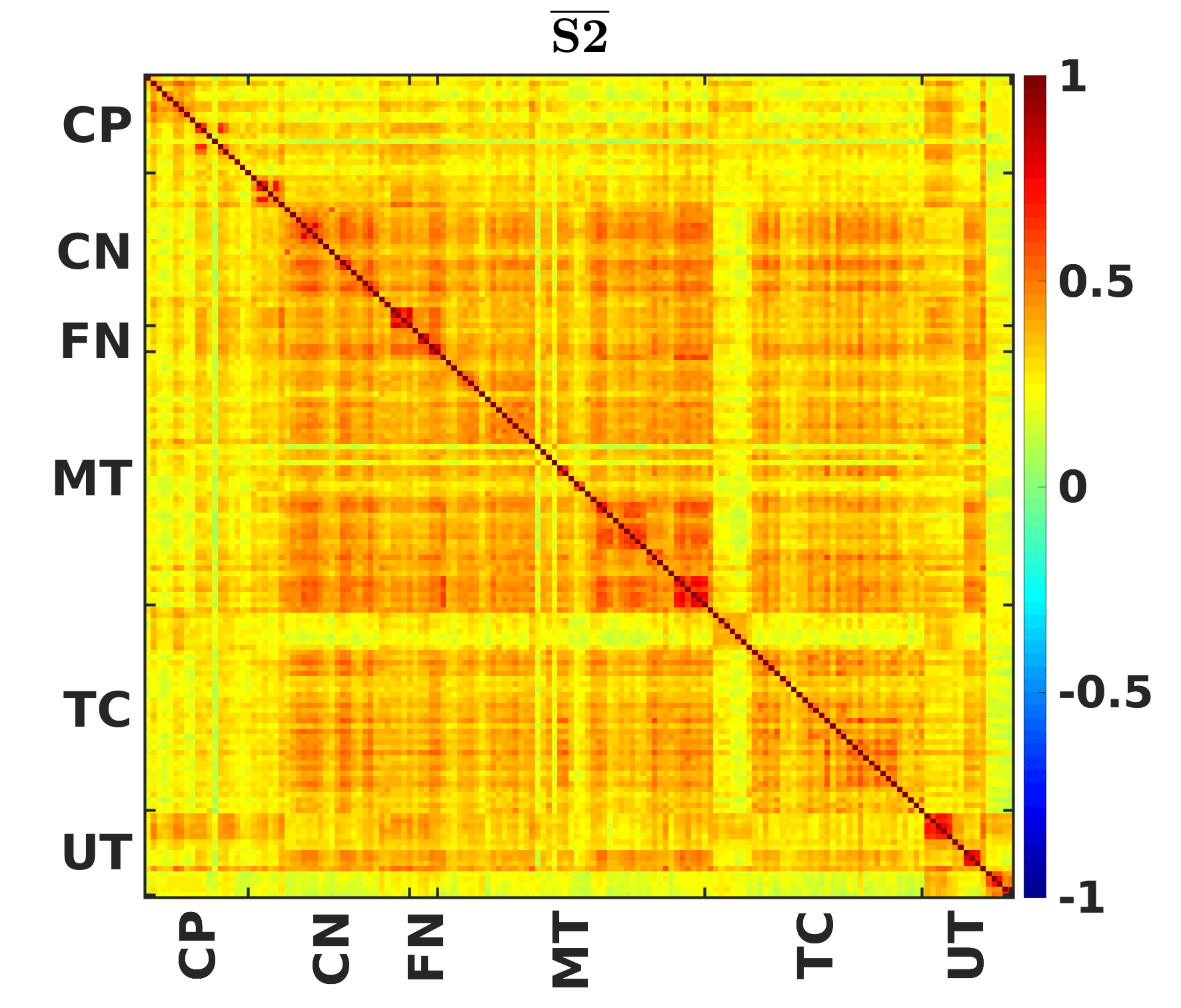}\llap{\parbox[b]{2.in}{\textbf{{\large (b)}}\\\rule{0ex}{1.6in}}}
\includegraphics[width=5cm]{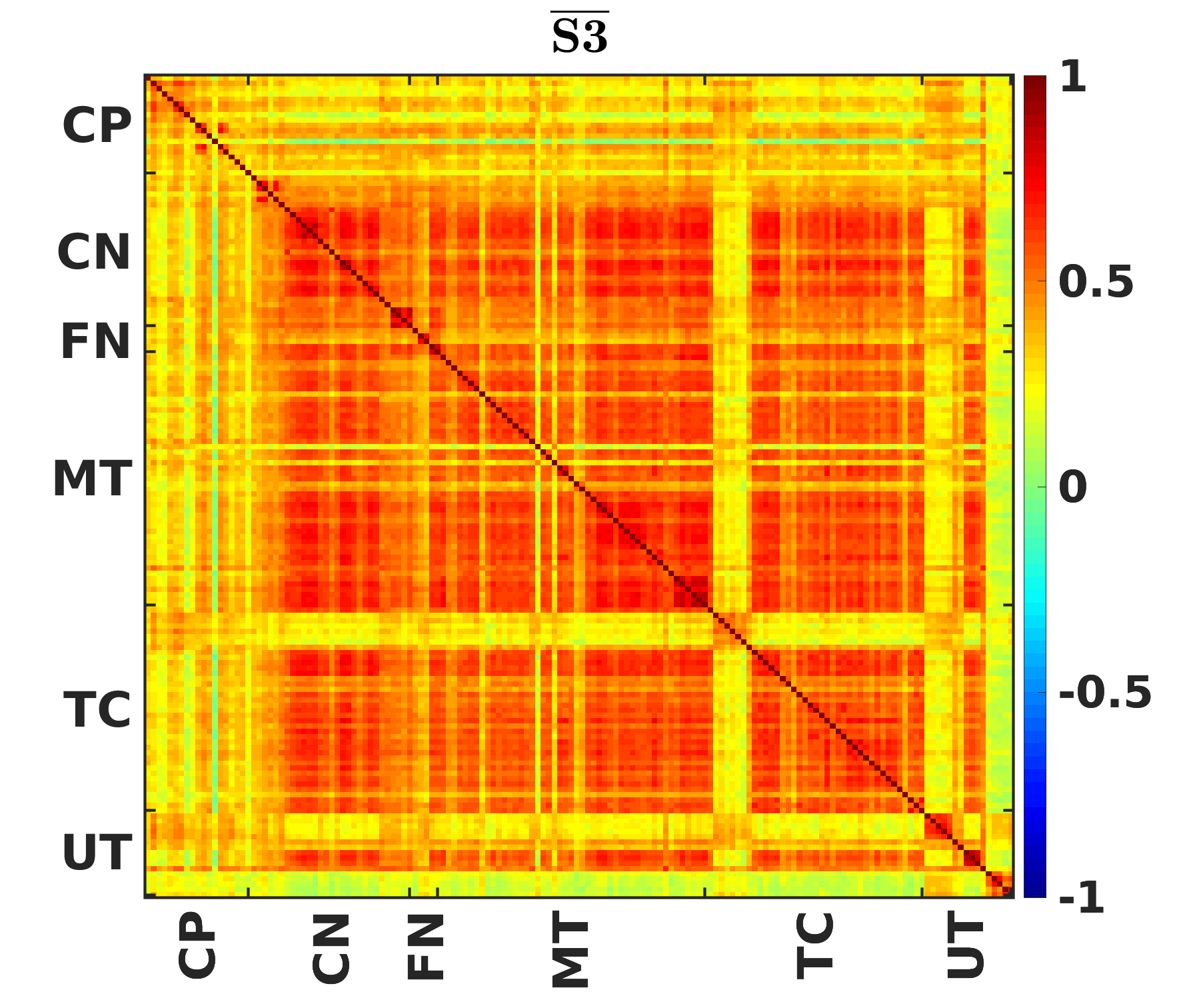}\llap{\parbox[b]{2.in}{\textbf{{\large (c)}}\\\rule{0ex}{1.6in}}}\\
\includegraphics[width=5cm]{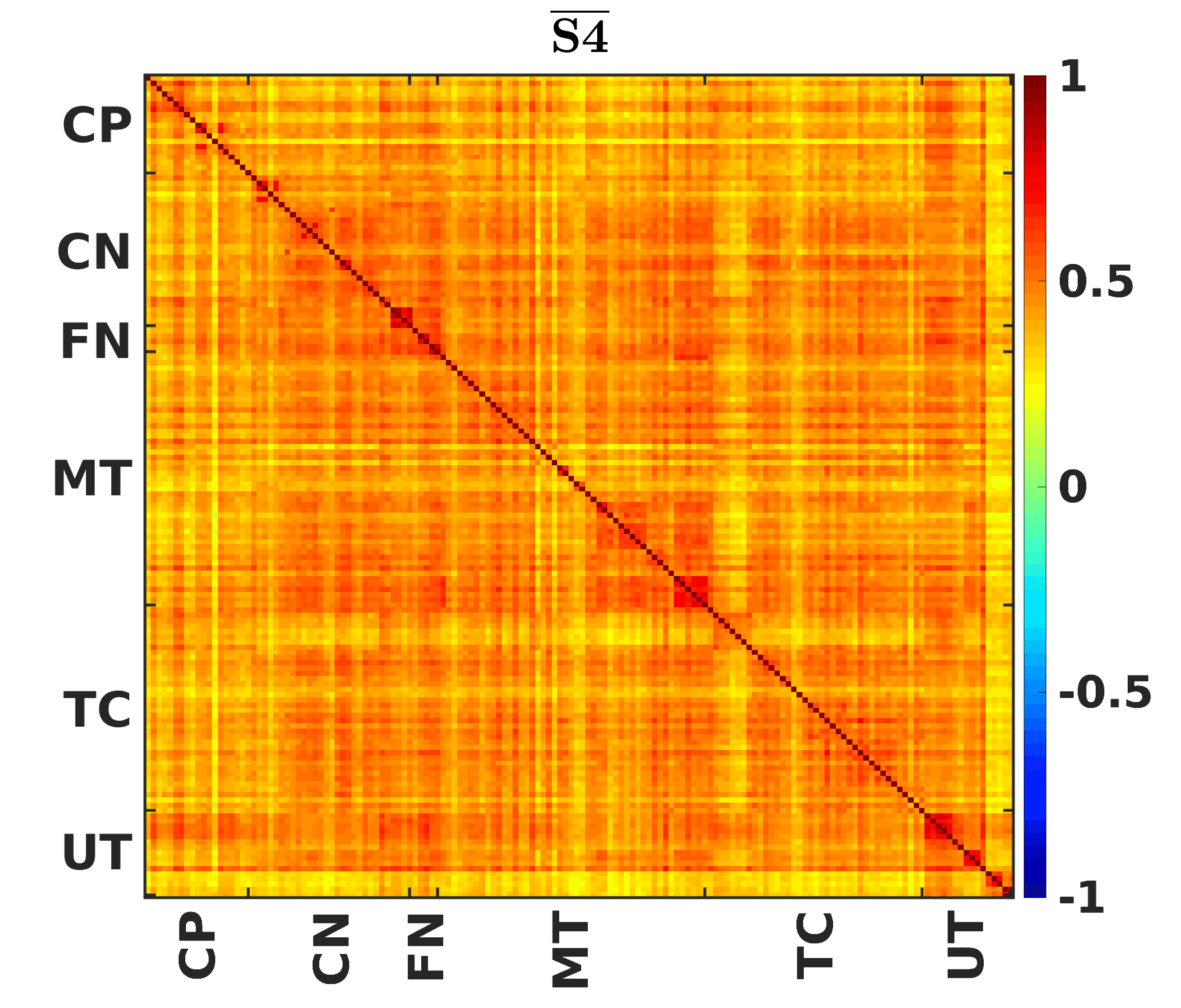}\llap{\parbox[b]{2.in}{\textbf{{\large (d)}}\\\rule{0ex}{1.6in}}}
\includegraphics[width=5cm]{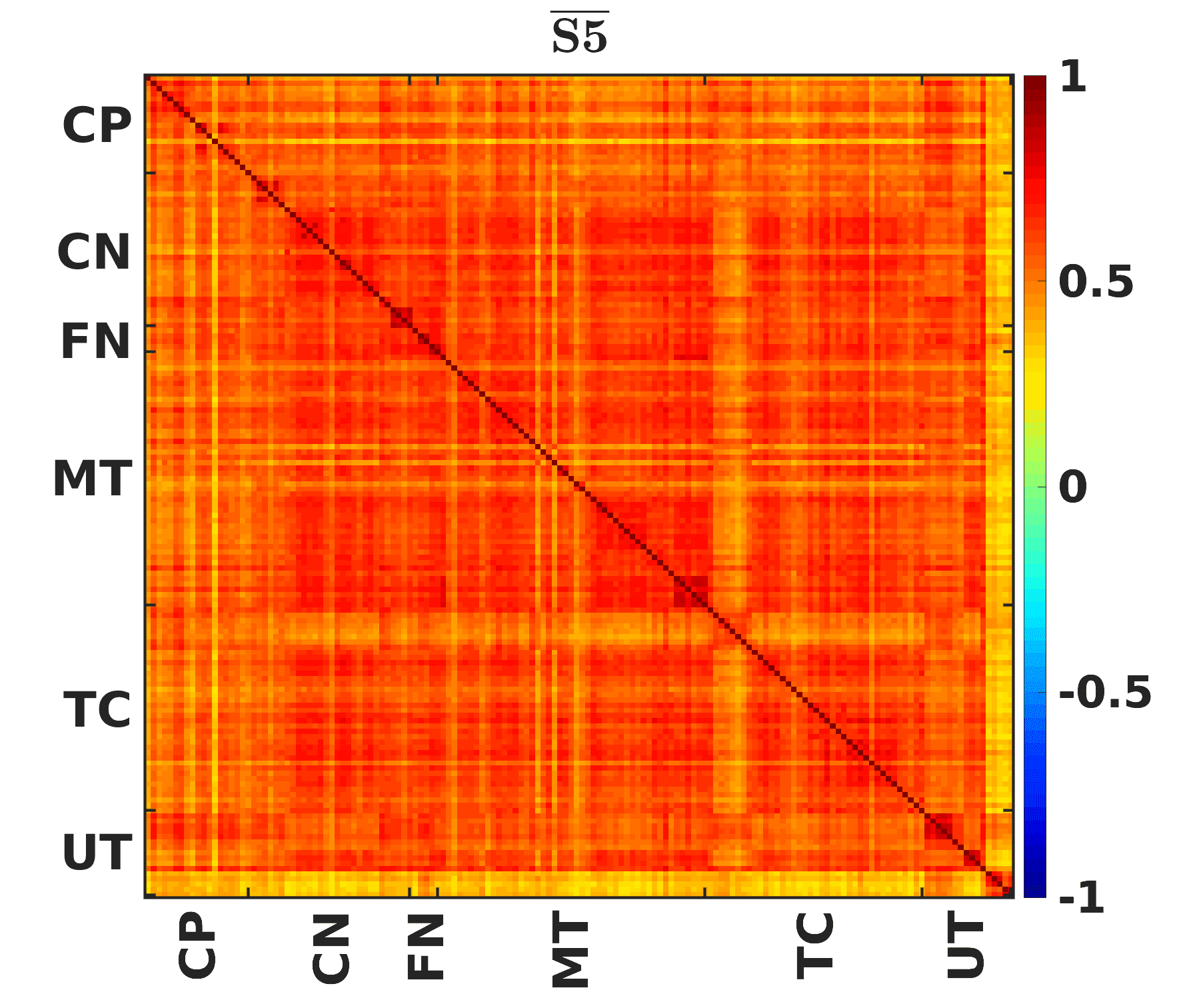}\llap{\parbox[b]{2.in}{\textbf{{\large (e)}}\\\rule{0ex}{1.6in}}}
\includegraphics[width=5cm]{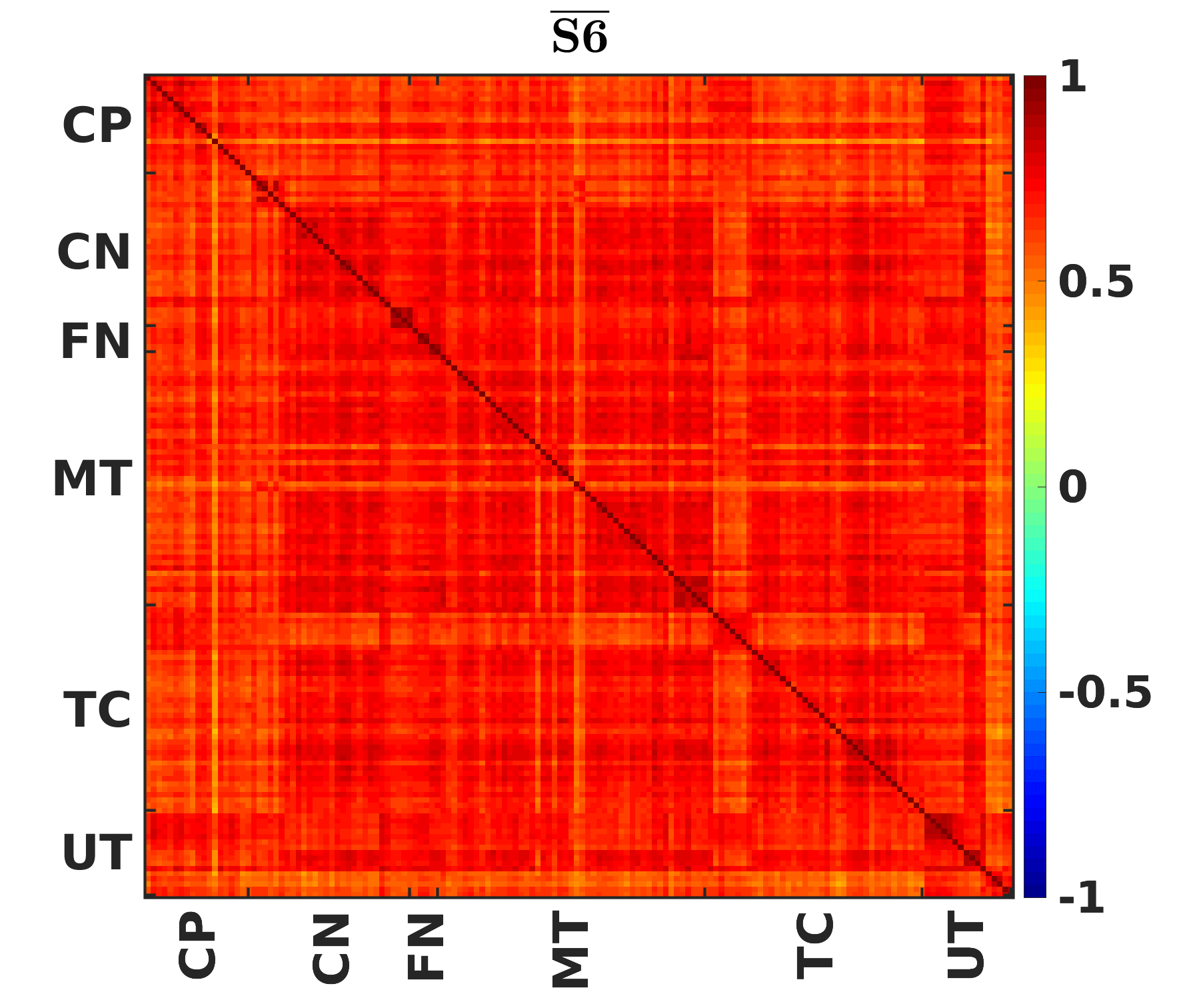}\llap{\parbox[b]{2.in}{\textbf{{\large (f)}}\\\rule{0ex}{1.6in}}}
\includegraphics[width=5cm]{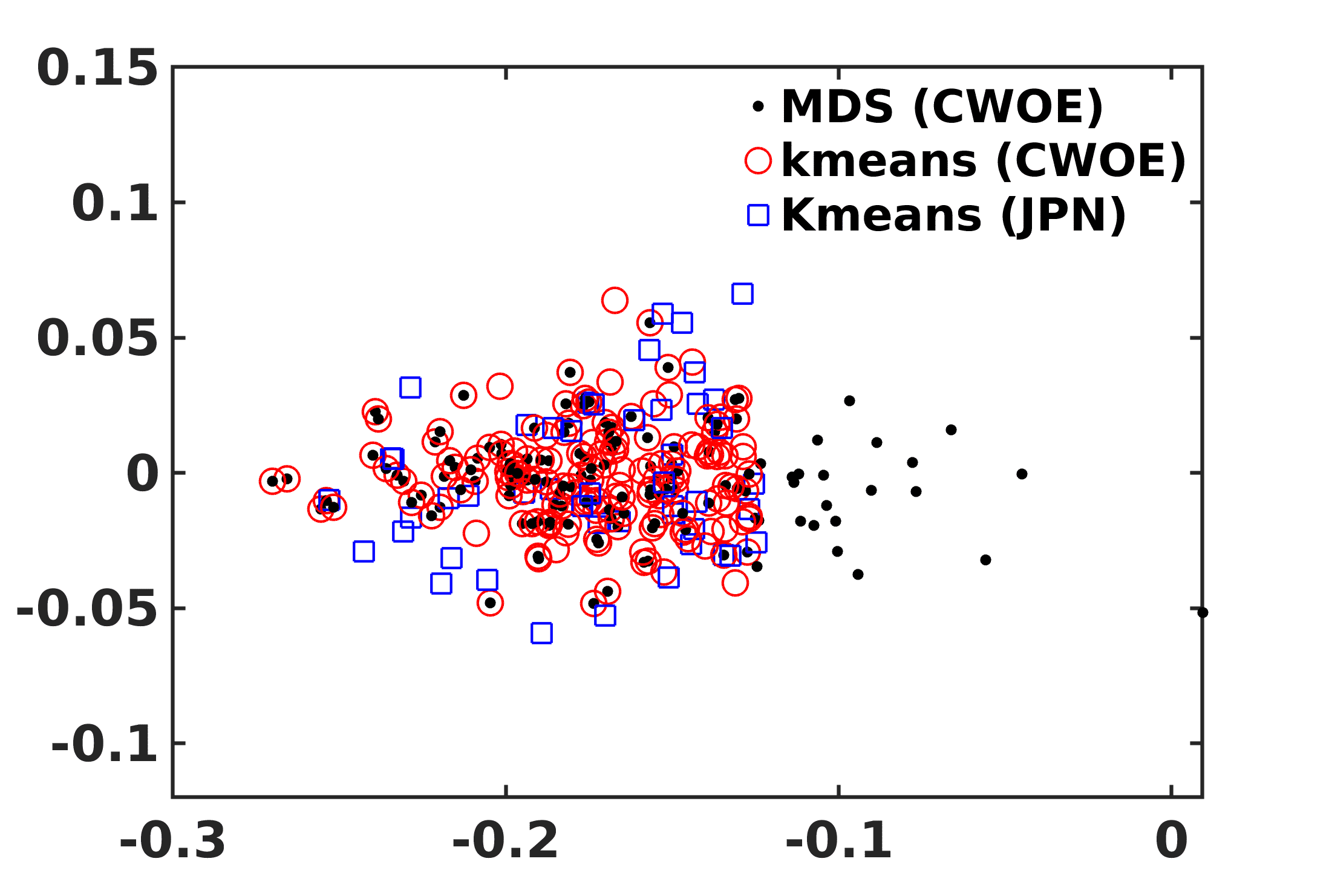}\llap{\parbox[b]{2in}{\textbf{{\large (g)}}\\\rule{0ex}{1.3in}}}
\includegraphics[width=5cm]{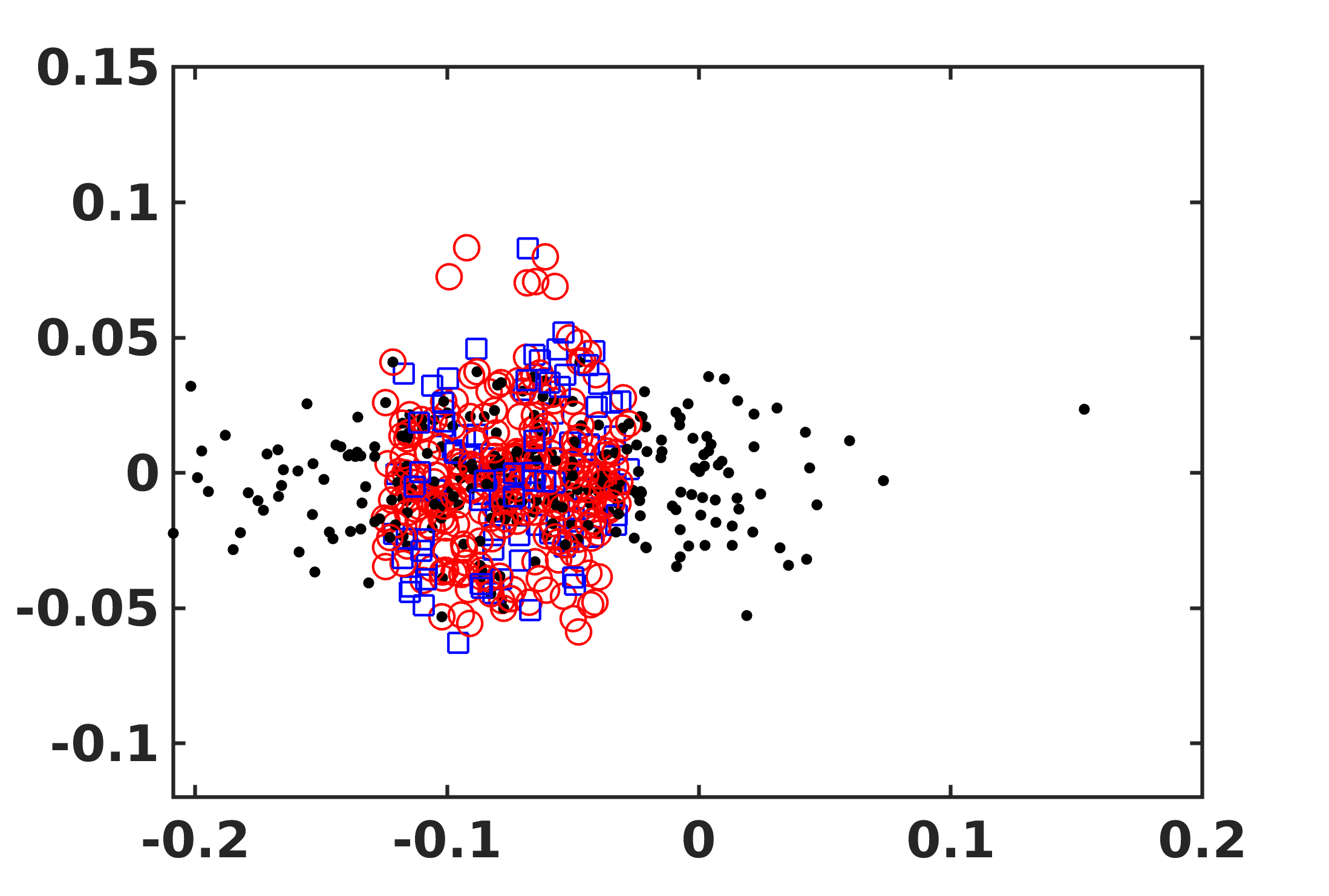}\llap{\parbox[b]{2in}{\textbf{{\large (h)}}\\\rule{0ex}{1.3in}}}
\includegraphics[width=5cm]{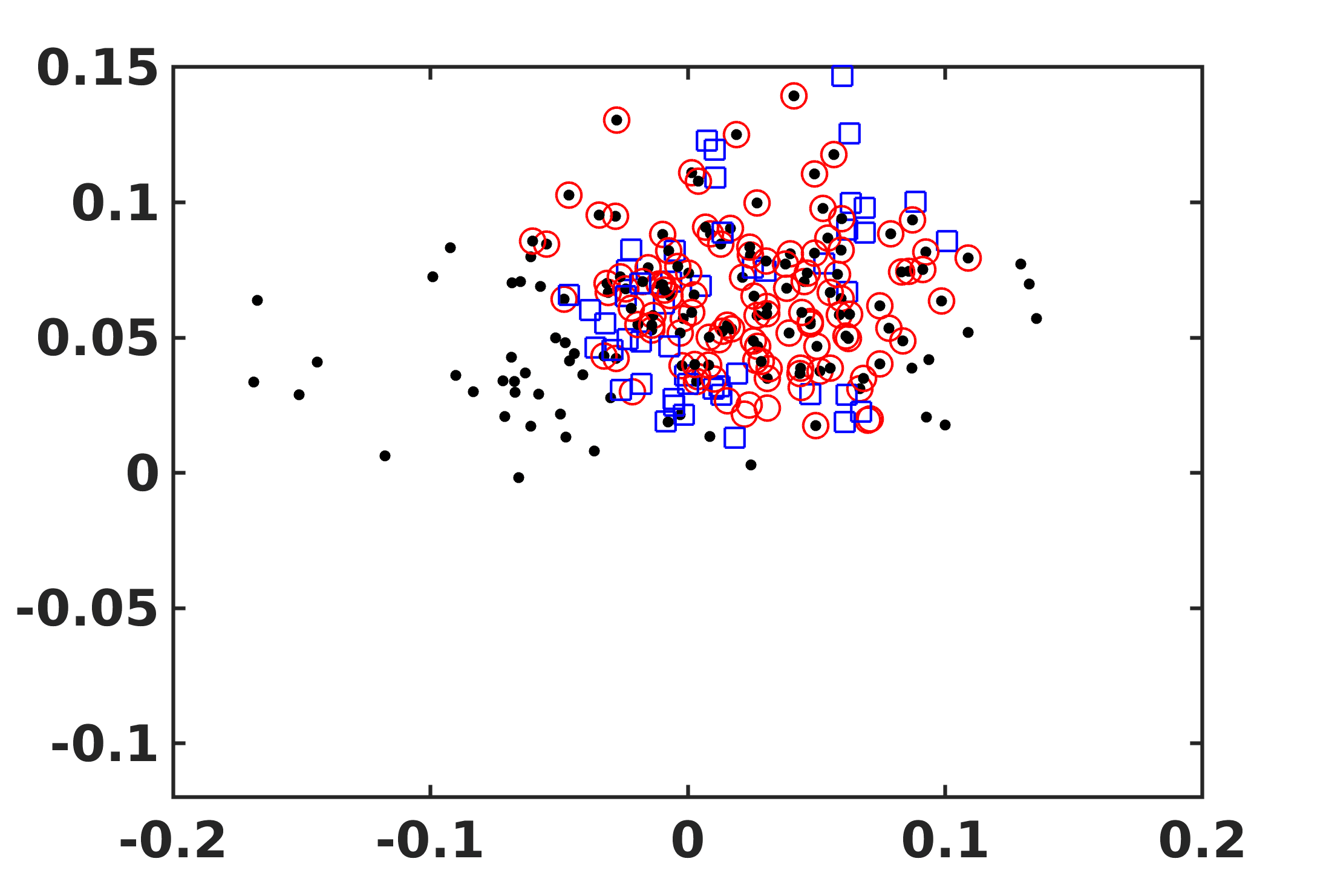}\llap{\parbox[b]{2in}{\textbf{{\large (i)}}\\\rule{0ex}{1.3in}}}\\
\includegraphics[width=5cm]{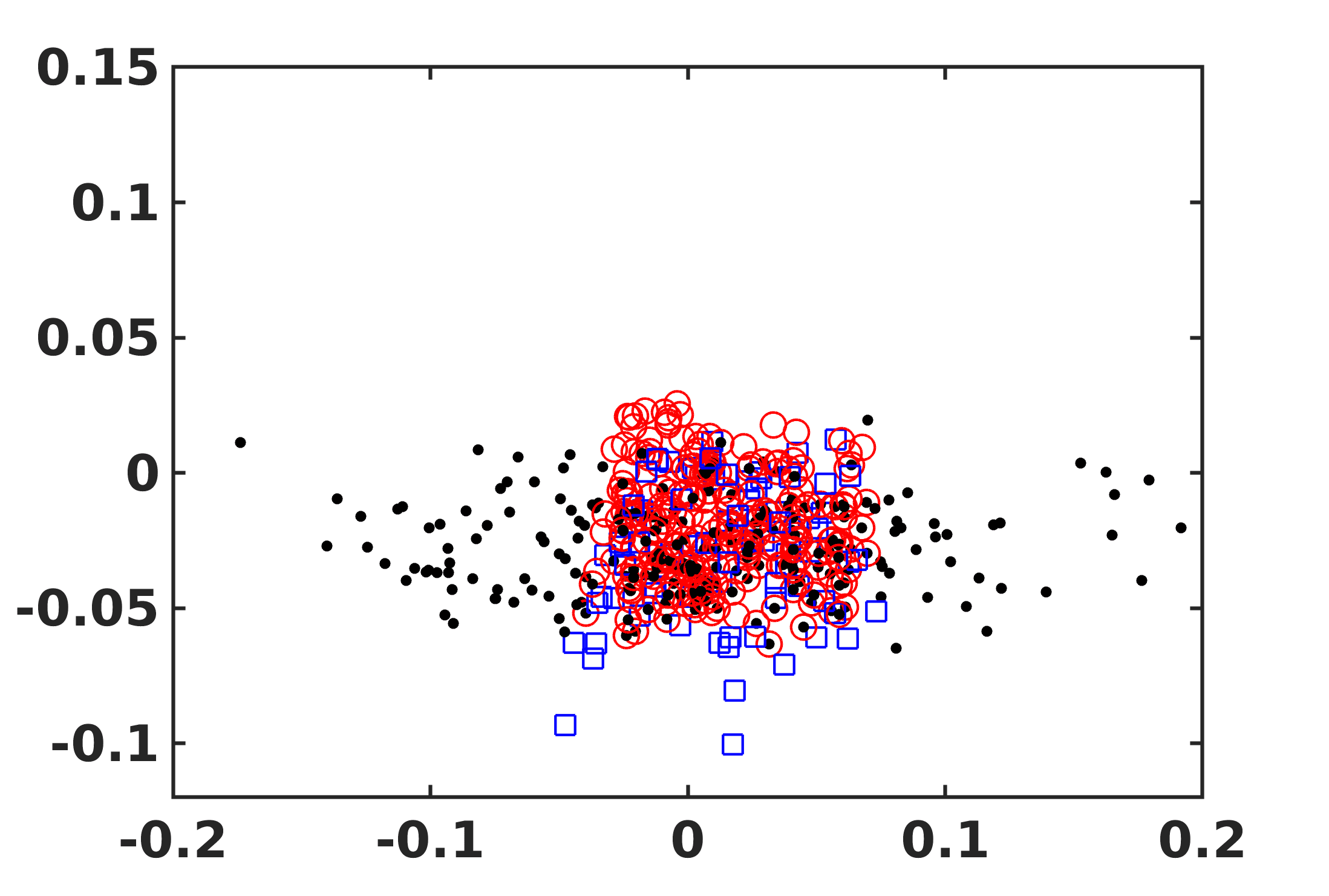}\llap{\parbox[b]{2in}{\textbf{{\large (j)}}\\\rule{0ex}{1.3in}}}
\includegraphics[width=5cm]{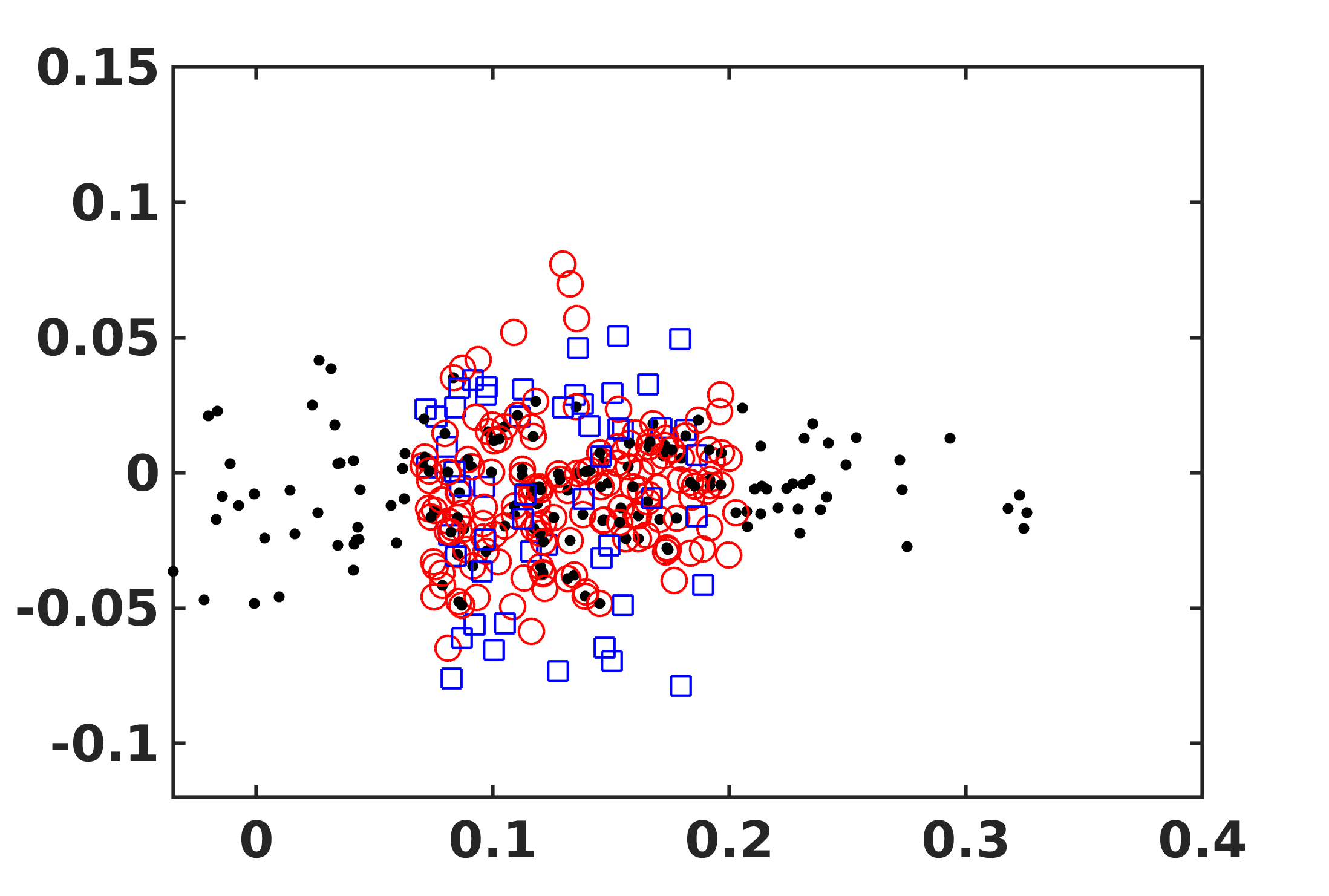}\llap{\parbox[b]{2in}{\textbf{{\large (k)}}\\\rule{0ex}{1.3in}}}
\includegraphics[width=5cm]{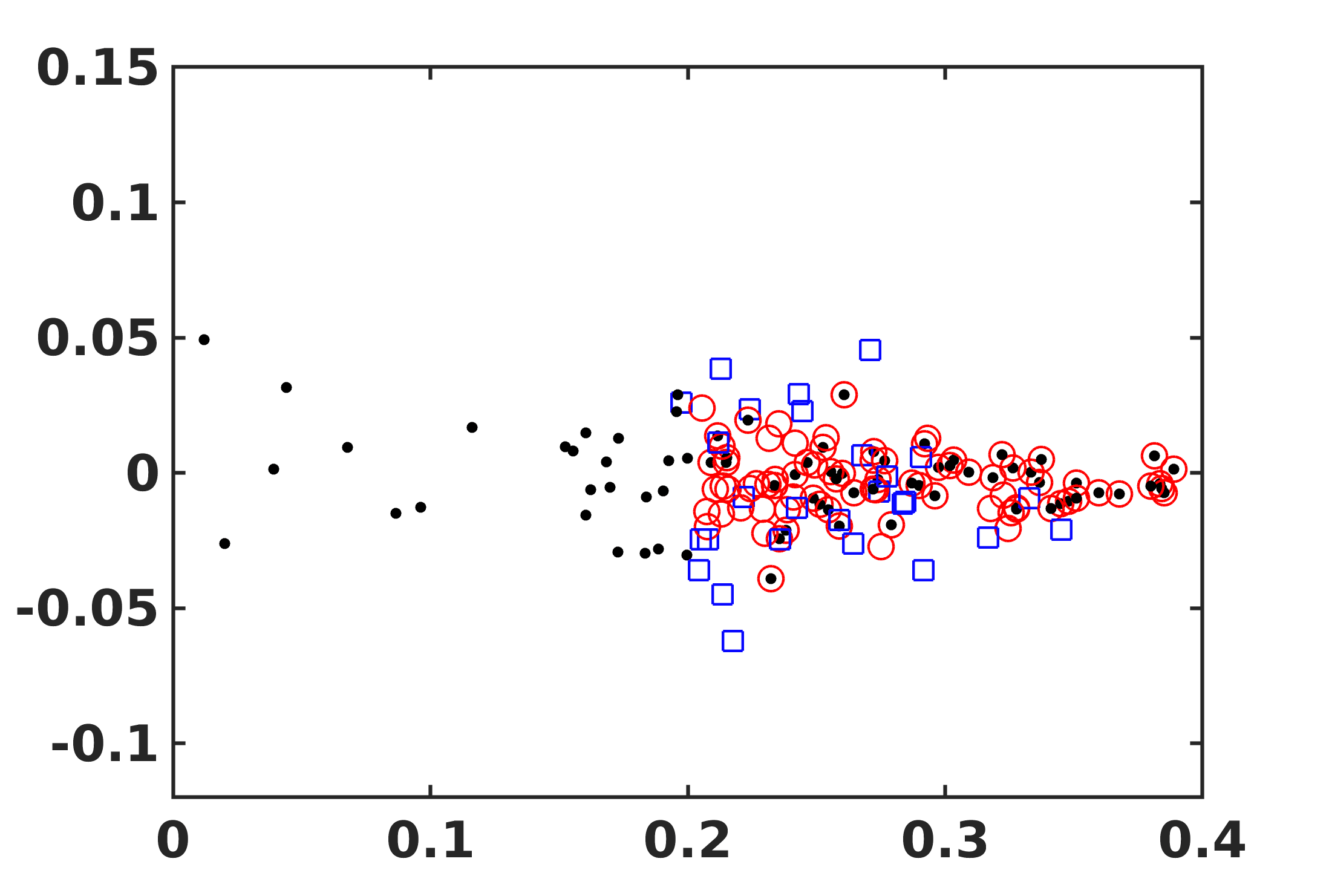}\llap{\parbox[b]{2in}{\textbf{{\large (l)}}\\\rule{0ex}{1.3in}}}
\caption{Plots of average correlation matrics of each market state of Nikkei 225 market and clustering analysis on surrogate data (CWOE). (a-f) The mean correlation  $\overline{Si}$ evaluated over all the frames correspond to each market state $S1,S2,S3,\&~S4$ but for original correlation frames ($\epsilon=0$). It shows the average behavior of all six market states of S\&P 500 over a period of 14 years (2006-2019). (e) Black dots (MDS (CWOE)) in plot show the MDS map of CWOE using mean correlations martix as $\overline{S1}$ and construction of three times bigger ensemble than $S1$ market state of Nikkei 225 market (see, Fig.~\ref{MS_evolution} (b)) with the same noise-suppression $\epsilon=0.3$. Red circles (k-means (CWOE)) in plot show the points of the first cluster of $k$-means clustering performed on CWOE (dots). Blue squares ($k$-means (USA)) in the plot show the $k$-means clustering on the emperical data of the Nikkei market. $k$-means clustering on the CWOE and S\&P 500 data shows a qualitative similar behavior. (g), (h), (i), (j), (k), and (l) show the same for mean correlation matrices $\overline{S2}, \overline{S3}, \overline{S4}, \overline{S5}$, and $\overline{S6}$, respectively.}\label{mean_MS_jpn}
\end{figure*}
We now define similarity measure for two correlation matrices $C(\tau_1)$ and $C(\tau_2)$ evaluated at different time $\tau_1$ and $\tau_2$ by the distance: $\zeta(C,C') \equiv \langle \mid C_{ij} (\tau_1)- C_{ij} (\tau_2) \mid \rangle_{ij}$, where $\langle ...\rangle_{ij}$ denotes the average over all components. The similarity matrices for S\&P 500 and Nikkei 225 are shown in supplementary Figs.~\ref{fig:similarity_USA_JPN_2019} and \ref{fig:similarity_USA_JPN_2018} for periods $2006-2019$ and $2006-2018$, respectively.
The (dis-)similarity measure is thus a distance in a $N(N-1)/2 $ dimensional space which is quite inconvenient particularly because the data set, i.e., the number of correlation matrices we have is smaller than the dimensionality. Multidimensional Scaling (MDS) is a method to accommodate the distances between a set of vectors in higher dimension to a lower dimensional space. This will not always be possible exactly. Thus a statistical method with random initial conditions and subsequent optimization is used as given in the code~\cite{mds_2009} to reproduce the distances as well as possible with a given tolerance. In our case we can do this by projection into a three dimensional (3D) space, which will allow visualization of projections or animated 3D arrangements. We use these suppressed data in a $k$-means procedure where we slightly improve the techniques of Ref.~\cite{Pharasi_2018} to show the clustering of the recent data set. We use the noise suppressed correlation matrix as a starting point, and explicitly discard cluster numbers smaller than 4! As before we choose a noise-suppression, and a cluster number, which minimizes the variance of the intracluster distances generated by the random selections involved. The results are seen in Figs.~\ref{intracluster} (a, c) for the US and Figs.~\ref{intracluster} (b, d) for the Japanese market using $\epsilon = 0.5$ and $0.3$, respectively. 
The differences mentioned above concerning indices and average correlations now become much more striking. Compared to earlier data, the Japanese market shows one more cluster and while the clusters $3$ and $4$ are not entirely clear in the older data set. 
Now we have six clusters for a shorter time period and we encounter the phenomenon that the clusters by no means are in a quasi-linear arrangement which looks somewhat trivial, though it actually is not. Indeed we find that the alignment is along the average correlation, as already pointed out in the very first paper~\cite{Munnix_2012}.

Figs.~\ref{MS_evolution} (a) and (b) show the evolution of S\&P 500 through the transitions among these four clusters named as market states $S1, S2, S3,$ and $S4$ and Nikkei 225 through six market states $S1, S2, S3$, $S4$, $S5$ and $S6$ for the period of $2006-2019$, respectively.  For the US market we find average correlations correspond to each market state $\mu(S1, S2, S3, S4) =  (0.19, 0.31, 0.46, 0.63)$ are lined up according to the cluster index $S1, S2, S3, S4$, while we get for the Japanese market $\mu(S1, S2, S3, S4, S5, S6) = (0.21, 0.33, 0,43, 0.44, 0.57, 0.68)$, where states $S3$ and $S4$ have practically the same average correlation.  Note also that the transition probabilities of the Japanese market between states $S3$ and $S4$  are very small, and the absolute numbers ($S3\rightarrow S4 =3$, $S4\rightarrow S3=1$) are not visible in the bar plot Fig.~\ref{transition_probabilities} (b). What properties of the two markets can reflect this nature should be open for discussion.

If the differences in variances of the intracluster measure $d_{intra}$ are small, we suggest  adding additional criteria  relates to Fig.~\ref{transition_probabilities}, where bar-plots for absolute transition  numbers (see, Fig.~\ref{transition_probabilities} (a) and (b)) and probability graphs for the dynamics (see, Fig.~\ref{transition_probabilities} (c) and (d)) of both markets are shown. On one hand select a clustering that minimizes transitions between clusters, which is particularly apparent for the split pathways to high correlations seen above for the Japanese market. On the other hand a clustering that looks consistent with surrogate data, obtained as CWOE from the average correlation obtained for each cluster, as we shall discuss now. In the supplementary material we again show that the probabilities are consistent with a master equation as seen already in Ref.~\cite{Pharasi_2018}.

We can consider that the average correlation matrix obtained for each cluster to large extent characterizes the states. This assumption implies that the cluster arises from noise around this average as given by a CWOE constructed from the average correlation matrix. Average correlation matrices of each market state of the S\&P 500 market are shown in Fig.~\ref{mean_MS_usa} (a-d). Fig.~\ref{mean_MS_usa} (e), (f) (g), and (h) show the comparison of clustering analysis of surrogate data CWOE versus each state $S1, S2, S3$, and  $S4$ of the S\&P 500 market, respectively. $k$-means clustering (red circles) on the surrogate data (black dots) shows quantitatively similar behavior with the market states $S1, S2, S3, S4$ of S\&P 500. Fig.~\ref{mean_MS_jpn} shows the same for six market states $S1, S2, S3$, $S4$, $S5$ and $S6$ of Japanese market. Clustering has a strong dependence on epoch size. 

\section*{Discussions and Summary}
We present improved criteria for the definition of clusters and thus market states and apply these to data sets starting in 2006 and ending in 2019 for both the US and the Japanese stock market as represented by the stocks in the S\&P 500 and the Nikkei 225 indices.  In general terms, we find that the turbulence that started in 2008 are awaiting to some extent. On the technical side we have improved the selection of precursors of market shocks and found a more detailed structure in the Japanese market that will allow a better understanding of market dynamics beyond the dominating influence of average correlation. This brings the idea of market states and their dynamics one step nearer to real market conditions and practical use. As shares are not the entire market at some point both derivative and bond prices as well as possible connections to intra-day trading will have to be included in such studies.
\section*{Acknowledgment}
The authors are grateful to Anirban Chakraborti and Francois Leyvraz for their critical inputs and suggestions. H.K.P. is grateful for financial support provided by UNAM-DGAPA and CONACYT Proyecto Fronteras 952. T.H.S. and H.K.P. acknowledges the support grant by CONACyT through Project Fronteras 201 and UNAM-DGAPA-PAPIIT AG100819 and IN113620. 

\newpage
\section*{SUPPLEMENTARY INFORMATION }
\noindent
\setcounter{figure}{0}
\setcounter{table}{0}
\renewcommand{\thefigure}{S\arabic{figure}}
\renewcommand{\thetable}{S\arabic{table}}
\section*{Data considered for the analysis over a period of 2006-2019}
\noindent Transition probabilities of the S\&P 500 market between four states ($S1, S2, S3,$ and $S4$) and Nikkei 225 between six states ($S1, S2, S3$, $S4$, $S5$, and $S6$) over the period of $2006-2019$ is given in Table~\ref{table:USA_freq} and Table~\ref{table:JPN_freq}, respectively. The data considered for the analysis are given in Table~\ref{table:usa_2019} and Table~\ref{table:jpn_2019} for the same period.
\vskip 0.5in
\begin{table*}[!htb]
    \caption{Transition probability of market states for\\ USA: Four market states (MS)}\label{table:USA_freq}
    \centering
\begin{tabular}{|p{2.cm}|p{2.cm}|p{2.cm}|p{2.cm}|p{2.cm}|}
\hline
   & S1     & S2     & S3     & S4     \\ \hline
S1 & 0.703 & 0.243 & 0.036 & 0.018 \\ \hline
S2 & 0.232 & 0.455 & 0.313 & 0 \\ \hline
S3 & 0.077 & 0.33 & 0.484 & 0.11 \\ \hline
S4 & 0      & 0.111     & 0.222 & 0.667 \\ \hline
\end{tabular}
\end{table*}
\begin{table}[!htb]
 \centering
 \caption{Transition probability of market states for JPN: Six market states (MS)}\label{table:JPN_freq}
\begin{tabular}{|p{2.cm}|p{2.cm}|p{2.cm}|p{2.cm}|p{2.cm}|p{2.cm}|p{2.cm}|}
\hline
   & S1     & S2     & S3     & S4     & S5    & S6 \\ \hline
S1 & 0.531 & 0.327 & 0.082 & 0.041 & 0.0 & 0.02\\ \hline
S2 & 0.193 & 0.42 & 0.159 & 0.114 & 0.102 & 0.011\\ \hline
S3 & 0.026 & 0.299 & 0.494 & 0.013 & 0.156 & 0.013\\ \hline
S4 & 0.02 & 0.204 & 0.061 & 0.612 & 0.082 & 0.02\\ \hline
S5 & 0.038 & 0.057 & 0.264 & 0.075 & 0.396 & 0.17\\ \hline
S6 & 0.037 & 0.0 & 0.111 & 0.074 & 0.259 & 0.519\\ \hline
\end{tabular}
\end{table}

\begin{center}
	\begin{figure}[!h]
	\centering
		\includegraphics[width=8cm]{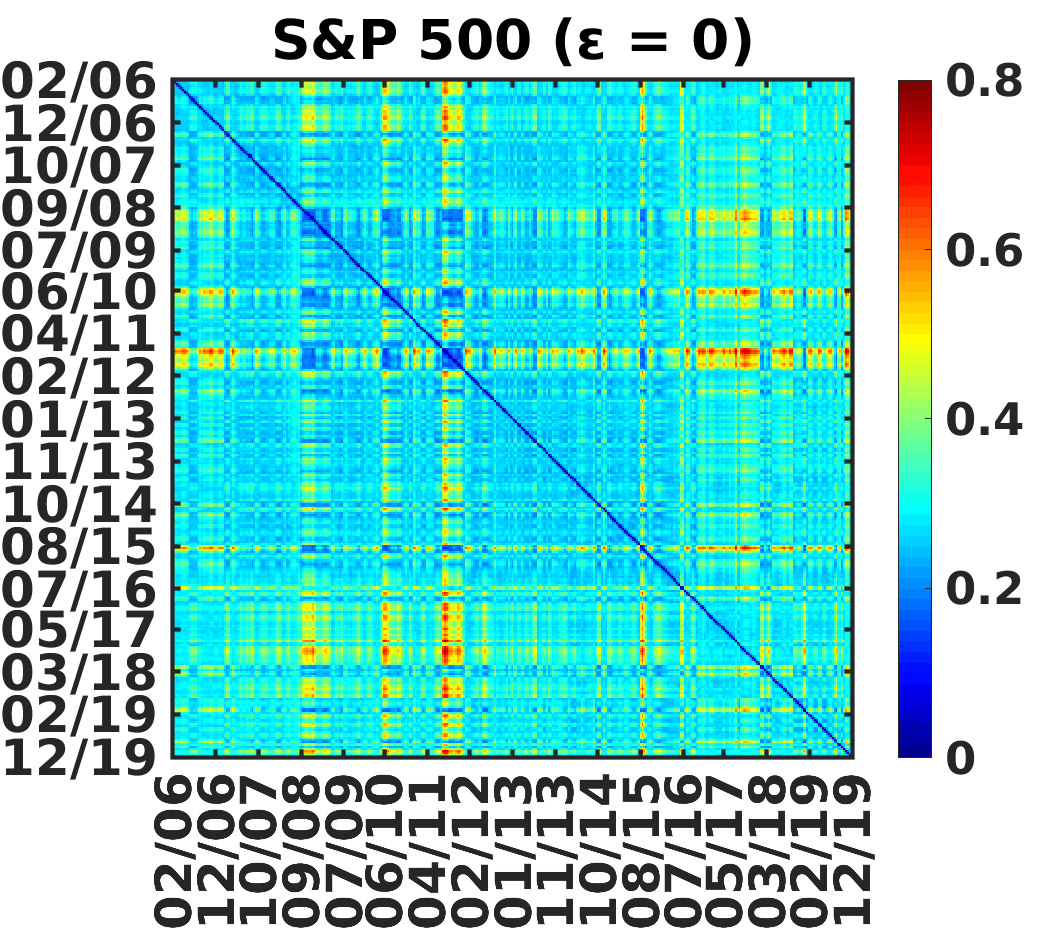}\llap{\parbox[b]{3.3in}{\textbf{{\Large (a)}}\\\rule{0ex}{2.7in}}}
		\includegraphics[width=8cm]{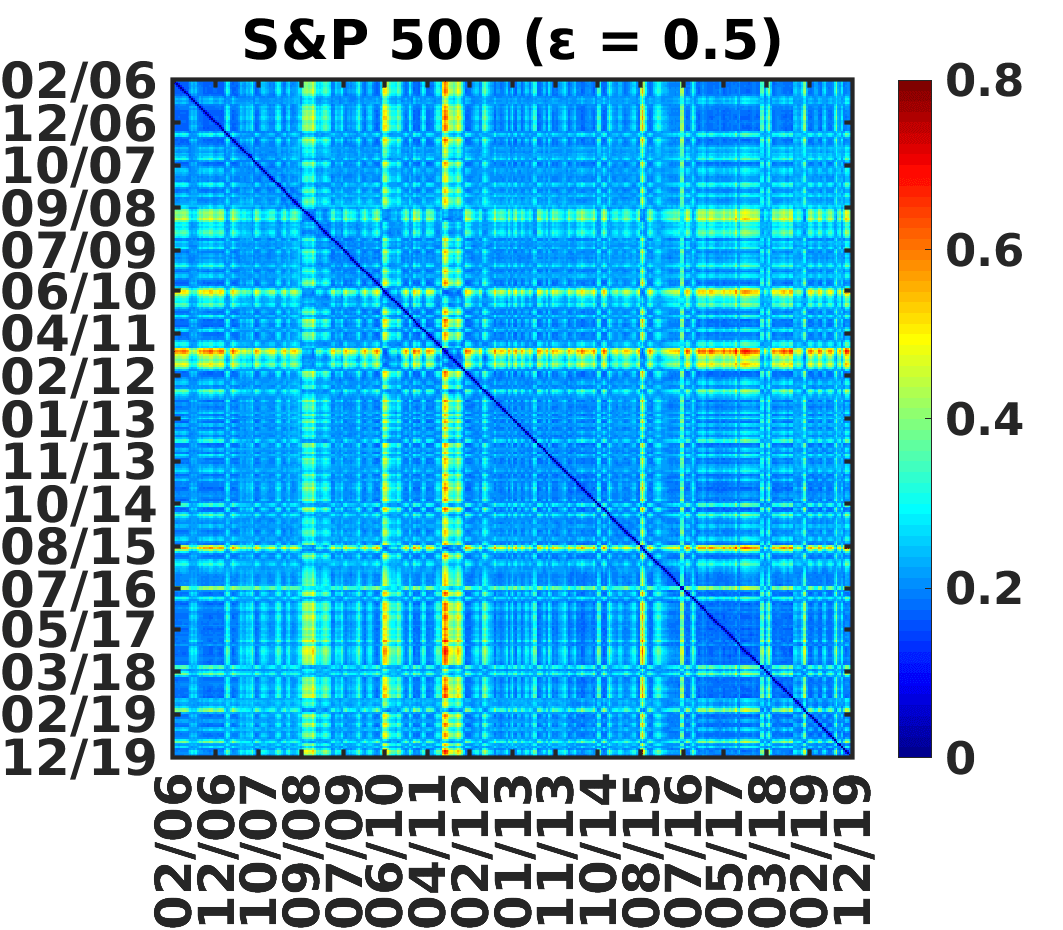}\llap{\parbox[b]{3.3in}{\textbf{{\Large (b)}}\\\rule{0ex}{2.7in}}}
		\includegraphics[width=8cm]{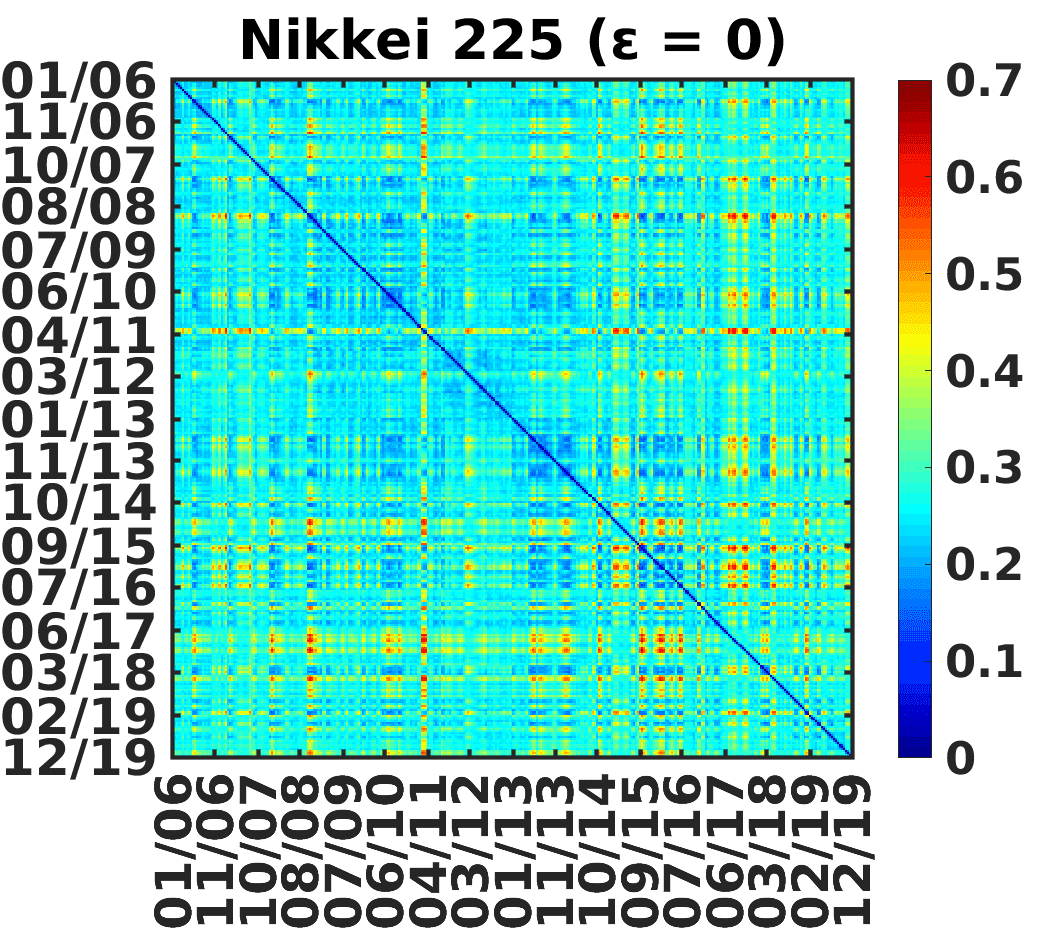}\llap{\parbox[b]{3.3in}{\textbf{{\Large (c)}}\\\rule{0ex}{2.7in}}}
		\includegraphics[width=8cm]{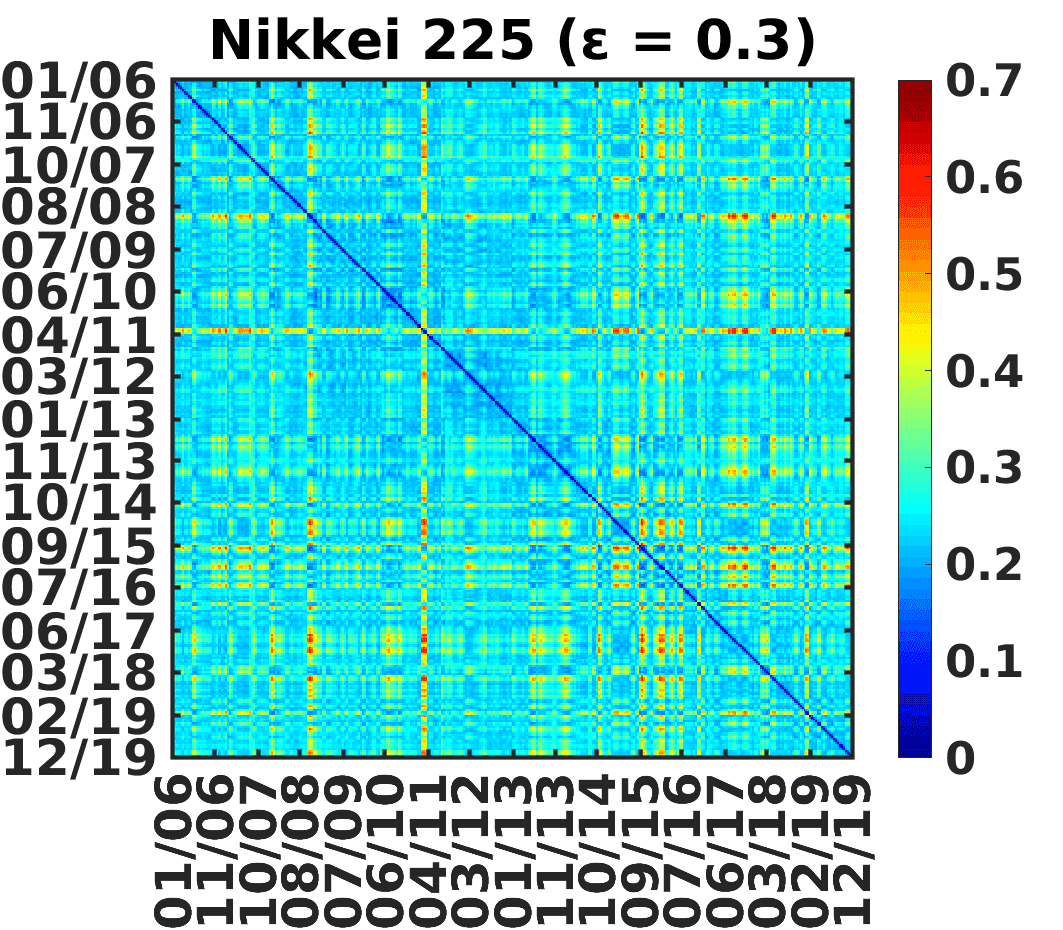}\llap{\parbox[b]{3.3in}{\textbf{{\Large (d)}}\\\rule{0ex}{2.7in}}}
		\caption{Similarity measure $\zeta (C,C')$ among $351$ correlation matrices of S\&P 500 (top row) and $344$ correlation matrices of Nikkei 225 (bottom row).  (a) and (c) are without noise-suppression $\epsilon=0$ and (b) and (d) are with noise-suppression $\epsilon=0.5$ and  $\epsilon=0.3$, respectively. The evolution of the stock market over 14 years (2006--2019) can be understood from the similarity matrices. Critical events of the markets  in red-yellow strips are shown in the similarity matrices. S\&P 500 market is less volatile than Nikkei 225 and  shows less fluctuations after 2012.} \label{fig:similarity}
	 \end{figure}
\end{center}
\newpage
\section*{Data considered for the analysis over a period of 2006-2018}
\noindent We show here similar results with data running up to 2018. The results for Japanese market differ substantially indicating a change in the general market situation. The data considered for the analysis are given in Table~\ref{table:usa_2018} and Table~\ref{table:jpn_2018} for the period 2006-2018.

\begin{center}
	\begin{figure}[h!]
	\centering
		\includegraphics[width=8cm]{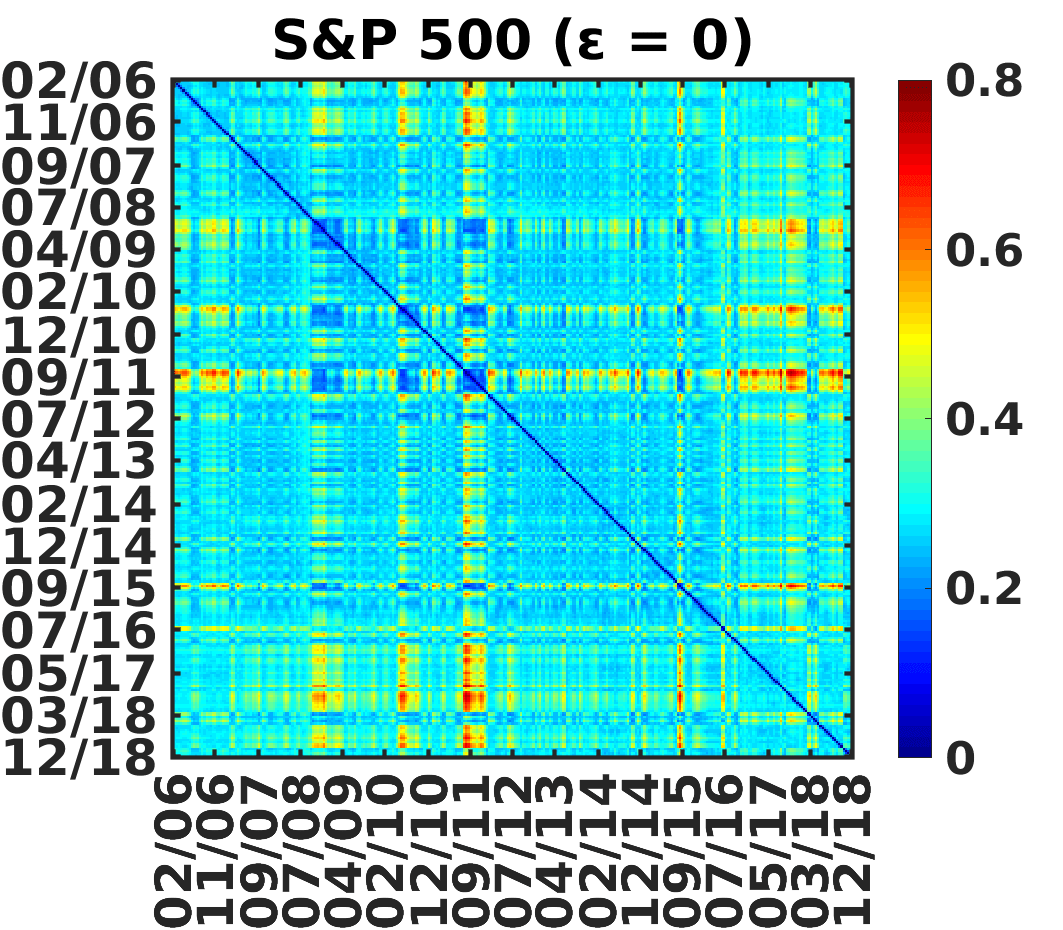}\llap{\parbox[b]{3.3in}{\textbf{{\Large (a)}}\\\rule{0ex}{2.7in}}}
		\includegraphics[width=8cm]{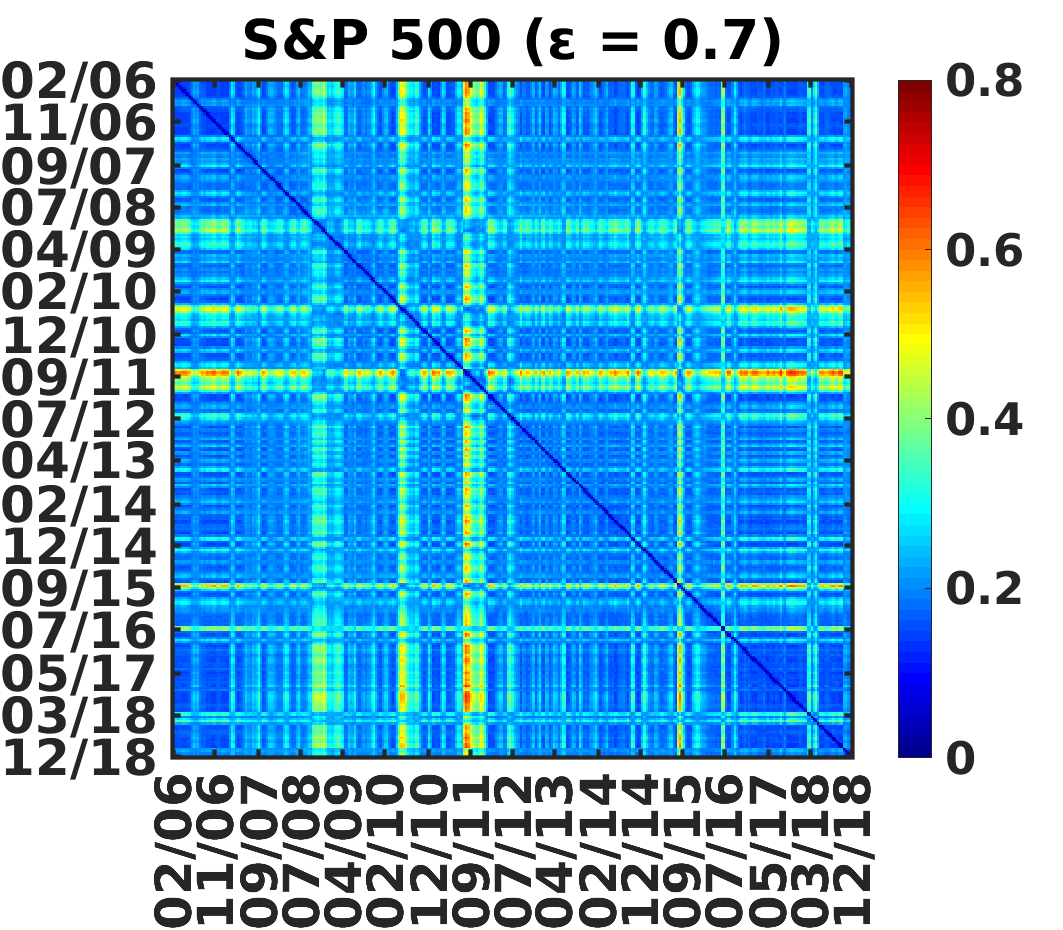}\llap{\parbox[b]{3.3in}{\textbf{{\Large (b)}}\\\rule{0ex}{2.7in}}}\\
		\includegraphics[width=8cm]{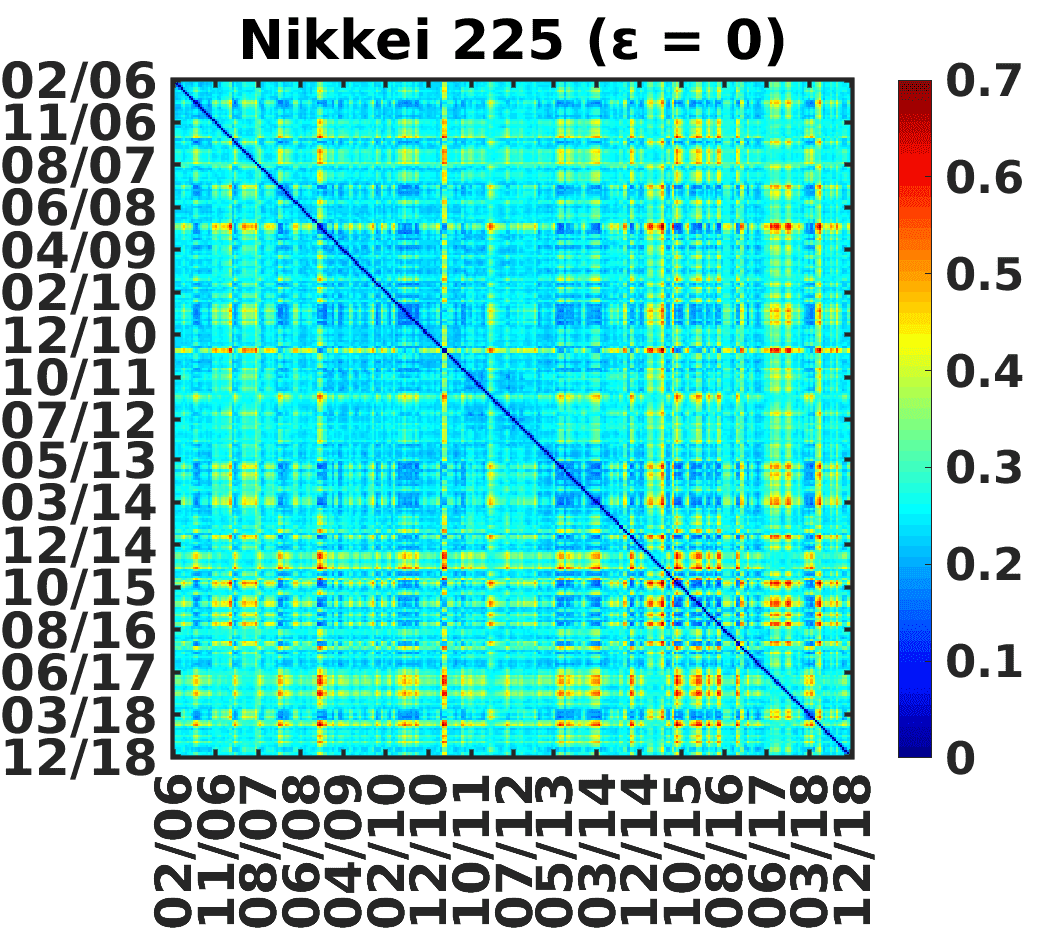}\llap{\parbox[b]{3.3in}{\textbf{{\Large (c)}}\\\rule{0ex}{2.75in}}}
		\includegraphics[width=8cm]{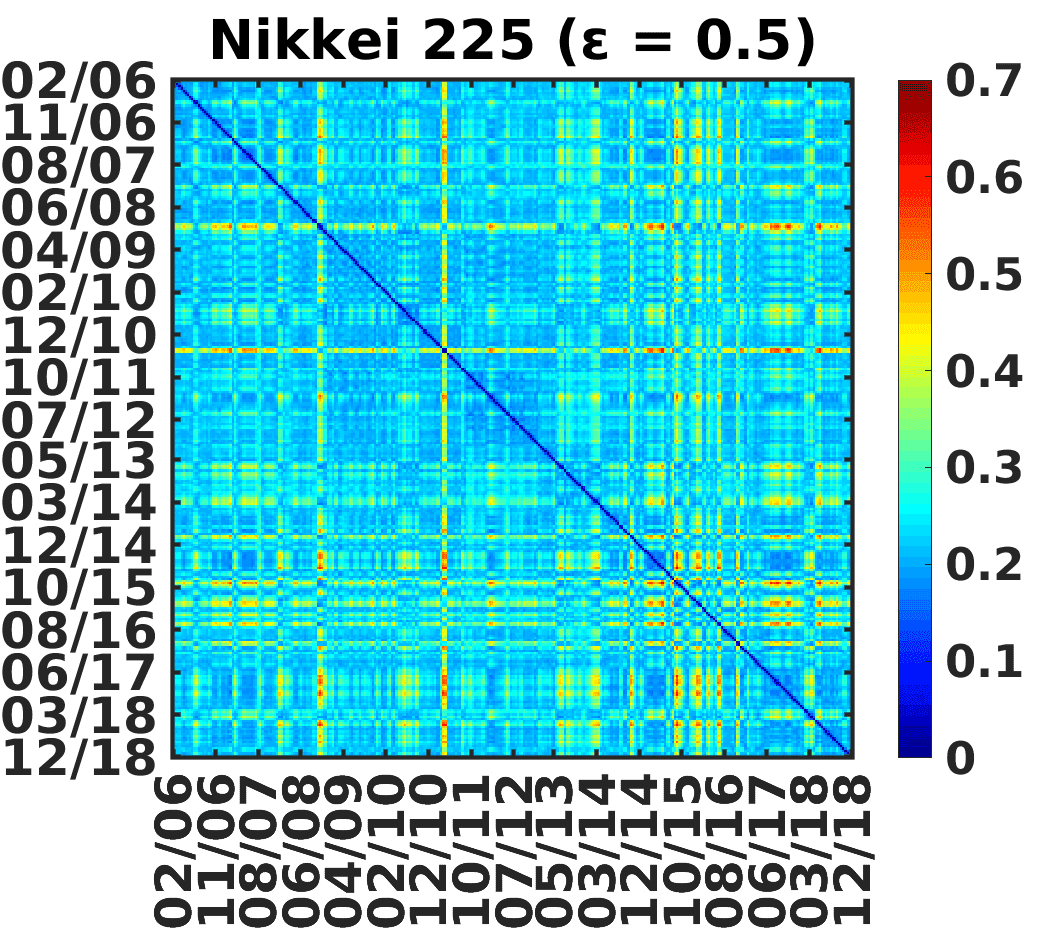}\llap{\parbox[b]{3.3in}{\textbf{{\Large (d)}}\\\rule{0ex}{2.7in}}}
		\caption{\text{Plots of similarity measure $\zeta (C,C')$ of S\&P 500 (top row) and Nikkei 225 (bottom row).} Similarity measure among $325$ correlation matrices of S\&P 500 (top row) and $320$ of Nikkei 225 (bottom row): (a) and (c) without noise-reduction $\epsilon=0$, and (b) and (d) with noise-reduction $\epsilon=0.7$ and $\epsilon=0.5$ over a period of 2006-2018, respectively.} \label{fig:similarity_USA_JPN_2019}
	 \end{figure}
\end{center}
\begin{center}
	\begin{figure}[t!]
	\centering
		\includegraphics[width=8cm]{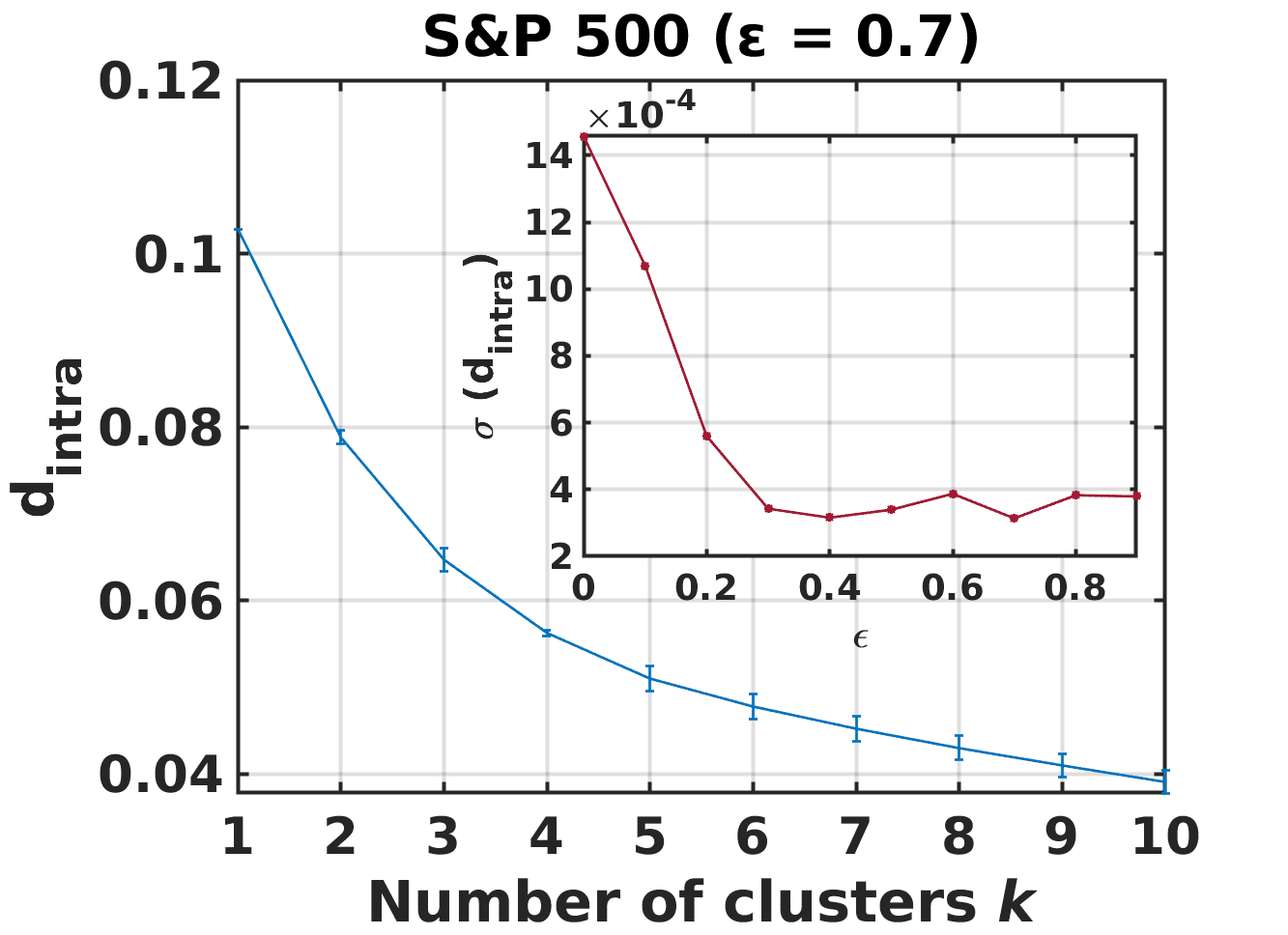}\llap{\parbox[b]{3.2in}{\textbf{({\Large a)}}\\\rule{0ex}{2.2in}}}
		\includegraphics[width=8cm]{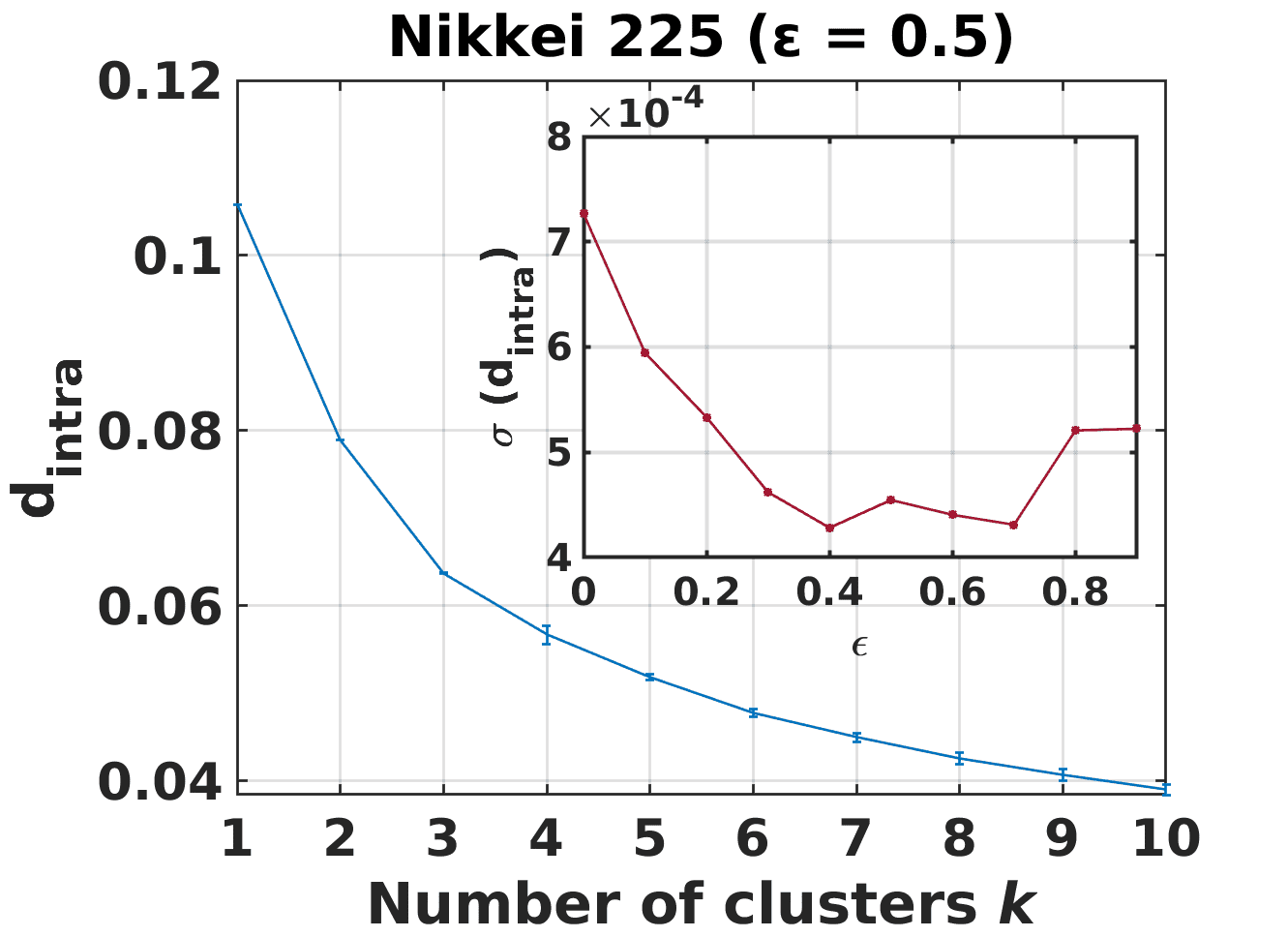}\llap{\parbox[b]{3.2in}{\textbf{{\Large (b)}}\\\rule{0ex}{2.2in}}}
		\includegraphics[width=8cm]{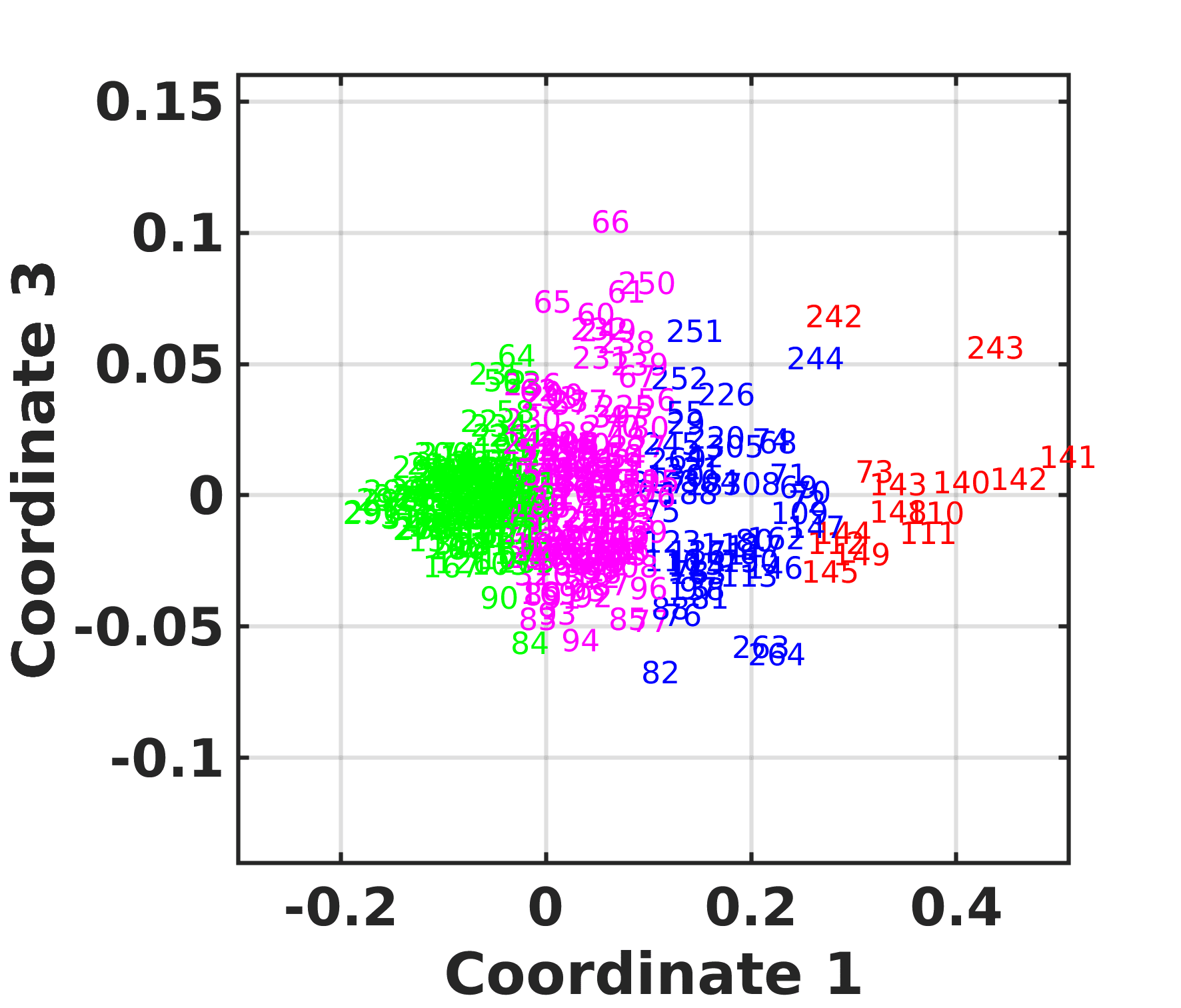}\llap{\parbox[b]{3.2in}{\textbf{{\Large (c)}}\\\rule{0ex}{2.5in}}}
		\includegraphics[width=8cm]{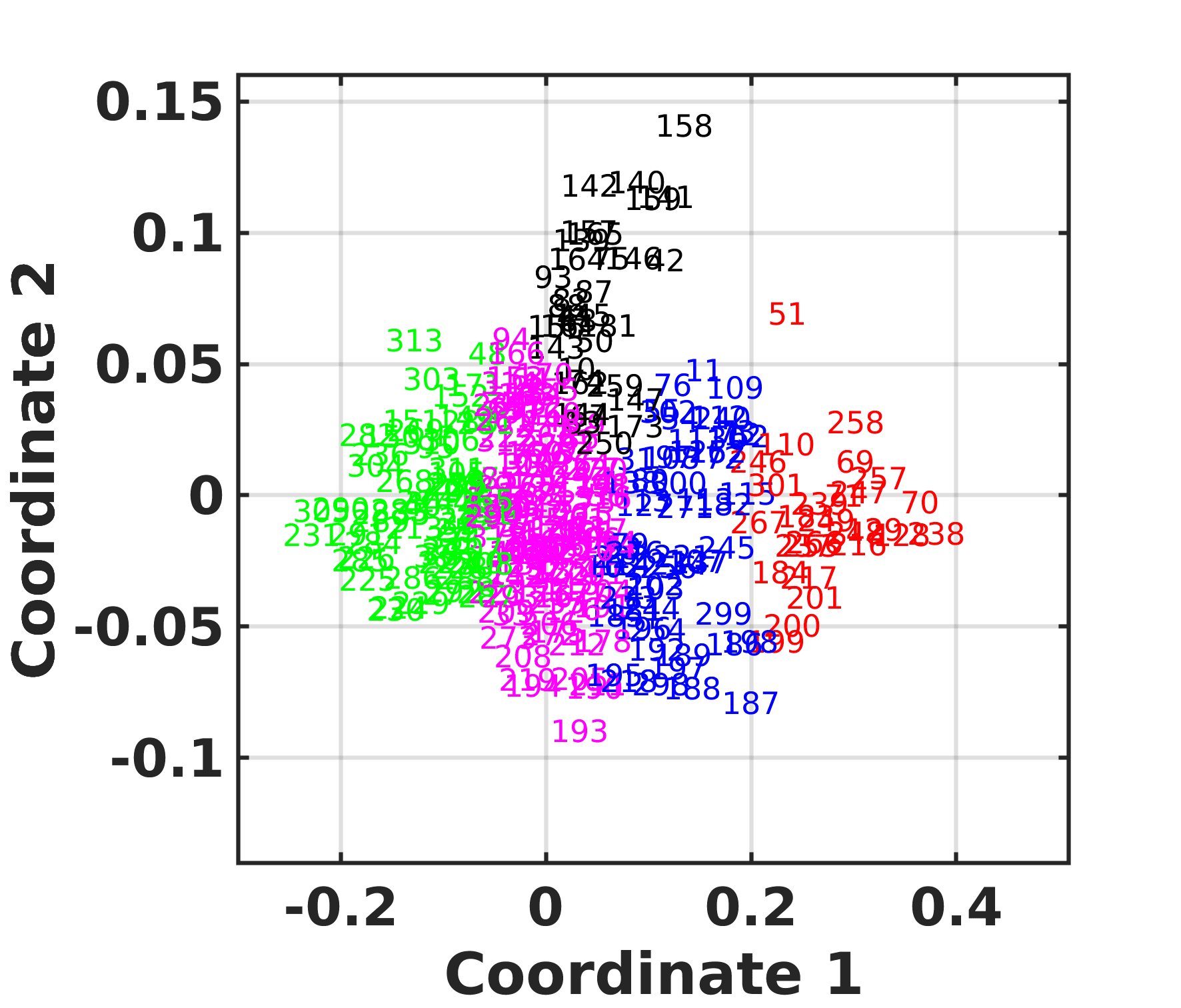}\llap{\parbox[b]{3.2in}{\textbf{{\Large (d)}}\\\rule{0ex}{2.5in}}}\\
		\caption{Classification of market states using optimal intra cluster distances  $d_{intra}$ for the S\&P 500 and Nikkei 225. The measure of the $d_{intra}$ calculated for different number of clusters is shown in (a) and (b) for the S\&P 500 and Nikkei 225 over a period of 2006-2018, respectively. The plots  show the minima of standard deviations of intracluster distance $\sigma (d_{intra})$ at $k=4$ for S\&P 500 and $k=5$ for Nikkei 225, respectively, which correspond to the ``optimal'' number of clusters. $k$-means clustering is used to cluster the S\&P 500 and Nikkei 225 in four and five number of cluster, respectively. The alignment of the clusters is different in two markets: clusters in (c) S\&P 500 market are aligned along the average correlation and trivial but for (d) Nikkei 225 market one more cluster forms between second and fourth cluster with nearly same average correlation. }
		\label{fig:similarity_USA_JPN_2018}
	 \end{figure}
\end{center}
\begin{center}
	\begin{figure}[t!]
	\centering
		\includegraphics[width=.99\linewidth]{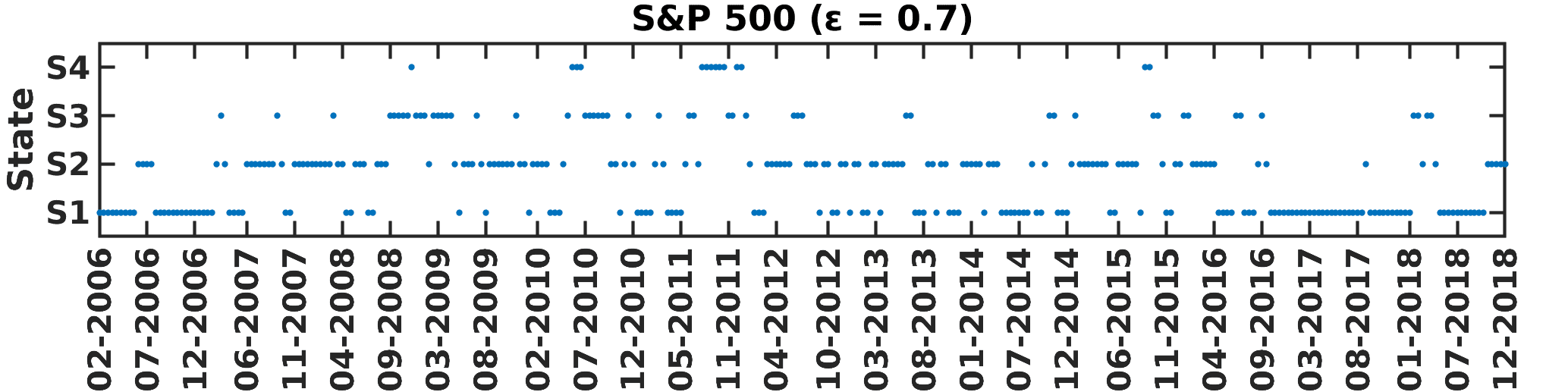}\llap{\parbox[b]{6.8in}{\textbf{{\Large (a)}}\\\rule{0ex}{1.7in}}}\\
		\includegraphics[width=17cm]{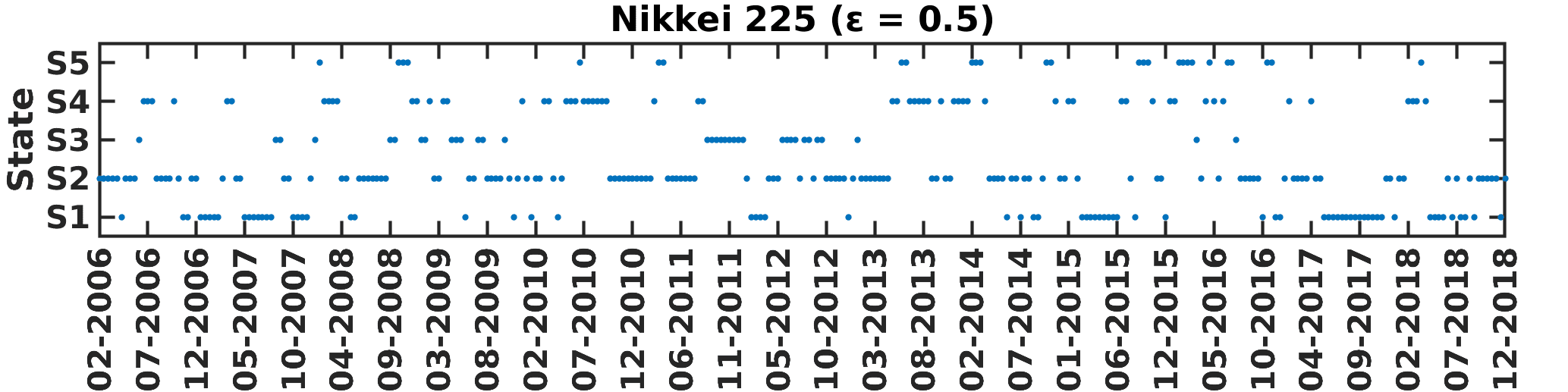}\llap{\parbox[b]{6.8in}{\textbf{{\Large (b)}}\\\rule{0ex}{1.7in}}}\\
		\caption{Market state dynamics of stock markets: (a) Evolution of S\&P 500 market through the transitions among four different characterized states ($S1, S2, S3$, and $S4$) for the period of $2006-2018$ and (b) Evolution of Nikkei 225 market though the transitions among five different characterized states ($S1, S2, S3$, $S4$, and $S5$) for the period of $2006-2018$. The probability to remaining in the same state or transition to nearby states are relatively high.} \label{fig:market_dynamics_Supp}
	 \end{figure}
\end{center}
\begin{center}
	\begin{figure}[h!]
	\centering
	\includegraphics[width=8cm]{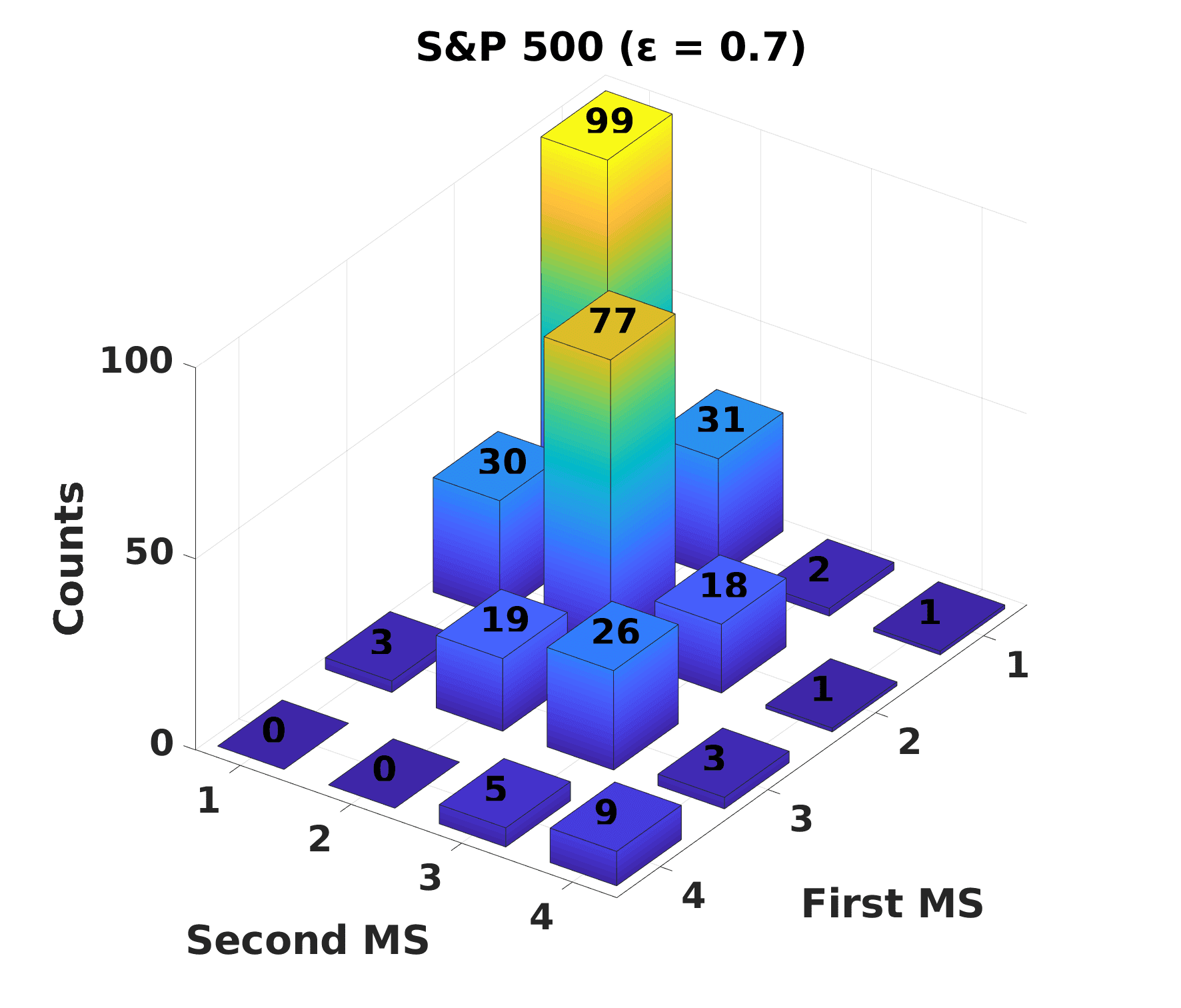}\llap{\parbox[b]{3.2in}{\textbf{{\Large (a)}}\\\rule{0ex}{2.4in}}}
	\includegraphics[width=8cm]{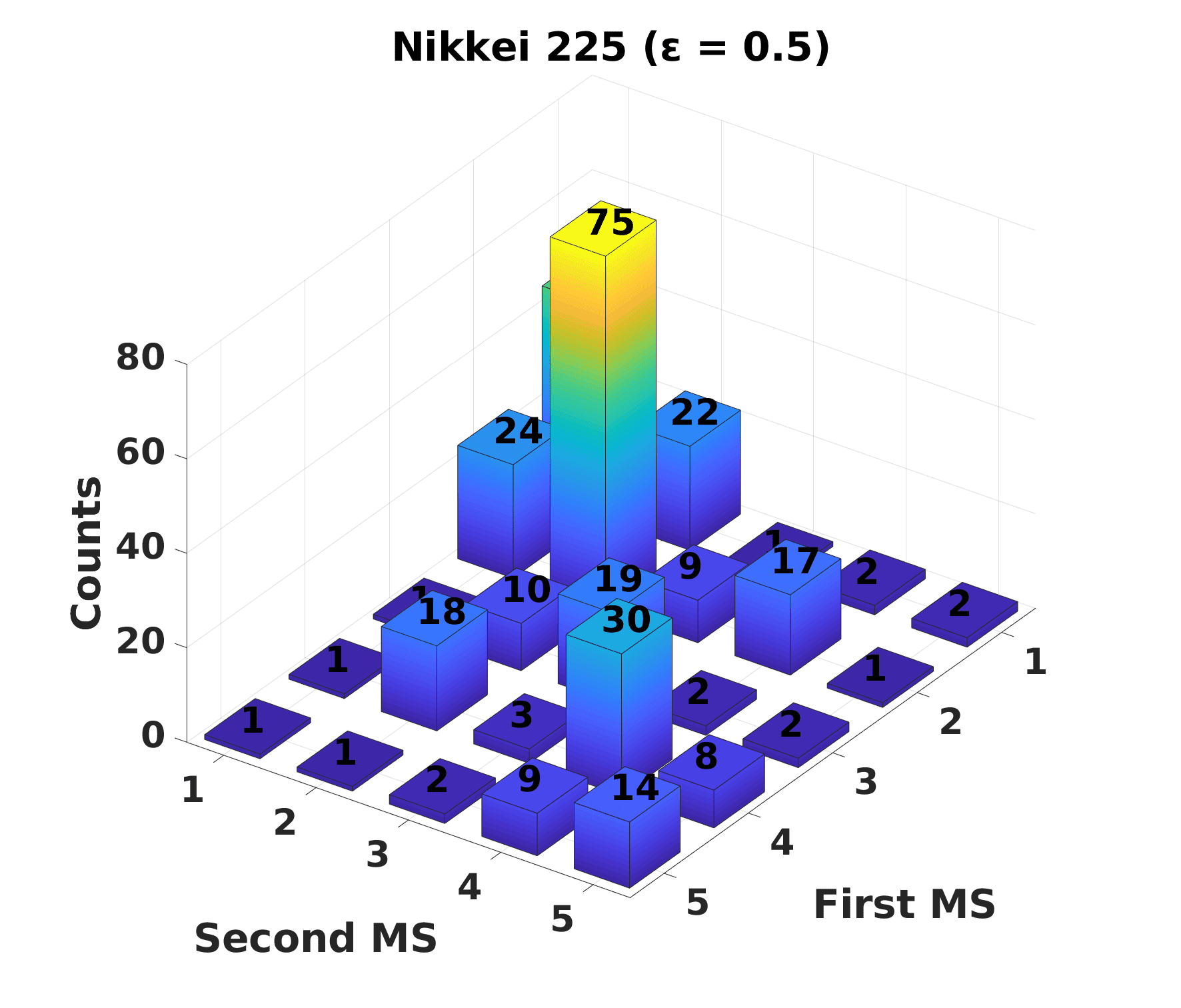}\llap{\parbox[b]{3.2in}{\textbf{{\Large (b)}}\\\rule{0ex}{2.4in}}}	\\
	\includegraphics[width=10cm]{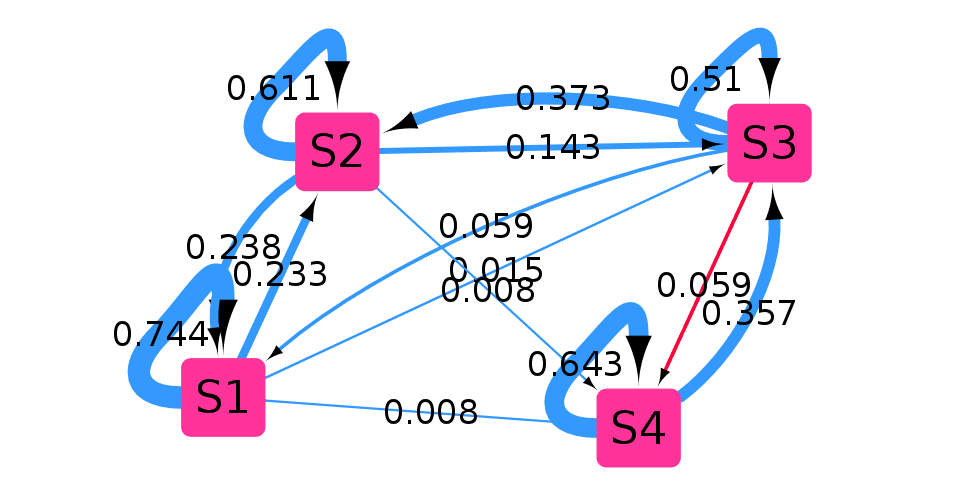}\llap{\parbox[b]{3.4in}{\textbf{{\Large (c)}}\\\rule{0ex}{2.in}}}
	\includegraphics[width=10cm]{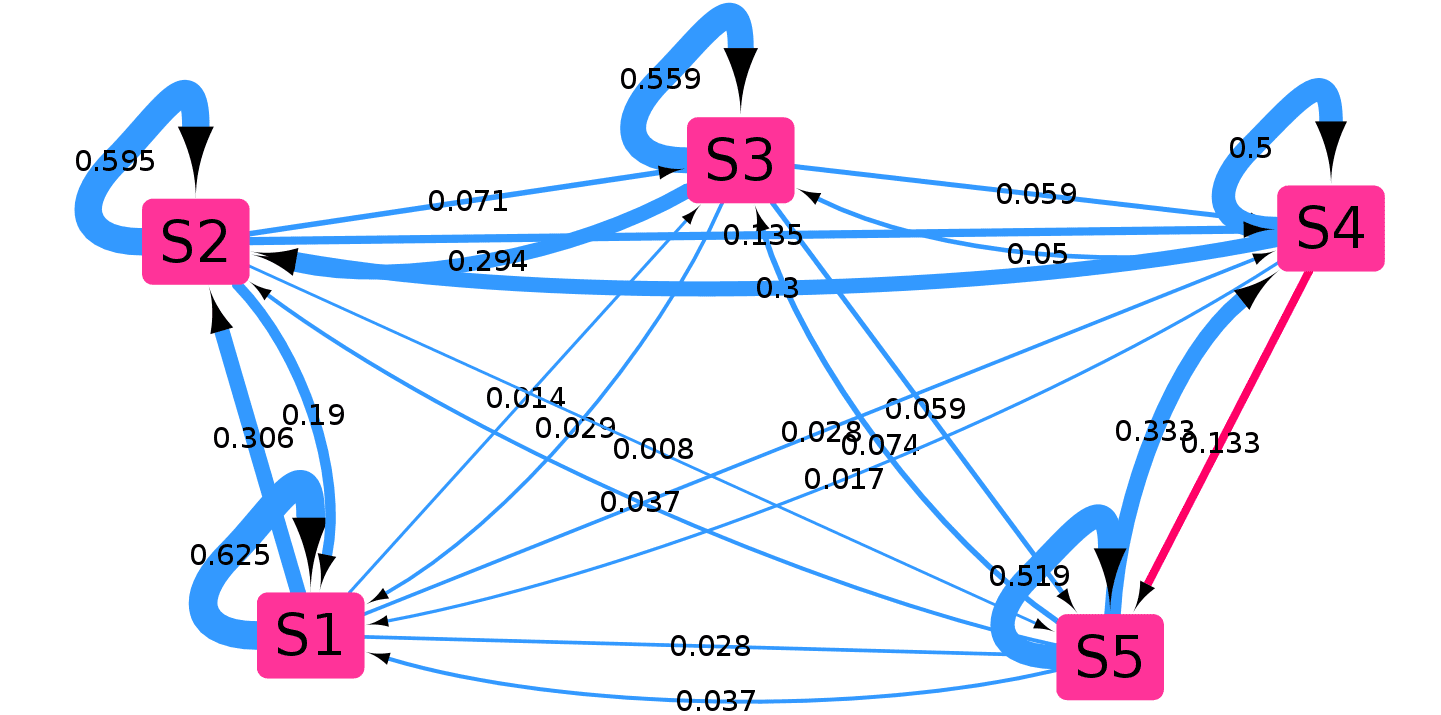}\llap{\parbox[b]{3.4in}{\textbf{{\Large (d)}}\\\rule{0ex}{2.in}}}	
	\caption{Bar plots of transition counts (frequencies) of paired market states (MS) for S\&P 500 and Nikkei 225 are shown in (a) and (b), respectively. The networks plots of transition probabilities between different states of (c) S\&P 500 and (d) Nikkei 225, respectively. The transition probability of market state transition of $S3$ to $S4$ is $6\%$ for S\&P 500 market, and similarly, for Nikkei 225, the probability of market state transition of $S4$ to $S5$ is $13\%$.}\label{transition probabilities}
	\end{figure}
\end{center}
\begin{figure*}[ht!]
\centering
\includegraphics[width=5cm]{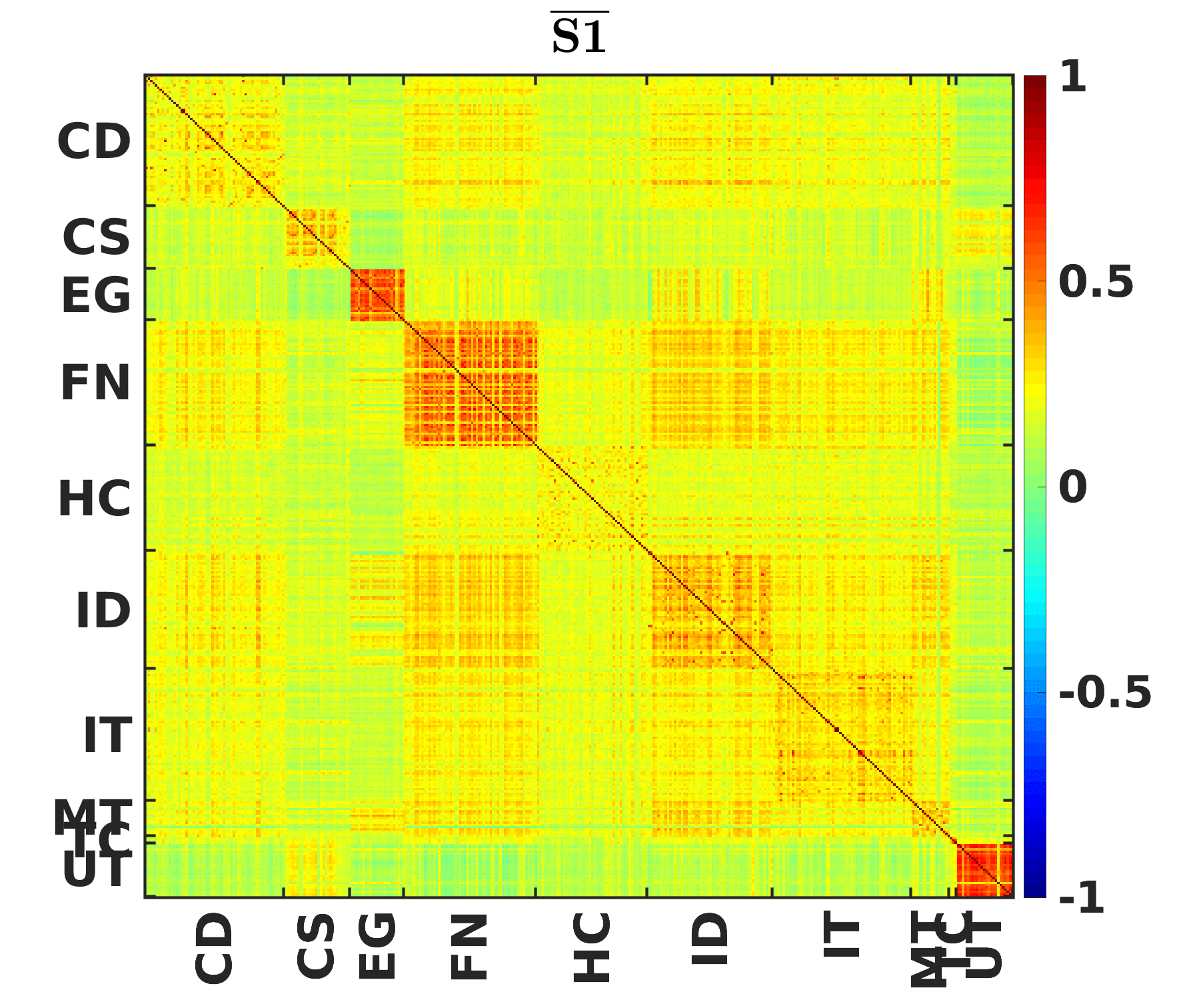}\llap{\parbox[b]{2.1in}{\textbf{{\Large (a)}}\\\rule{0ex}{1.7in}}}
\includegraphics[width=5cm]{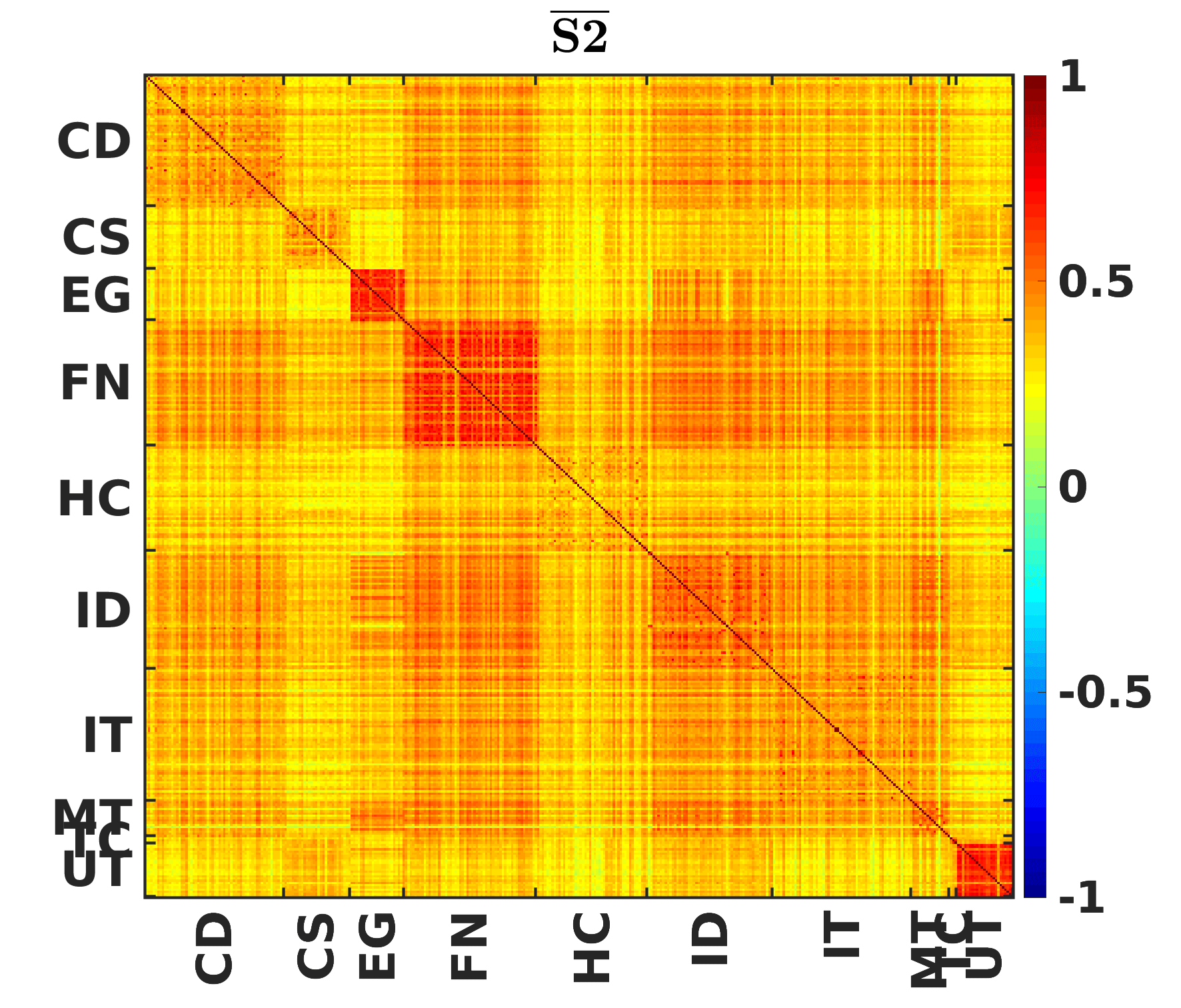}\llap{\parbox[b]{2.1in}{\textbf{{\Large (b)}}\\\rule{0ex}{1.7in}}}\\
\includegraphics[width=5cm]{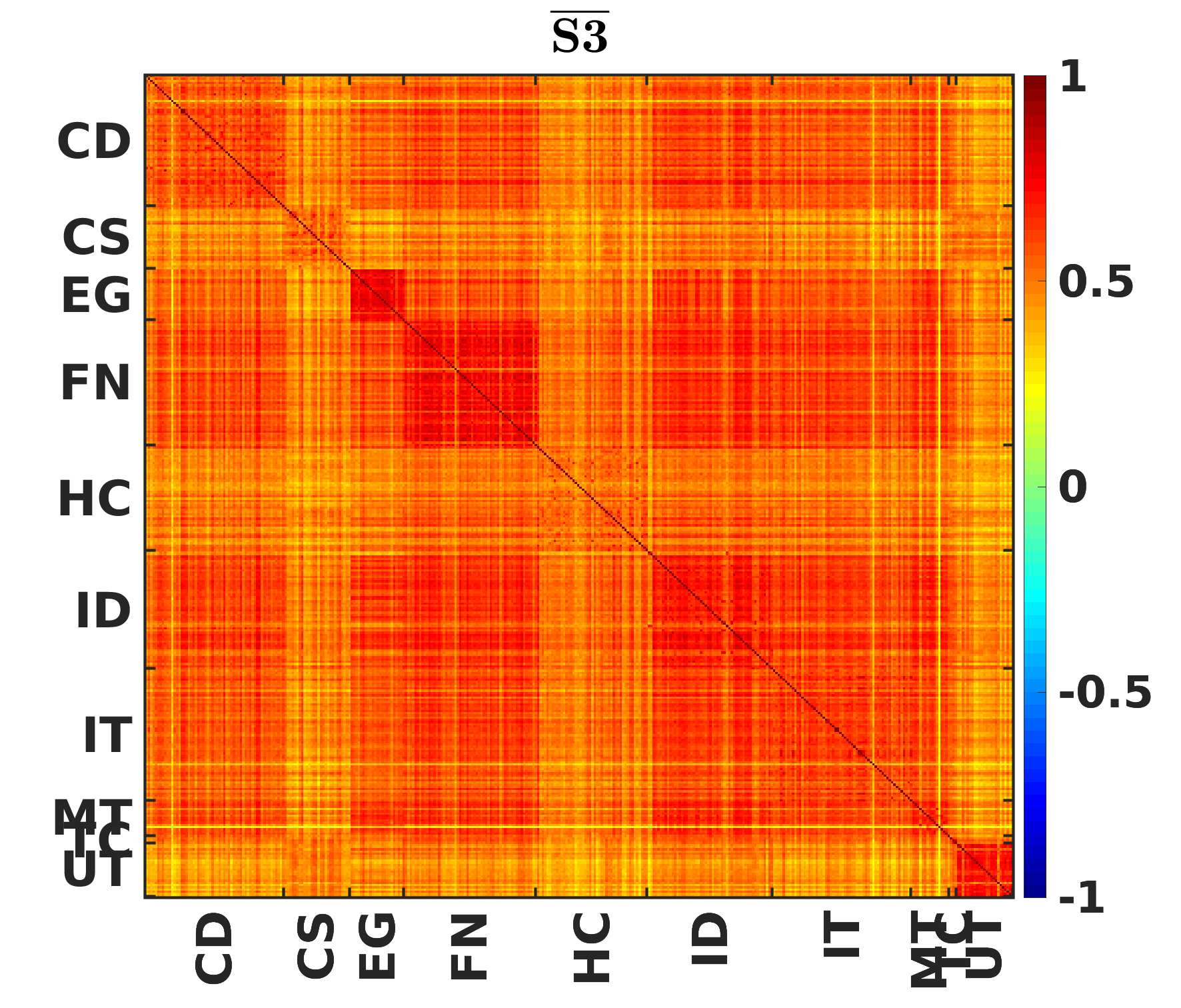}\llap{\parbox[b]{2.1in}{\textbf{{\Large (c)}}\\\rule{0ex}{1.7in}}}
\includegraphics[width=5cm]{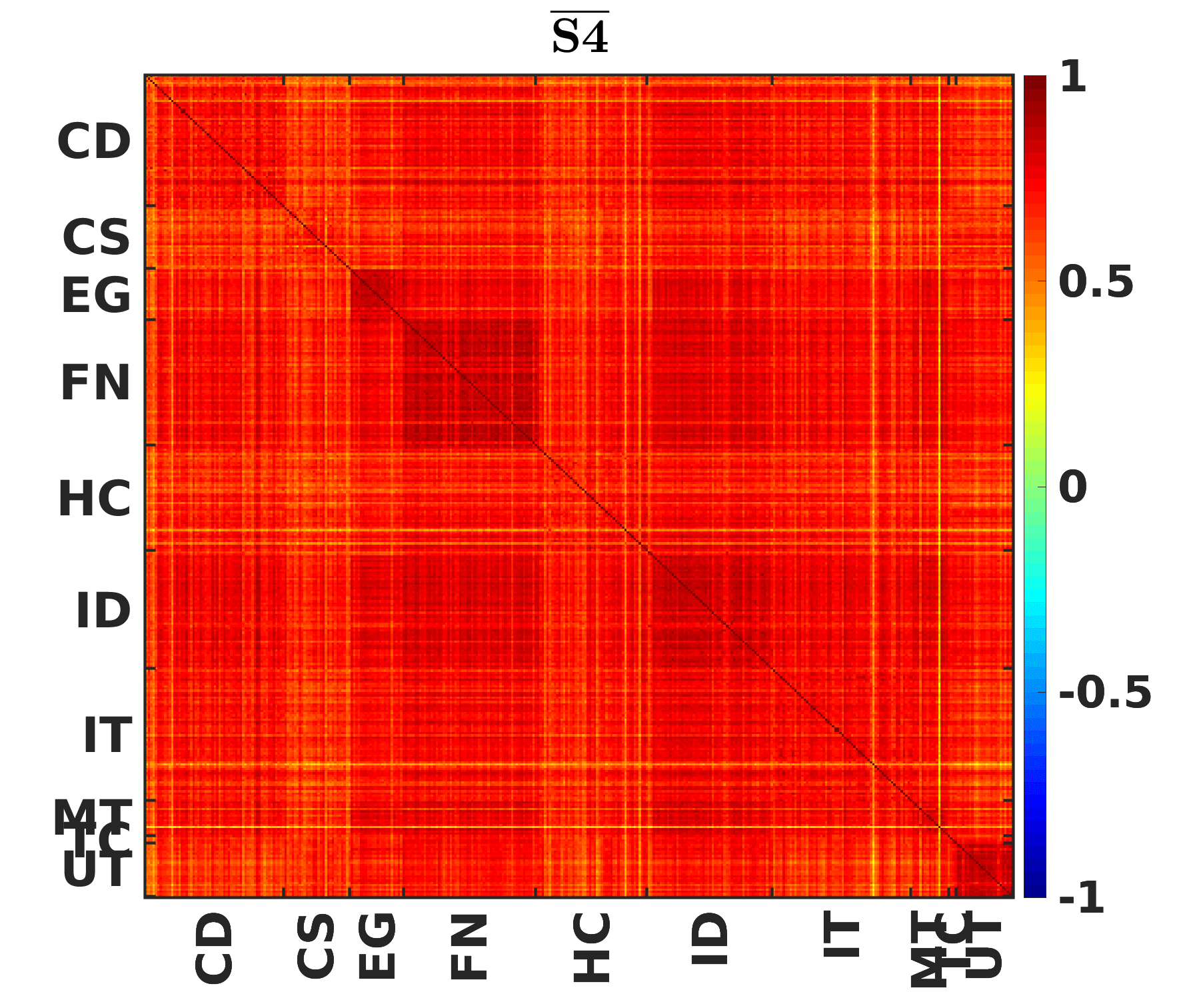}\llap{\parbox[b]{2.1in}{\textbf{{\Large (d)}}\\\rule{0ex}{1.7in}}}\\
\includegraphics[width=6cm]{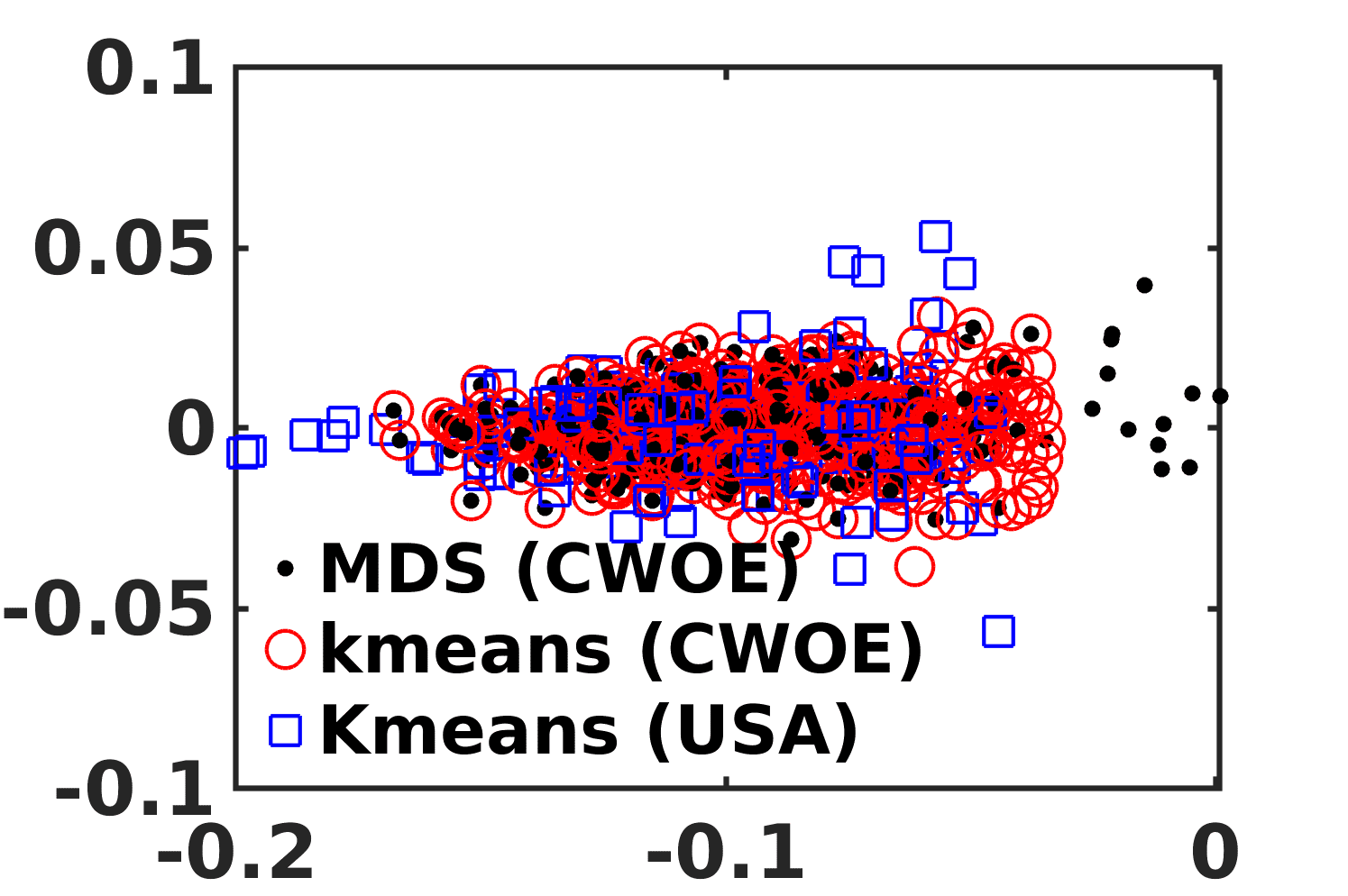}\llap{\parbox[b]{2.5in}{\textbf{{\Large (e)}}\\\rule{0ex}{1.6in}}}
\includegraphics[width=6cm]{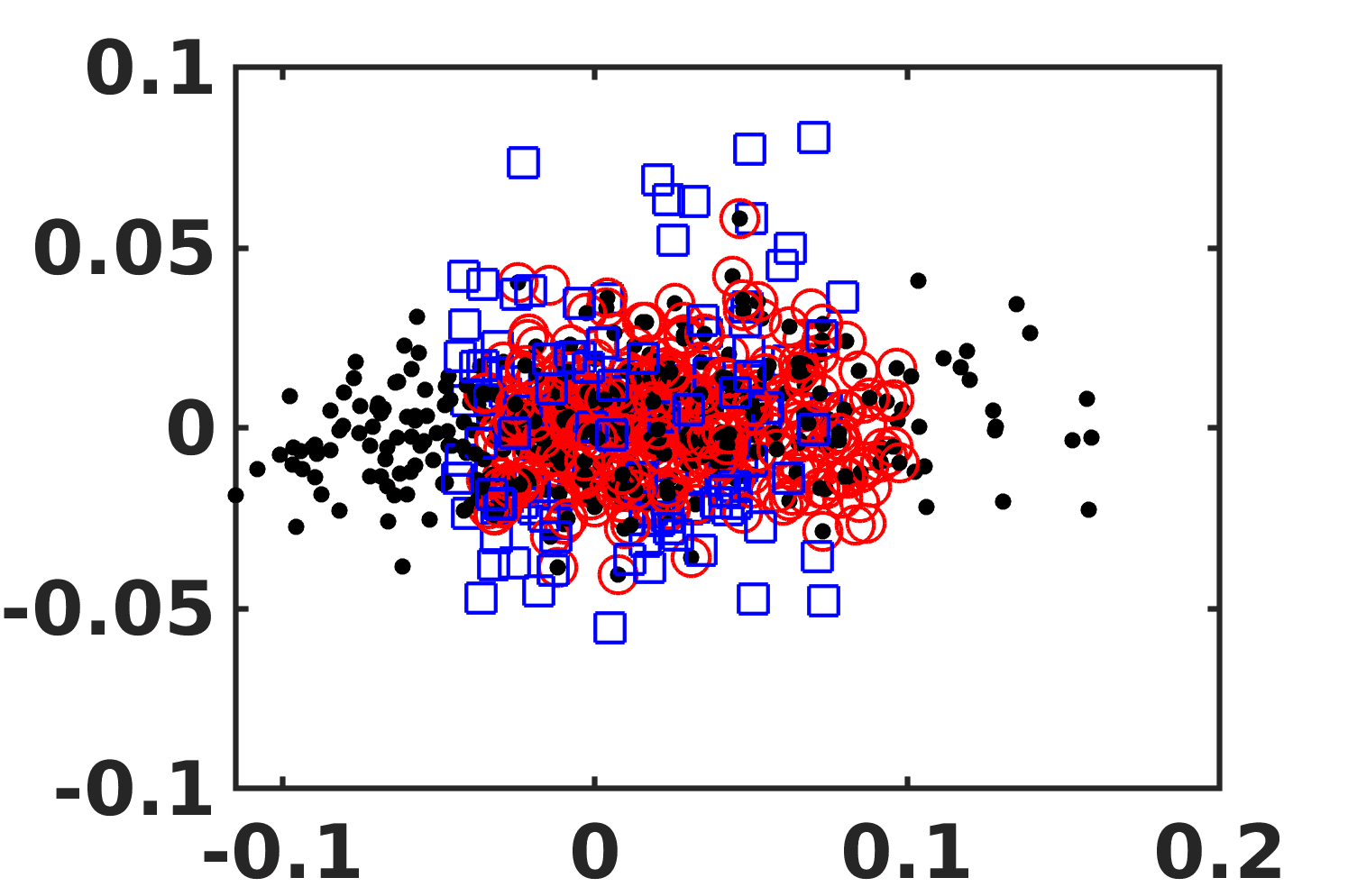}\llap{\parbox[b]{2.5in}{\textbf{{\Large (f)}}\\\rule{0ex}{1.6in}}}\\
\includegraphics[width=6cm]{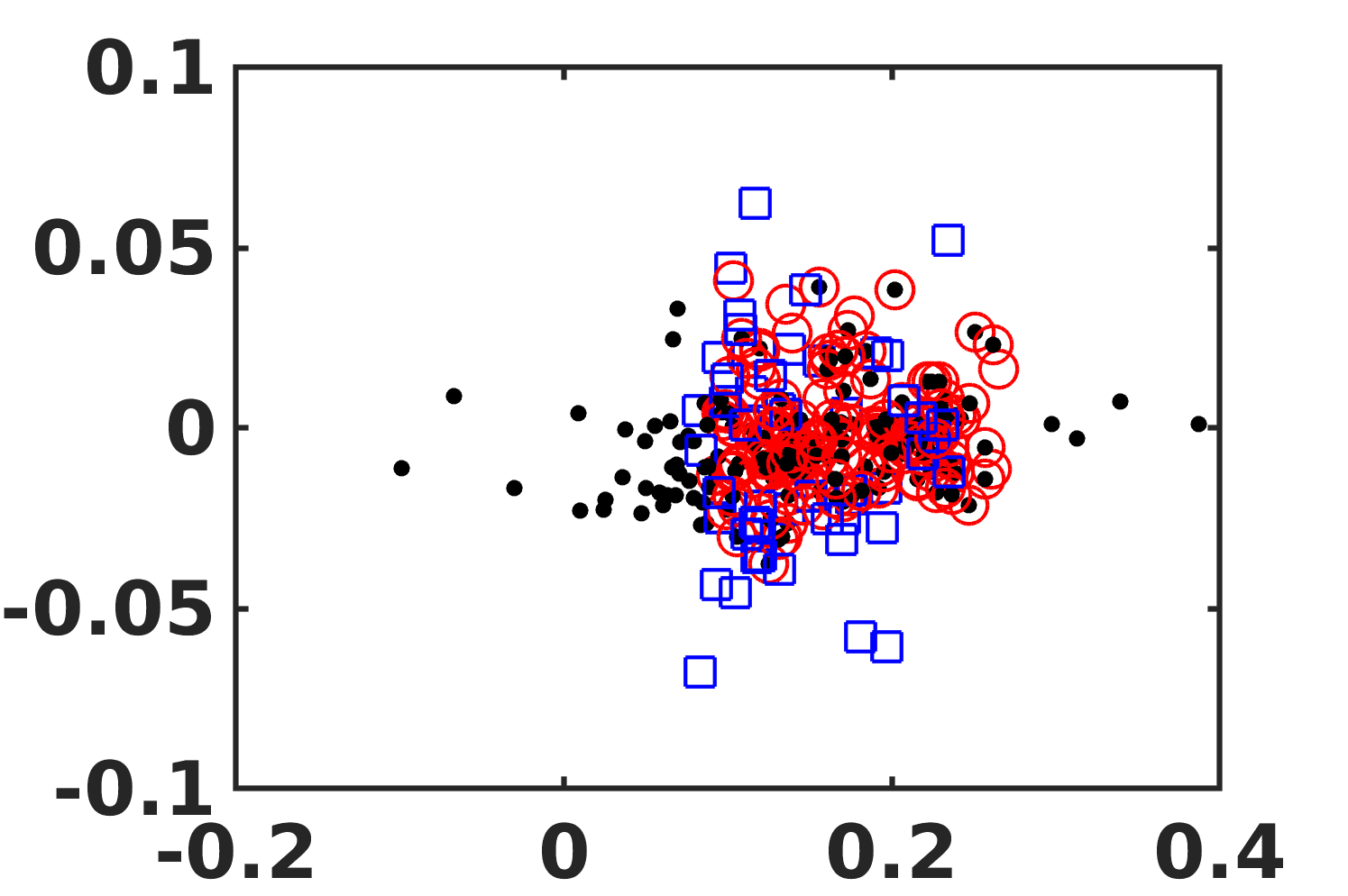}\llap{\parbox[b]{2.5in}{\textbf{{\Large (g)}}\\\rule{0ex}{1.6in}}}
\includegraphics[width=6cm]{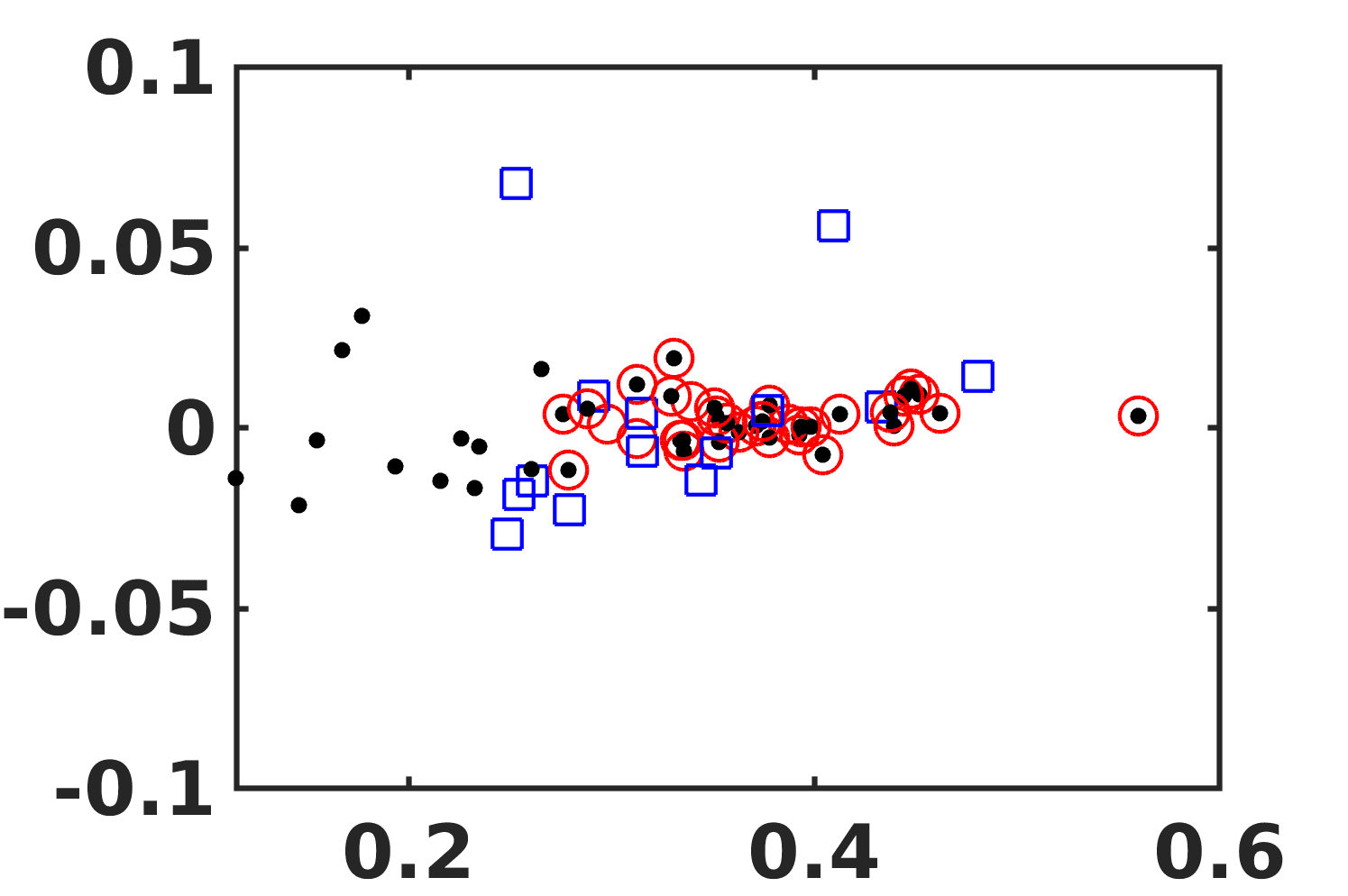}\llap{\parbox[b]{2.5in}{\textbf{{\Large (h)}}\\\rule{0ex}{1.6in}}}
\caption{Average correlation of each market state of S\&P 500 market and clustering of correlated Wishart orthogonal ensembles (CWOE). (a-d) show mean correlation matrices $\overline{Si}$ evaluated over all the frames correspond to each market state $S1,S2,S3,\&~S4$ but for the original correlation matrices ($\epsilon=0$). It shows the average behavior of each market states of S\&P 500 over a period of 13 years (2006-2018). (e) Dots (MDS (CWOE)) in the plot show the multidimensional scaling map of CWOE using mean correlations martix as $\overline{S1}$ and constructing three times bigger ensemble than $S1$ market state of the S\&P 500 market with same noise-suppression $\epsilon=0.7$. Red circles ($k$-means (CWOE)) in plot show the points of the $S1$ cluster of $k$-means clustering performed on CWOE (dots). Blue squares (k-means (USA)) in the plot show the $k$-means clustering on the emperical data of S\&P 500. $k$-means clustering on the CWOE and S\&P 500 data shows qualitatively similar behavior. (f), (g), and (h) show the same using mean correlation matrices $\overline{S2}, \overline{S3}$, and $\overline{S4}$ states, respectively.}\label{mean_MS}
\end{figure*}
\begin{figure*}[ht!]
\centering
\includegraphics[width=3.cm]{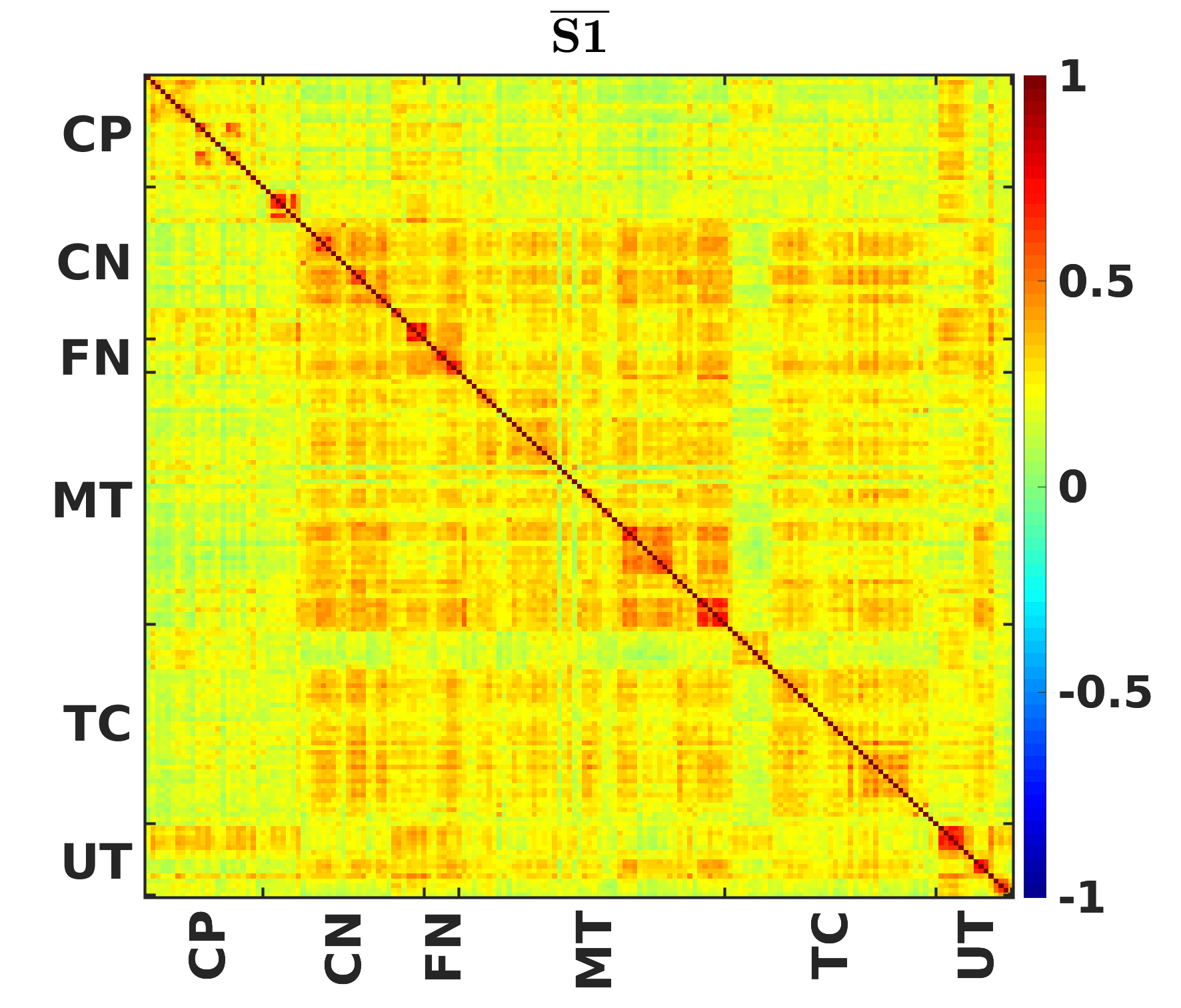}\llap{\parbox[b]{1.2in}{\textbf{{\Large (a)}}\\\rule{0ex}{1in}}}
\includegraphics[width=3.cm]{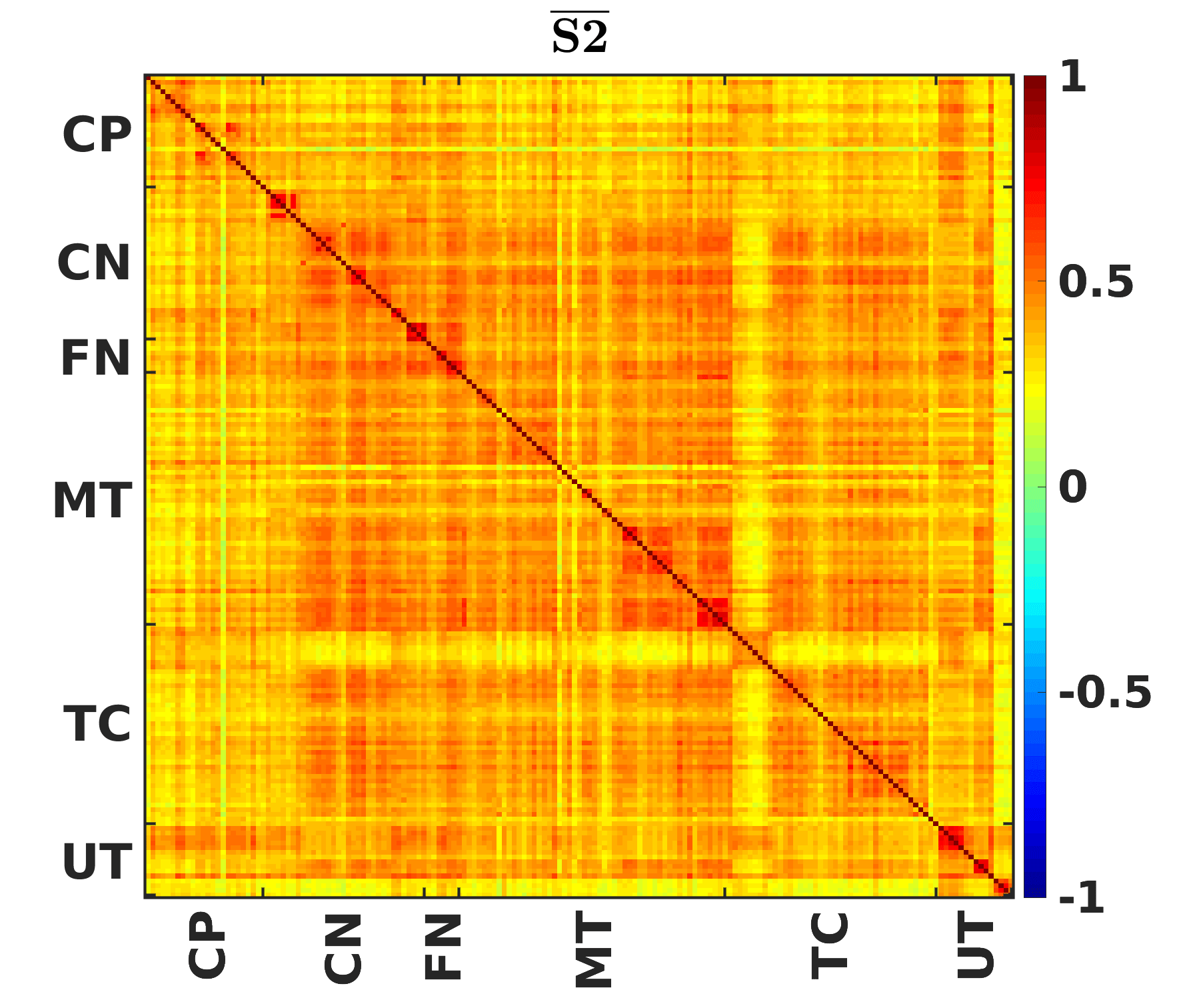}\llap{\parbox[b]{1.2in}{\textbf{{\Large (b)}}\\\rule{0ex}{1in}}}
\includegraphics[width=3.cm]{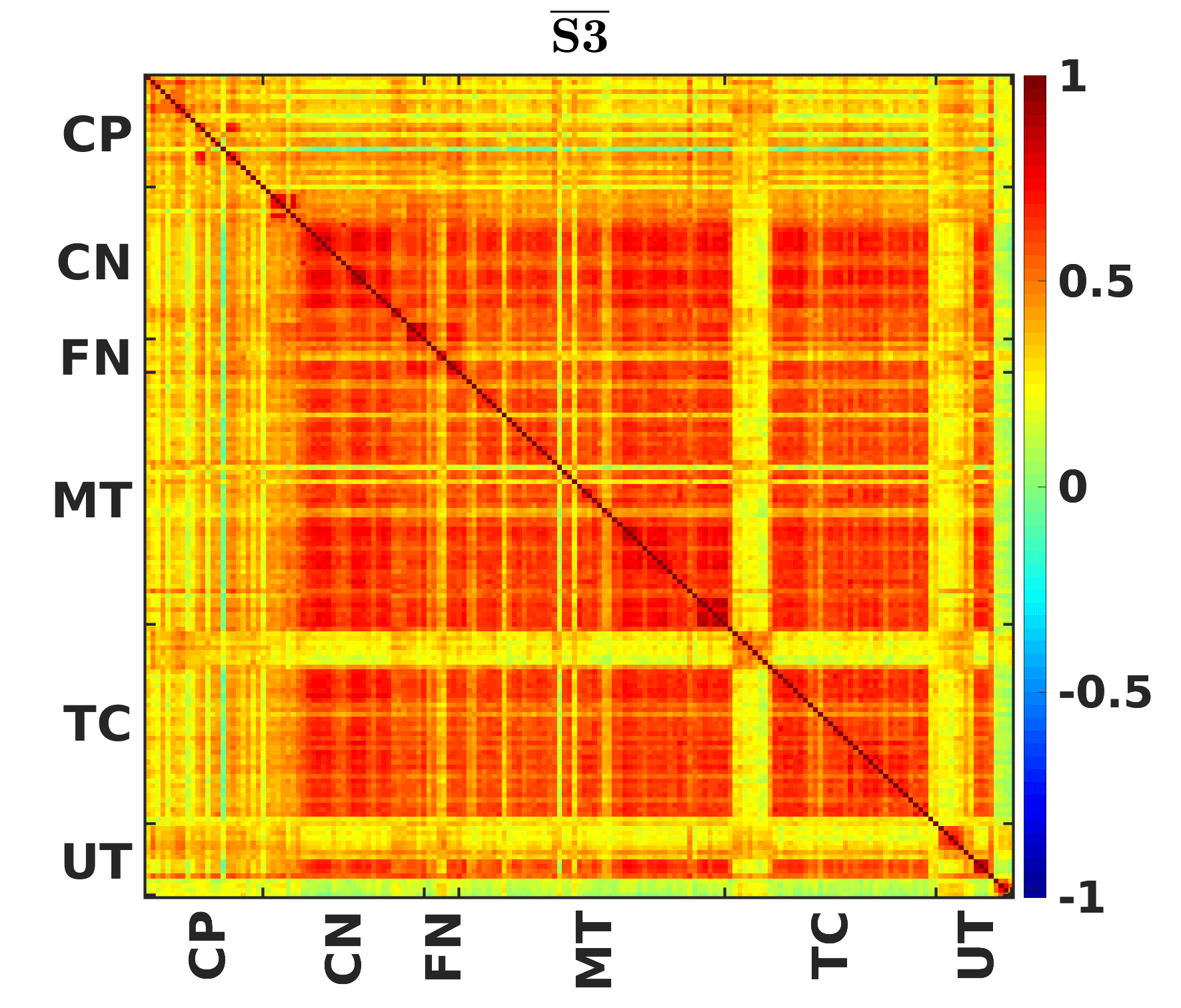}\llap{\parbox[b]{1.2in}{\textbf{{\Large (c)}}\\\rule{0ex}{1in}}}
\includegraphics[width=3.cm]{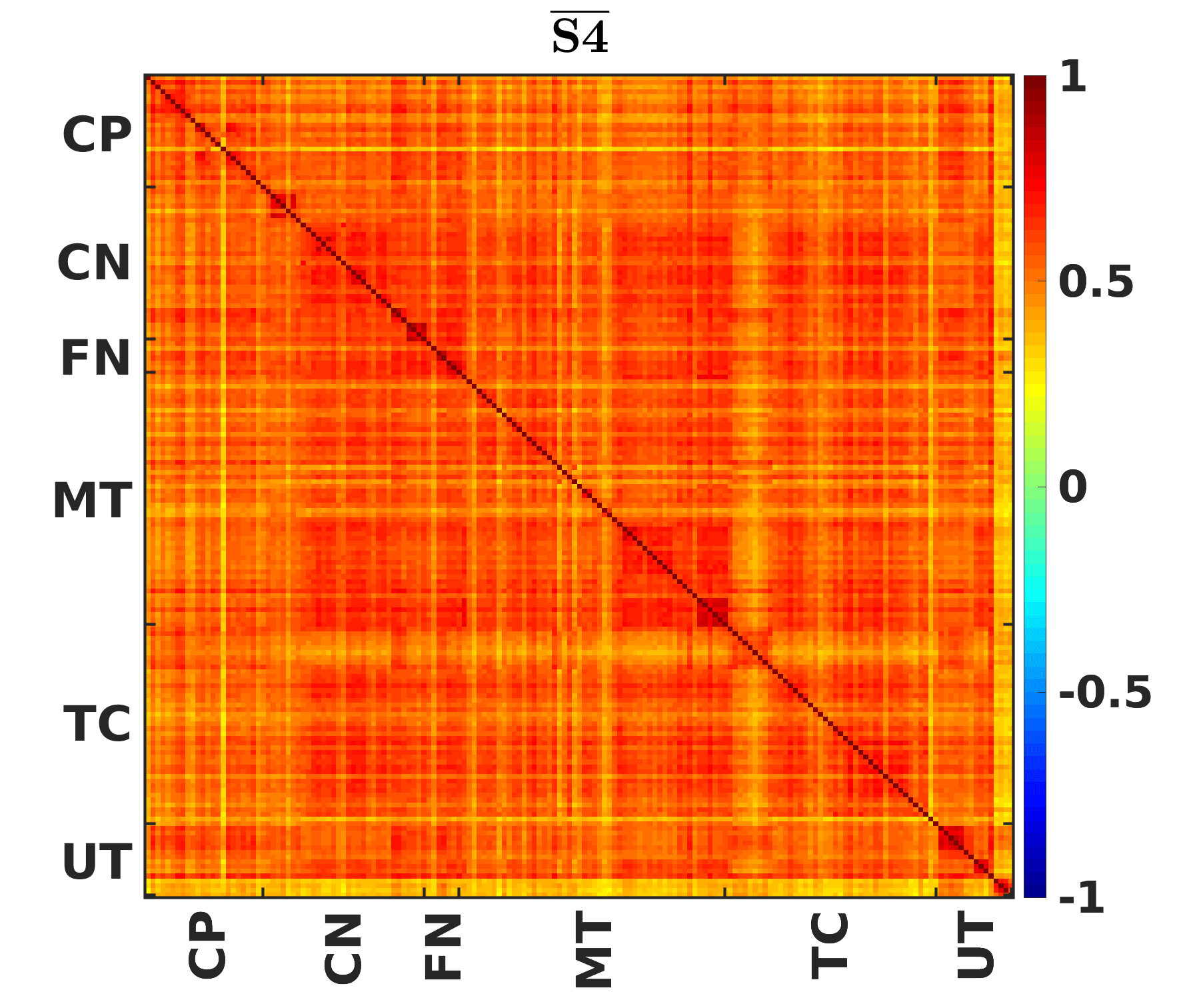}\llap{\parbox[b]{1.2in}{\textbf{{\Large (d)}}\\\rule{0ex}{1in}}}
\includegraphics[width=3.cm]{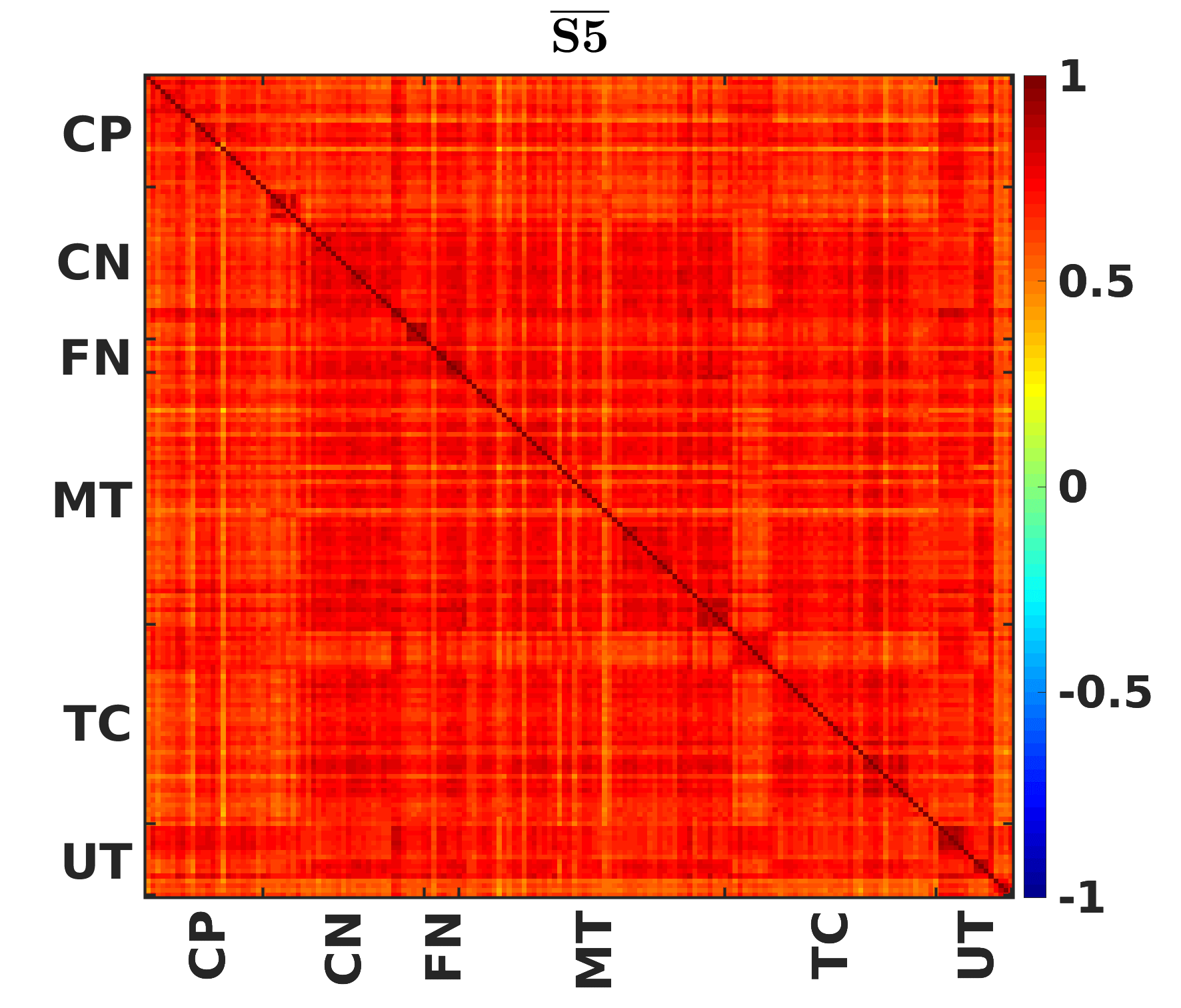}\llap{\parbox[b]{1.2in}{\textbf{{\Large (e)}}\\\rule{0ex}{1in}}}
\includegraphics[width=7cm]{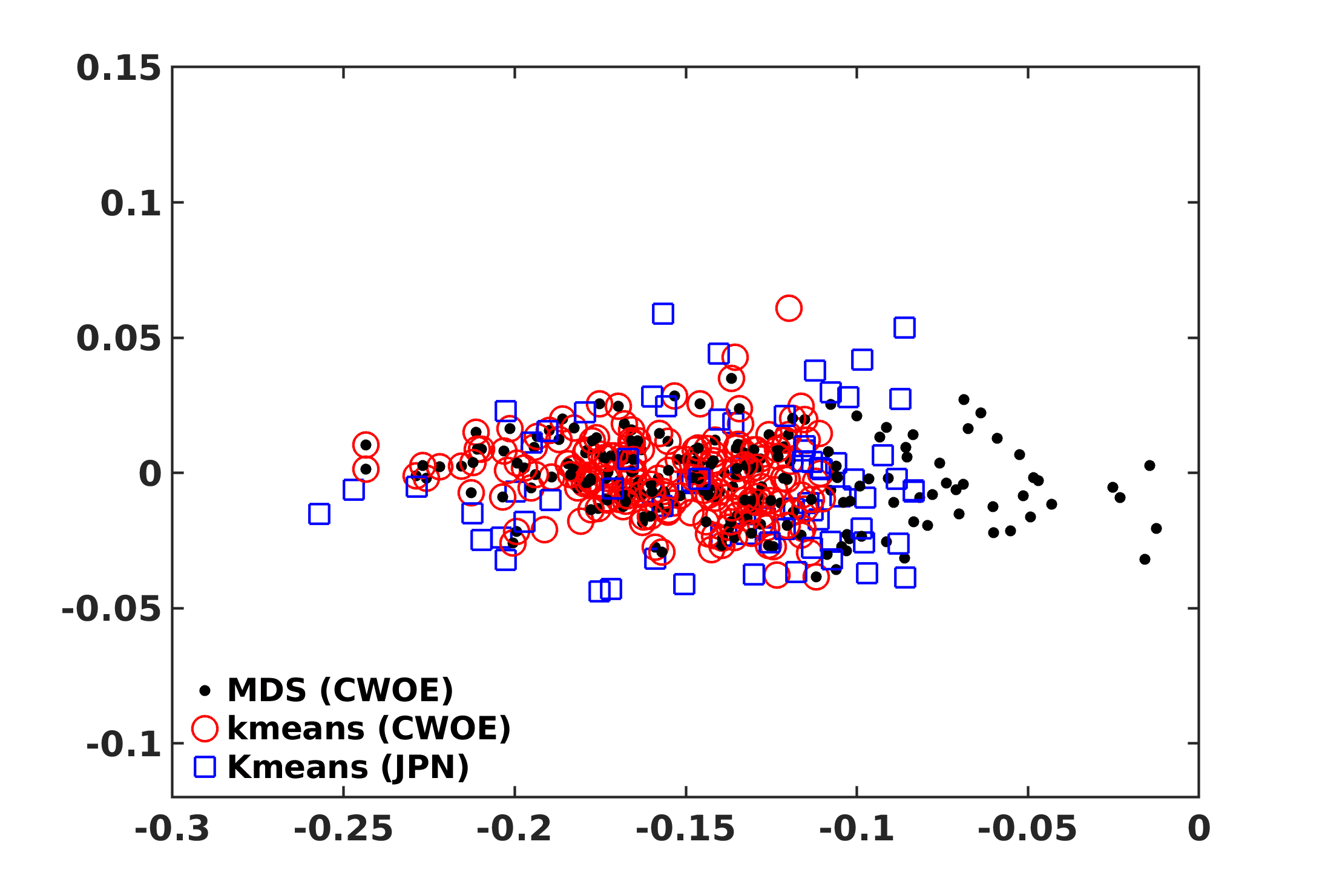}\llap{\parbox[b]{2.8in}{\textbf{{\Large (f)}}\\\rule{0ex}{1.8in}}}
\includegraphics[width=7cm]{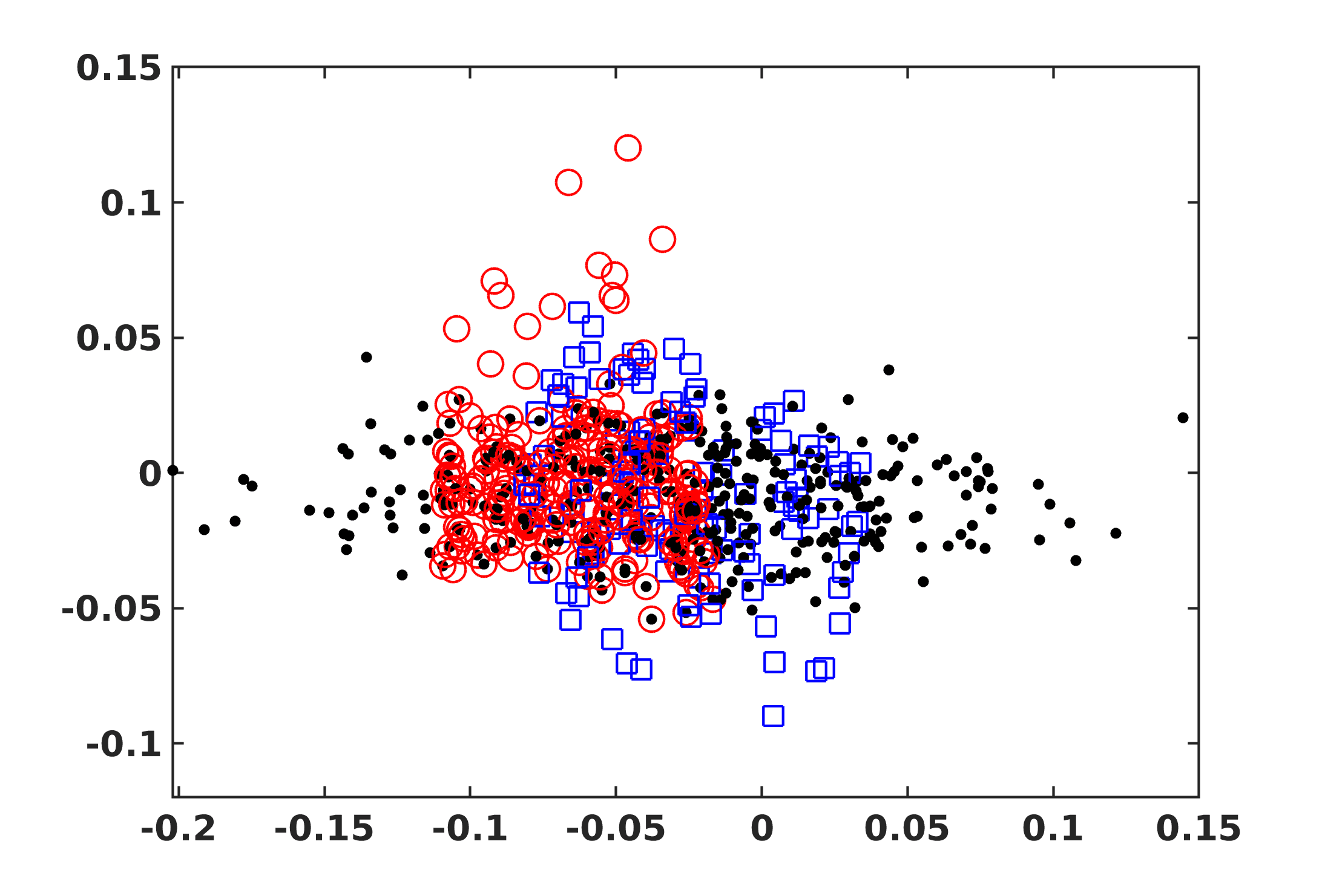}\llap{\parbox[b]{2.8in}{\textbf{{\Large (g)}}\\\rule{0ex}{1.8in}}}
\includegraphics[width=7cm]{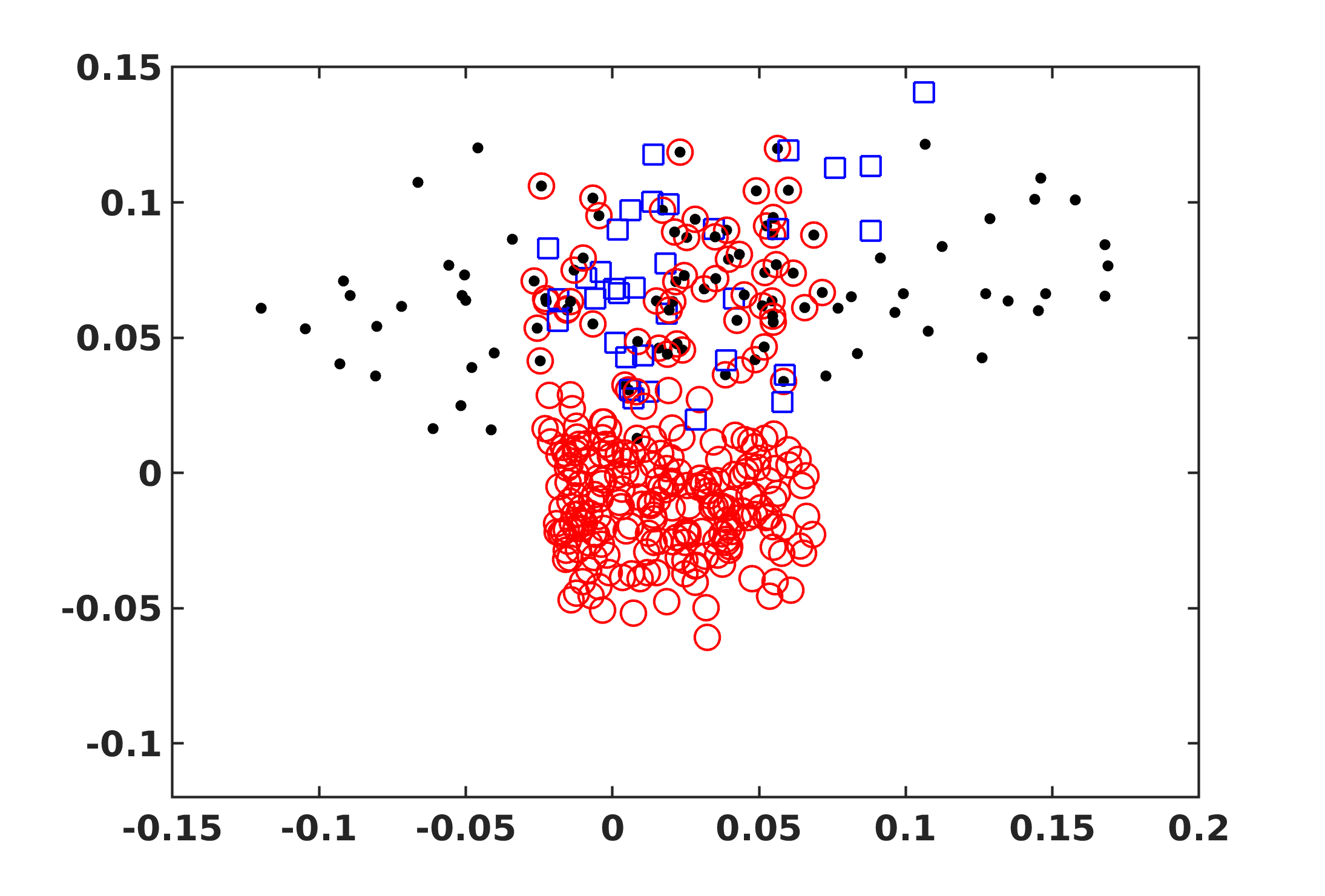}\llap{\parbox[b]{2.8in}{\textbf{{\Large (h)}}\\\rule{0ex}{1.8in}}}
\includegraphics[width=7cm]{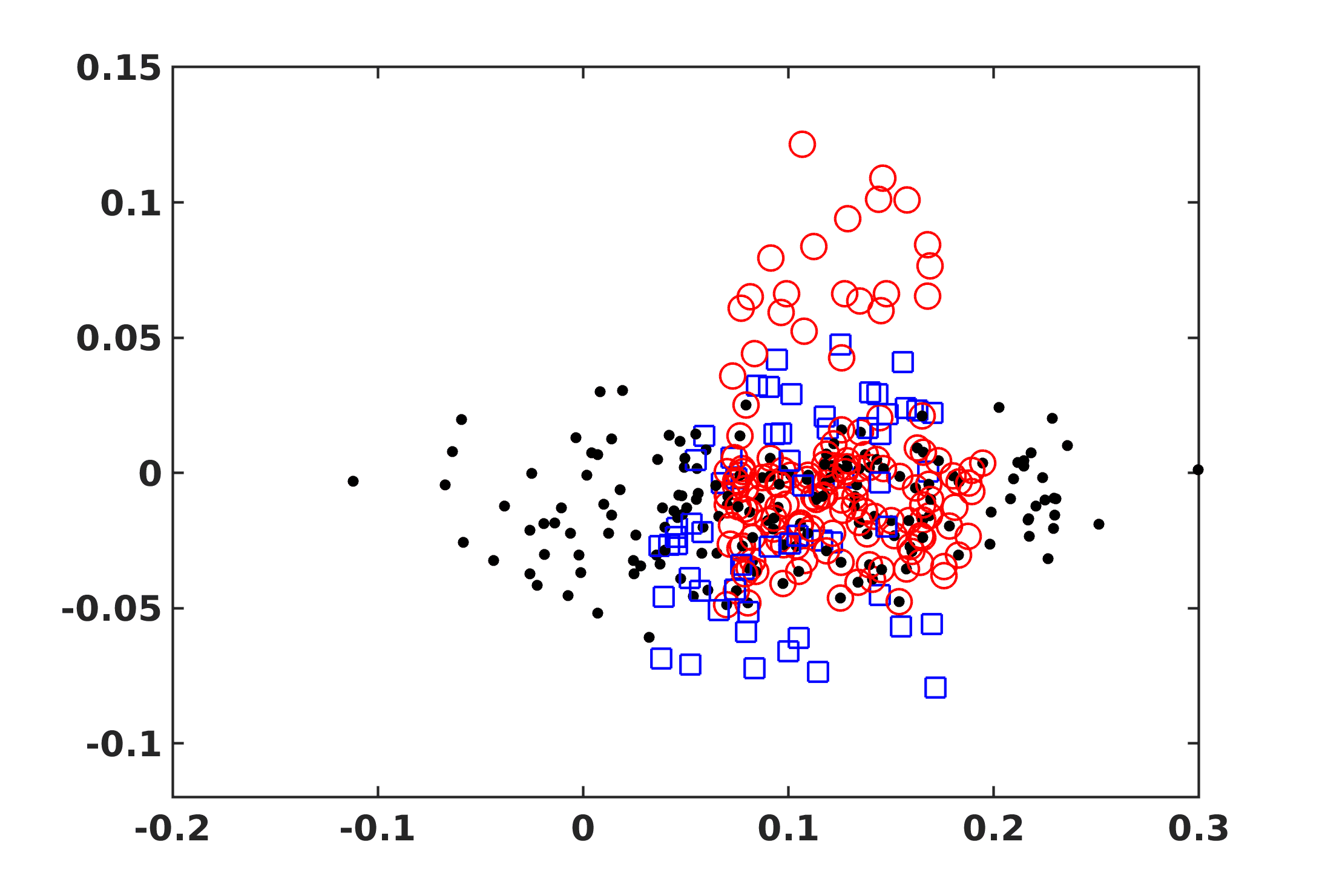}\llap{\parbox[b]{2.8in}{\textbf{{\Large (i)}}\\\rule{0ex}{1.8in}}}
\includegraphics[width=7cm]{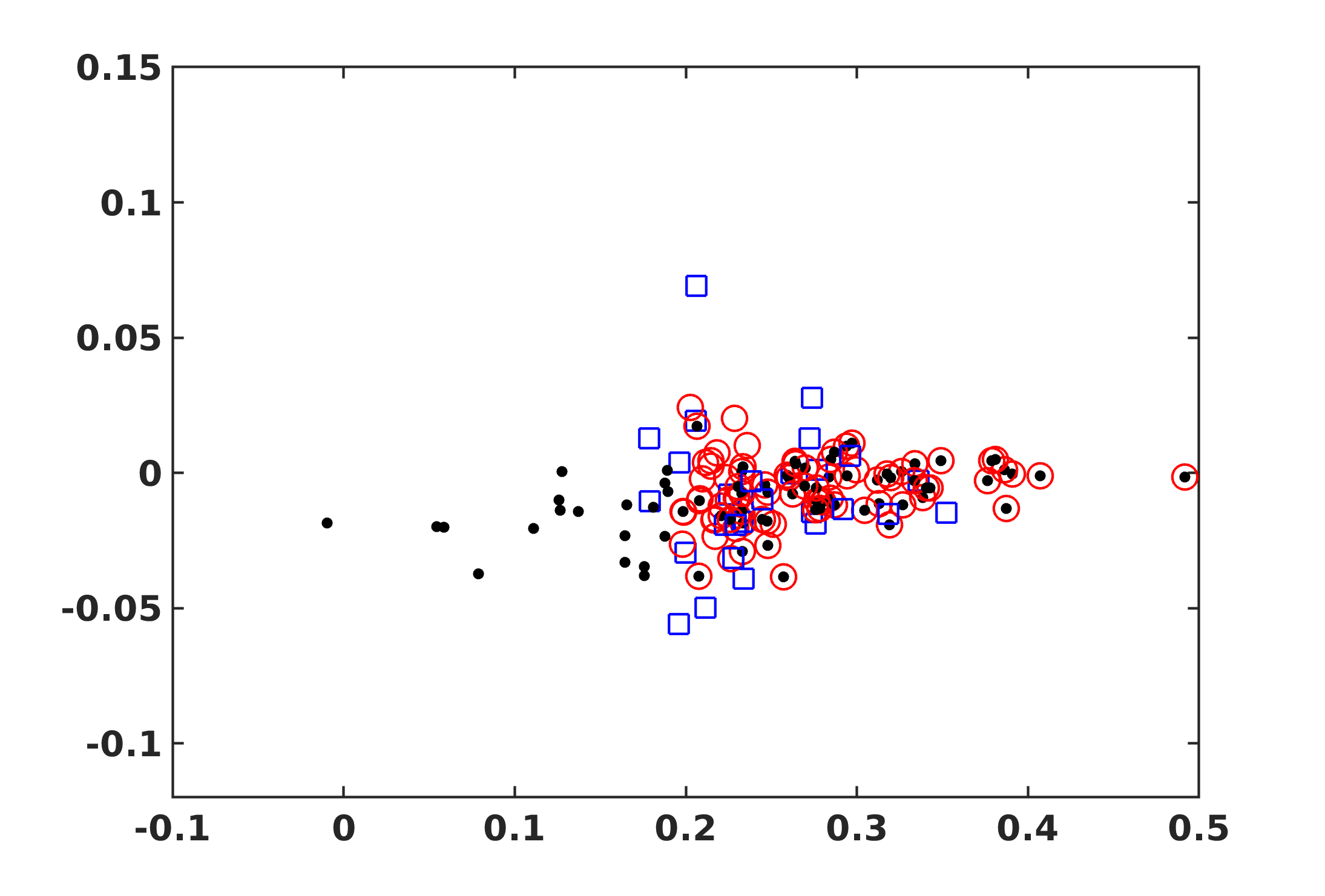}\llap{\parbox[b]{2.8in}{\textbf{{\Large (j)}}\\\rule{0ex}{1.8in}}}
\caption{Average correlation of each market state of Nikkei 225 market and clustering of correlated Wishart orthogonal ensembles (CWOE). (a-e) show mean correlation matrices $\overline{Si}$ evaluated over all the frames correspond to each market state $S1,S2,S3,S4,\&~S5$ but for the original correlation matrices ($\epsilon=0$). It shows the average behavior of each market states of S\&P 500 over a period of 13 years (2006-2018). (f) Dots (MDS (CWOE)) in the plot show the multidimensional scaling map of CWOE using mean correlations martix as $\overline{S1}$ and constructing three times bigger ensemble than $S1$ market state of the S\&P 500 market with same noise-suppression $\epsilon=0.7$. Red circles ($k$-means (CWOE)) in plot show the points of the $S1$ cluster of $k$-means clustering performed on CWOE (dots). Blue squares (k-means (USA)) in the plot show the $k$-means clustering on the emperical data of S\&P 500. $k$-means clustering on the CWOE and S\&P 500 data shows qualitatively similar behavior. (g), (h), (i), and (j) show the same using mean correlation matrices $\overline{S2}, \overline{S3}$, $\overline{S4}$, and $\overline{S5}$ states, respectively.}\label{mean_MS}
\end{figure*}

\begin{table}[hbt!]
\caption{List of 350 stocks of USA S\&P 500 considered for the analysis present over the period of 2006-2019.}\label{table:usa_2019}
\begin{tabular}{|p{1cm}|p{1.5cm}|p{6cm}|p{5cm}|p{1.2cm}|}
\hline
\textbf{S.No.} & \textbf{Code} & \textbf{Company Name} & \textbf{Sector} & \textbf{Abbrv} \\ \hline
1     & AAP   & Advance Auto Parts                     & Consumer Discretionary     & CD     \\ \hline
2     & AMZN  & Amazon.com Inc.                        & Consumer Discretionary     & CD     \\ \hline
3     & AZO   & AutoZone Inc                           & Consumer Discretionary     & CD     \\ \hline
4     & BBY   & Best Buy Co. Inc.                      & Consumer Discretionary     & CD     \\ \hline
5     & BKNG  & Booking Holdings Inc                   & Consumer Discretionary     & CD     \\ \hline
6     & CCL   & Carnival Corp.                         & Consumer Discretionary     & CD     \\ \hline
7     & CMCSA & Comcast Corp.                          & Consumer Discretionary     & CD     \\
\hline
\end{tabular}
\end{table}
\begin{table}[]
\centering
\begin{tabular}{|p{1cm}|p{1.5cm}|p{6cm}|p{5cm}|p{1.2cm}|}
\hline
\textbf{S.No.} & \textbf{Code} & \textbf{Company Name} & \textbf{Sector} & \textbf{Abbrv} \\ \hline

51    & WYNN  & Wynn Resorts Ltd                       & Consumer Discretionary     & CD     \\ \hline
52    & YUM   & Yum! Brands Inc                        & Consumer Discretionary     & CD     \\ \hline
53    & ADM   & Archer-Daniels-Midland Co              & Consumer Staples           & CS     \\ \hline
54    & CAG   & Conagra Brands                         & Consumer Staples           & CS     \\ \hline
55    & CHD   & Church \& Dwight                       & Consumer Staples           & CS     \\ \hline
56    & CL    & Colgate-Palmolive                      & Consumer Staples           & CS     \\ \hline
57    & CLX   & The Clorox Company                     & Consumer Staples           & CS     \\
\hline
\end{tabular}
\end{table}
\begin{table}[htb!]
\centering
\begin{tabular}{|p{1cm}|p{1.5cm}|p{6cm}|p{5cm}|p{1.2cm}|}
\hline
\textbf{S.No.} & \textbf{Code} & \textbf{Company Name} & \textbf{Sector} & \textbf{Abbrv} \\ \hline
103   & AIG   & American International Group, Inc.     & Financials                 & FN     \\ \hline
104   & AIZ   & Assurant Inc.                          & Financials                 & FN     \\ \hline
105   & AJG   & Arthur J. Gallagher \& Co.             & Financials                 & FN     \\ \hline
106   & AMG   & Affiliated Managers Group Inc          & Financials                 & FN     \\ \hline
107   & AMP   & Ameriprise Financial                   & Financials                 & FN     \\ \hline
108   & AON   & Aon plc                                & Financials                 & FN     \\ \hline
109   & AXP   & American Express Co                    & Financials                 & FN     \\
\hline

\end{tabular}
\end{table}
\begin{table}[]
\centering
\begin{tabular}{|p{1cm}|p{1.5cm}|p{6cm}|p{5cm}|p{1.2cm}|}
\hline
\textbf{S.No.} & \textbf{Code} & \textbf{Company Name} & \textbf{Sector} & \textbf{Abbrv} \\ \hline
156   & A     & Agilent Technologies Inc               & Health Care                & HC     \\ \hline
157   & ABC   & AmerisourceBergen Corp                 & Health Care                & HC     \\ \hline
158   & ABT   & Abbott Laboratories                    & Health Care                & HC     \\ \hline
159   & AGN   & Allergan, Plc                          & Health Care                & HC     \\ \hline
160   & ALGN  & Align Technology                       & Health Care                & HC     \\ \hline
161   & ALXN  & Alexion Pharmaceuticals                & Health Care                & HC     \\ \hline
162   & AMGN  & Amgen Inc.                             & Health Care                & HC     \\
\hline

\end{tabular}
\end{table}
\begin{table}[]
\centering
\begin{tabular}{|p{1cm}|p{1.5cm}|p{6cm}|p{5cm}|p{1.2cm}|}
\hline
\textbf{S.No.} & \textbf{Code} & \textbf{Company Name} & \textbf{Sector} & \textbf{Abbrv} \\ \hline

207   & AOS   & A.O. Smith Corp                        & Industrials                & ID     \\ \hline
208   & ARNC  & Arconic Inc.                           & Industrials                & ID     \\ \hline
209   & BA    & Boeing Company                         & Industrials                & ID     \\ \hline
210   & CAT   & Caterpillar Inc.                       & Industrials                & ID     \\ \hline
211   & CHRW  & C. H. Robinson Worldwide               & Industrials                & ID     \\ \hline
212   & CMI   & Cummins Inc.                           & Industrials                & ID     \\ \hline
213   & CSX   & CSX Corp.                              & Industrials                & ID     \\
\hline

\end{tabular}
\end{table}
\begin{table}[]
\centering
\begin{tabular}{|p{1cm}|p{1.5cm}|p{6cm}|p{5cm}|p{1.2cm}|}
\hline
\textbf{S.No.} & \textbf{Code} & \textbf{Company Name} & \textbf{Sector} & \textbf{Abbrv} \\ \hline
259   & AKAM  & Akamai Technologies Inc                & Information Technology     & IT     \\ \hline
260   & AMAT  & Applied Materials Inc.                 & Information Technology     & IT     \\ \hline
261   & AMD   & Advanced Micro Devices Inc             & Information Technology     & IT     \\ \hline
262   & ANSS  & ANSYS                                  & Information Technology     & IT     \\ \hline
263   & APH   & Amphenol Corp                          & Information Technology     & IT     \\ \hline
264   & ATVI  & Activision Blizzard                    & Information Technology     & IT     \\ \hline
265   & CDNS  & Cadence Design Systems                 & Information Technology     & IT     \\
\hline
\end{tabular}
\end{table}
\begin{table}[]
\centering
\begin{tabular}{|p{1cm}|p{1.5cm}|p{6cm}|p{5cm}|p{1.2cm}|}
\hline
\textbf{S.No.} & \textbf{Code} & \textbf{Company Name} & \textbf{Sector} & \textbf{Abbrv} \\ \hline
310   & CF    & CF Industries Holdings Inc             & Materials                  & MT     \\ \hline
311   & ECL   & Ecolab Inc.                            & Materials                  & MT     \\ \hline
312   & FCX   & Freeport-McMoRan Inc.                  & Materials                  & MT     \\ \hline
313   & FMC   & FMC Corporation                        & Materials                  & MT     \\ \hline
314   & IFF   & Intl Flavors \& Fragrances             & Materials                  & MT     \\ \hline
315   & IP    & International Paper                    & Materials                  & MT     \\ \hline
316   & MOS   & The Mosaic Company                     & Materials                  & MT     \\
\hline
\end{tabular}
\end{table}
\begin{table}[h!]
\caption{List of 156 stocks of JPN Nikkei 225 considered for the analysis present over the period of 2006-2019.}
\label{table:jpn_2019}
\begin{tabular}{|p{1cm}|p{1.5cm}|p{8cm}|p{4cm}|p{1.2cm}|}
\hline
\textbf{S.No.} & \textbf{Code} & \textbf{Company Name} & \textbf{Sector} & \textbf{Abbrv} \\ \hline
1     & 1332.T & NIPPON SUISAN KAISHA, LTD.               & Consumer Goods               & CG    \\ \hline
2     & 2002.T & NISSHIN SEIFUN GROUP INC.                & Consumer Goods               & CG    \\ \hline
3     & 2282.T & NH FOODS LTD.                            & Consumer Goods               & CG    \\ \hline
4     & 2502.T & ASAHI GROUP HOLDINGS, LTD.               & Consumer Goods               & CG    \\ \hline
5     & 2503.T & KIRIN HOLDINGS CO., LTD.                 & Consumer Goods               & CG    \\ \hline
6     & 2531.T & TAKARA HOLDINGS INC.                     & Consumer Goods               & CG    \\ \hline
7     & 2801.T & KIKKOMAN CORP.                           & Consumer Goods               & CG    \\
\hline

\end{tabular}
\end{table}
\begin{table}[]
\centering
\begin{tabular}{|p{1cm}|p{1.5cm}|p{8cm}|p{4cm}|p{1.2cm}|}
\hline
\textbf{S.No.} & \textbf{Code} & \textbf{Company Name} & \textbf{Sector} & \textbf{Abbrv} \\ \hline
52    & 8601.T & DAIWA SECURITIES GROUP INC.              & Financials                   & FN    \\ \hline
53    & 8604.T & NOMURA HOLDINGS, INC.                    & Financials                   & FN    \\ \hline
54    & 2768.T & SOJITZ CORP.                             & Materials                    & MT    \\ \hline
55    & 3101.T & TOYOBO CO., LTD.                         & Materials                    & MT    \\ \hline
56    & 3103.T & UNITIKA, LTD.                            & Materials                    & MT    \\ \hline
57    & 3401.T & TEIJIN LTD.                              & Materials                    & MT    \\ \hline
58    & 3402.T & TORAY INDUSTRIES, INC.                   & Materials                    & MT    \\
\hline

\end{tabular}
\end{table}
\begin{table}[]
\centering
\begin{tabular}{|p{1cm}|p{1.5cm}|p{8cm}|p{4cm}|p{1.2cm}|}
\hline
\textbf{S.No.} & \textbf{Code} & \textbf{Company Name} & \textbf{Sector} & \textbf{Abbrv} \\ \hline
105   & 4503.T & ASTELLAS PHARMA INC.                     & Technology                   & TC    \\ \hline
106   & 4506.T & SUMITOMO DAINIPPON PHARMA CO., LTD.      & Technology                   & TC    \\ \hline
107   & 4507.T & SHIONOGI \& CO., LTD.                    & Technology                   & TC    \\ \hline
108   & 4519.T & CHUGAI PHARMACEUTICAL CO., LTD.          & Technology                   & TC    \\ \hline
109   & 4543.T & TERUMO CORP.                             & Technology                   & TC    \\ \hline
110   & 4902.T & KONICA MINOLTA, INC.                     & Technology                   & TC    \\ \hline
111   & 6479.T & MINEBEA MITSUMI INC.                     & Technology                   & TC    \\
\hline
\end{tabular}
\end{table}

\begin{table}[]
\caption{List of 368 stocks of USA S\&P 500 considered for the analysis present over the period of 2006-2018.}\label{table:usa_2018}
\centering
\begin{tabular}{|p{1cm}|p{1.5cm}|p{7cm}|p{5cm}|p{1.2cm}|}
\hline
\textbf{S.No.} & \textbf{Code} & \textbf{Company Name} & \textbf{Sector} & \textbf{Abbrv} \\ \hline
1     & AAP   & Advance Auto Parts                     & Consumer Discretionary     & CD     \\ \hline
2     & AMZN  & Amazon.com Inc.                        & Consumer Discretionary     & CD     \\ \hline
3     & AZO   & AutoZone Inc                           & Consumer Discretionary     & CD     \\ \hline
4     & BBY   & Best Buy Co. Inc.                      & Consumer Discretionary     & CD     \\ \hline
5     & BKNG  & Booking Holdings Inc                   & Consumer Discretionary     & CD     \\ \hline
6     & CBS   & CBS Corp.                              & Consumer Discretionary     & CD     \\ \hline
7     & CCL   & Carnival Corp.                         & Consumer Discretionary     & CD     \\
\hline
\end{tabular}
\end{table}
\begin{table}[]
\centering
\begin{tabular}{|p{1cm}|p{1.5cm}|p{7cm}|p{5cm}|p{1.2cm}|}
\hline
\textbf{S.No.} & \textbf{Code} & \textbf{Company Name} & \textbf{Sector} & \textbf{Abbrv} \\ \hline
52    & TJX   & TJX Companies Inc.                     & Consumer Discretionary     & CD     \\ \hline
53    & TPR   & Tapestry, Inc.                         & Consumer Discretionary     & CD     \\ \hline
54    & UAA   & Under Armour Class A                   & Consumer Discretionary     & CD     \\ \hline
55    & VFC   & V.F. Corp.                             & Consumer Discretionary     & CD     \\ \hline
56    & VIAB  & Viacom Inc.                            & Consumer Discretionary     & CD     \\ \hline
57    & WHR   & Whirlpool Corp.                        & Consumer Discretionary     & CD     \\ \hline
58    & WYNN  & Wynn Resorts Ltd                       & Consumer Discretionary     & CD     \\
\hline

\end{tabular}
\end{table}
\begin{table}[]
\centering
\begin{tabular}{|p{1cm}|p{1.5cm}|p{7cm}|p{5cm}|p{1.2cm}|}
\hline
\textbf{S.No.} & \textbf{Code} & \textbf{Company Name} & \textbf{Sector} & \textbf{Abbrv} \\ \hline
105   & RRC   & Range Resources Corp.                  & Energy                     & EG     \\ \hline
106   & SLB   & Schlumberger Ltd.                      & Energy                     & EG     \\ \hline
107   & VLO   & Valero Energy                          & Energy                     & EG     \\ \hline
108   & WMB   & Williams Cos.                          & Energy                     & EG     \\ \hline
109   & XEC   & Cimarex Energy                         & Energy                     & EG     \\ \hline
110   & XOM   & Exxon Mobil Corp.                      & Energy                     & EG     \\ \hline
111   & AFL   & AFLAC Inc                              & Financials                 & FN     \\
\hline

\end{tabular}
\end{table}
\begin{table}[]
\centering
\begin{tabular}{|p{1cm}|p{1.5cm}|p{7cm}|p{5cm}|p{1.2cm}|}
\hline
\textbf{S.No.} & \textbf{Code} & \textbf{Company Name} & \textbf{Sector} & \textbf{Abbrv} \\ \hline
157   & STI   & SunTrust Banks                         & Financials                 & FN     \\ \hline
158   & STT   & State Street Corp.                     & Financials                 & FN     \\ \hline
159   & TMK   & Torchmark Corp.                        & Financials                 & FN     \\ \hline
160   & TROW  & T. Rowe Price Group                    & Financials                 & FN     \\ \hline
161   & TRV   & The Travelers Companies Inc.           & Financials                 & FN     \\ \hline
162   & UNM   & Unum Group                             & Financials                 & FN     \\ \hline
163   & USB   & U.S. Bancorp                           & Financials                 & FN     \\
\hline

\end{tabular}
\end{table}
\begin{table}[]
\centering
\begin{tabular}{|p{1cm}|p{1.5cm}|p{7cm}|p{5cm}|p{1.2cm}|}
\hline
\textbf{S.No.} & \textbf{Code} & \textbf{Company Name} & \textbf{Sector} & \textbf{Abbrv} \\ \hline
209   & UNH   & United Health Group Inc.               & Health Care                & HC     \\ \hline
210   & VRTX  & Vertex Pharmaceuticals Inc             & Health Care                & HC     \\ \hline
211   & WAT   & Waters Corporation                     & Health Care                & HC     \\ \hline
212   & XRAY  & Dentsply Sirona                        & Health Care                & HC     \\ \hline
213   & ZBH   & Zimmer Biomet Holdings                 & Health Care                & HC     \\ \hline
214   & AAL   & American Airlines Group                & Industrials                & ID     \\ \hline
215   & ALK   & Alaska Air Group Inc                   & Industrials                & ID     \\
\hline

\end{tabular}
\end{table}
\begin{table}[]
\centering
\begin{tabular}{|p{1cm}|p{1.5cm}|p{7cm}|p{5cm}|p{1.2cm}|}
\hline
\textbf{S.No.} & \textbf{Code} & \textbf{Company Name} & \textbf{Sector} & \textbf{Abbrv} \\ \hline
262   & UNP   & Union Pacific                          & Industrials                & ID     \\ \hline
263   & UPS   & United Parcel Service                  & Industrials                & ID     \\ \hline
264   & URI   & United Rentals, Inc.                   & Industrials                & ID     \\ \hline
265   & UTX   & United Technologies                    & Industrials                & ID     \\ \hline
266   & WM    & Waste Management Inc.                  & Industrials                & ID     \\ \hline
267   & AAPL  & Apple Inc.                             & Information Technology     & IT     \\ \hline
268   & ACN   & Accenture plc                          & Information Technology     & IT     \\
\hline
\end{tabular}
\end{table}
\begin{table}[]
\centering
\begin{tabular}{|p{1cm}|p{1.5cm}|p{7cm}|p{5cm}|p{1.2cm}|}
\hline
\textbf{S.No.} & \textbf{Code} & \textbf{Company Name} & \textbf{Sector} & \textbf{Abbrv} \\ \hline
315   & RHT   & Red Hat Inc.                           & Information Technology     & IT     \\ \hline
316   & SNPS  & Synopsys Inc.                          & Information Technology     & IT     \\ \hline
317   & STX   & Seagate Technology                     & Information Technology     & IT     \\ \hline
318   & SWKS  & Skyworks Solutions                     & Information Technology     & IT     \\ \hline
319   & SYMC  & Symantec Corp.                         & Information Technology     & IT     \\ \hline
320   & TSS   & Total System Services                  & Information Technology     & IT     \\ \hline
321   & TTWO  & Take-Two Interactive                   & Information Technology     & IT     \\
\hline
\end{tabular}
\end{table}
\begin{table}[]
\centering
\begin{tabular}{|p{1cm}|p{1.5cm}|p{7cm}|p{5cm}|p{1.2cm}|}
\hline
\textbf{S.No.} & \textbf{Code} & \textbf{Company Name} & \textbf{Sector} & \textbf{Abbrv} \\ \hline

365   & SO    & Southern Co.                           & Utilities                  & UT     \\ \hline
366   & SRE   & Sempra Energy                          & Utilities                  & UT     \\ \hline
367   & WEC   & Wec Energy Group Inc                   & Utilities                  & UT     \\ \hline
368   & XEL   & Xcel Energy Inc                        & Utilities                  & UT     \\ \hline
\end{tabular}
\end{table}
\begin{table}[]
\caption{List of 173 stocks of JPN Nikkei 225 considered for the analysis present over the period of 2006-2018.}\label{table:jpn_2018}
\centering
\begin{tabular}{|p{1cm}|p{1.5cm}|p{7cm}|p{5cm}|p{1.2cm}|}
\hline
\textbf{S.No.} & \textbf{Code} & \textbf{Company Name} & \textbf{Sector} & \textbf{Abbrv} \\ \hline
1     & 1332.T & NIPPON SUISAN KAISHA, LTD.               & Consumer Goods               & CG     \\ \hline
2     & 2002.T & NISSHIN SEIFUN GROUP INC.                & Consumer Goods               & CG     \\ \hline
3     & 2282.T & NH FOODS LTD.                            & Consumer Goods               & CG     \\ \hline
4     & 2501.T & SAPPORO HOLDINGS LTD.                    & Consumer Goods               & CG     \\ \hline
5     & 2502.T & ASAHI GROUP HOLDINGS, LTD.               & Consumer Goods               & CG     \\ \hline
6     & 2503.T & KIRIN HOLDINGS CO., LTD.                 & Consumer Goods               & CG     \\ \hline
7     & 2531.T & TAKARA HOLDINGS INC.                     & Consumer Goods               & CG     \\
\hline
\end{tabular}
\end{table}
\begin{table}[]
\centering
\begin{tabular}{|p{1cm}|p{1.5cm}|p{7cm}|p{5cm}|p{1.2cm}|}
\hline
\textbf{S.No.} & \textbf{Code} & \textbf{Company Name} & \textbf{Sector} & \textbf{Abbrv} \\ \hline
46    & 7004.T & HITACHI ZOSEN CORP.                      & Capital Goods/Others         & CP     \\ \hline
47    & 7011.T & MITSUBISHI HEAVY IND., LTD.              & Capital Goods/Others         & CP     \\ \hline
48    & 7012.T & KAWASAKI HEAVY IND., LTD.                & Capital Goods/Others         & CP     \\ \hline
49    & 7013.T & IHI CORP.                                & Capital Goods/Others         & CP     \\ \hline
50    & 7911.T & TOPPAN PRINTING CO., LTD.                & Capital Goods/Others         & CP     \\ \hline
51    & 7912.T & DAI NIPPON PRINTING CO., LTD.            & Capital Goods/Others         & CP     \\ \hline
52    & 7951.T & YAMAHA CORP.                             & Capital Goods/Others         & CP     \\
\hline
\end{tabular}
\end{table}
\begin{table}[]
\centering
\begin{tabular}{|p{1cm}|p{1.5cm}|p{7cm}|p{5cm}|p{1.2cm}|}
\hline
\textbf{S.No.} & \textbf{Code} & \textbf{Company Name} & \textbf{Sector} & \textbf{Abbrv} \\ \hline
96    & 5401.T & NIPPON STEEL CORP.                       & Materials                    & MT     \\ \hline
97    & 5406.T & KOBE STEEL, LTD.                         & Materials                    & MT     \\ \hline
98    & 5411.T & JFE HOLDINGS, INC.                       & Materials                    & MT     \\ \hline
99    & 5541.T & PACIFIC METALS CO., LTD.                 & Materials                    & MT     \\ \hline
100   & 5703.T & NIPPON LIGHT METAL HOLDINGS CO., LTD.    & Materials                    & MT     \\ \hline
101   & 5706.T & MITSUI MINING \& SMELTING CO.            & Materials                    & MT     \\ \hline
102   & 5707.T & TOHO ZINC CO., LTD.                      & Materials                    & MT     \\
\hline
\end{tabular}
\end{table}
\begin{table}[]
\centering
\begin{tabular}{|p{1cm}|p{1.5cm}|p{7cm}|p{5cm}|p{1.2cm}|}
\hline
\textbf{S.No.} & \textbf{Code} & \textbf{Company Name} & \textbf{Sector} & \textbf{Abbrv} \\ \hline
149   & 7261.T & MAZDA MOTOR CORP.                        & Technology                   & TC     \\ \hline
150   & 7269.T & SUZUKI MOTOR CORP.                       & Technology                   & TC     \\ \hline
151   & 7270.T & SUBARU CORP.                             & Technology                   & TC     \\ \hline
152   & 7272.T & YAMAHA MOTOR CO., LTD.                   & Technology                   & TC     \\ \hline
153   & 7731.T & NIKON CORP.                              & Technology                   & TC     \\ \hline
154   & 7735.T & SCREEN HOLDINGS CO., LTD.                & Technology                   & TC     \\ \hline
155   & 7752.T & RICOH CO., LTD.                          & Technology                   & TC     \\
\hline
\end{tabular}
\end{table}
\end{document}